\documentclass[10pt]{article}

\usepackage{authblk}
\usepackage{natbib}
\usepackage{mathtools}
\usepackage{amsmath}
\usepackage{amsthm}
\usepackage{amssymb}
\usepackage{amsbsy}
\usepackage{amsfonts}
\usepackage{amscd}
\usepackage{mathrsfs}
\usepackage{bm}
\usepackage{breqn}
\usepackage{color, soul}
\usepackage{enumerate}
\usepackage{empheq}
\usepackage{rotating}
\usepackage{ftnxtra}
\usepackage{fnpos}
\usepackage[titletoc,toc,title]{appendix}
\usepackage{euscript}
\usepackage{graphicx}
\usepackage{epsfig}
\usepackage{epstopdf}
\DeclareGraphicsExtensions{.pdf,.png,.jpg,.eps}
\usepackage{pstool}
\usepackage{upgreek}
\usepackage{mathrsfs}

\usepackage{psfrag}

\numberwithin{equation}{section}

\theoremstyle{plain}	
\newtheorem{thm}{Theorem}[section]

\newtheorem{prop}[thm]{Proposition}
\newtheorem*{prop*}{Proposition} 
\theoremstyle{definition}	
\newtheorem{defi}[thm]{Definition}

\usepackage{graphicx}
\usepackage{lipsum}

\usepackage{caption}
\usepackage{subcaption}

\usepackage{hyperref}
\hypersetup{colorlinks=true, linkcolor=blue}
\hypersetup{colorlinks=true,citecolor=blue}

\DeclareMathAlphabet{\mathpzc}{OT1}{pzc}{m}{it}

\usepackage{amsmath, amsthm, amssymb}

\usepackage{cleveref}

\DeclarePairedDelimiter\abs{\lvert}{\rvert}

\makeatletter
\newsavebox{\@brx}
\newcommand{\llangle}[1][]{\savebox{\@brx}{\(\m@th{#1\langle}\)}%
  \mathopen{\copy\@brx\mkern2mu\kern-0.9\wd\@brx\usebox{\@brx}}}
\newcommand{\rrangle}[1][]{\savebox{\@brx}{\(\m@th{#1\rangle}\)}%
  \mathclose{\copy\@brx\mkern2mu\kern-0.9\wd\@brx\usebox{\@brx}}}%
\let\oldabs\abs
\def\abs{\@ifstar{\oldabs}{\oldabs*}}
\makeatother

\usepackage{accents}

    {\end{bmatrix}}%

\usepackage{enumitem}

\usepackage[utf8]{inputenc}
\usepackage{dutchcal}
\usepackage{multirow}

\usepackage{xcolor}

\usepackage{setspace}

\begin{document}

\title{\textbf{On Universal Deformations and Material Preferred Directions in Anisotropic Cauchy Elasticity}}

\author[1]{Seyedemad Motaghian\thanks{e-mail: em.motaghian@ut.ac.ir}}
\author[2,3]{Arash Yavari\thanks{Corresponding author, e-mail: arash.yavari@ce.gatech.edu}}

\affil[1]{\small \textit{College of Engineering, University of Tehran, Tehran, Iran}}
\affil[2]{\small \textit{School of Civil and Environmental Engineering, Georgia Institute of Technology, Atlanta, GA 30332, USA}}
\affil[3]{\small \textit{The George W. Woodruff School of Mechanical Engineering, Georgia Institute of Technology, Atlanta, GA 30332, USA}}

\maketitle

\begin{abstract}
In this paper we study universal deformations in anisotropic Cauchy elasticity. We show that the universality constraints of hyperelasticity and Cauchy elasticity  for transversely isotropic, orthotropic, and monoclinic solids are equivalent. This implies that for each of these symmetry classes the universal deformations and the corresponding universal material preferred directions of hyperelastic and Cauchy elastic solids are identical. This is consistent with previous findings for isotropic solids.
Universal deformations and material preferred directions are therefore independent of the existence or absence of a strain energy function.
\end{abstract}

\begin{description}
\item[Keywords:] Universal deformations, Universal material preferred directions,  Cauchy elasticity, hyperelasticity, anisotropic elasticity.
\end{description}

\tableofcontents

\section{Introduction}

A \emph{universal deformation} is a deformation that can be maintained in the absence of body forces for every material in a given class. Equivalently, such a deformation can be supported by boundary tractions alone, independently of the specific constitutive equations within that class—for instance, homogeneous compressible isotropic solids or homogeneous anisotropic solids with prescribed symmetry. In nonlinear elasticity, universal deformations have long played an important role both experimentally \citep{Rivlin1951} and theoretically \citep{Tadmor2012,Goriely2017}.  

The concept of universal deformations was introduced by Jerry Ericksen in two seminal papers \citep{Ericksen1954,Ericksen1955}. In \citep{Ericksen1955}, he proved that in homogeneous compressible isotropic solids, all universal deformations must necessarily be homogeneous. His earlier study of incompressible isotropic solids \citep{Ericksen1954} was motivated by Rivlin’s pioneering work on special classes of deformations \citep{Rivlin1948,Rivlin1949a,Rivlin1949b}. Ericksen also conjectured that deformations with constant principal invariants must be homogeneous, a conjecture later shown to be false by \citet{Fosdick1966}. In fact, the fifth universal family discovered by \citet{SinghPipkin1965} and \citet{KlingbeilShield1966} provides examples of inhomogeneous universal deformations with constant principal invariants. Whether further inhomogeneous universal deformations with constant invariants exist remains unknown.  

Since Ericksen’s original contributions, the study of universal deformations has been extended to more general settings, including inhomogeneous isotropic elasticity \citep{Yavari2021}, anisotropic elasticity \citep{YavariGoriely2021,Yavari2022Universal}, and anelasticity \citep{YavariGoriely2016,Goodbrake2020}. In the linear theory, the analogue of universal deformations is that of \emph{universal displacements} \citep{Truesdell1966,Gurtin1972,Yavari2020,YavariGoriely2022}. In compressible anisotropic linear elasticity, these were classified for all eight symmetry classes in \citep{Yavari2020}, where it was shown that the higher the material symmetry, the larger the space of universal displacements. Thus, isotropic solids admit the largst set of universal displacements, while triclinic solids admit the smallest. This classification was later extended to inhomogeneous solids \citep{YavariGoriely2022} and to linear anelasticity \citep{Yavari2022Anelastic-Universality}.  

More recently, universal deformations have been studied in Cauchy elasticity, a broader framework that contains hyperelasticity as a special case but does not presuppose the existence of an energy function \citep{YavariGoriely2025}. For inhomogeneous isotropic Cauchy elastic solids, it was shown that the sets of universal deformations and universal inhomogeneities coincide with those of Green elasticity \citep{Yavari2024}. The universal displacements of anisotropic linear Cauchy elastic solids have also been systematically analyzed \citep{YavariSfyris2025}. Interestingly, despite the greater generality of Cauchy elasticity, for each of the eight symmetry classes the resulting set of universal displacements is identical to that of linear hyperelasticity.  

Universal deformations have also been examined in the setting of implicit elasticity, where constitutive relations take the form $\boldsymbol{\mathcal{F}}(\boldsymbol{\sigma},\mathbf{b})=\mathbf{0}$, with $\boldsymbol{\sigma}$ the Cauchy stress and $\mathbf{b}$ the Finger tensor \citep{Morgan1966,Rajagopal2003,Rajagopal2007}. It has been shown that in compressible isotropic implicit elasticity all universal deformations are homogeneous \citep{Yavari2024ImplicitElasticity}. However, unlike in Green or Cauchy elasticity, not every homogeneous deformation is admissible. The resulting set of universal deformations is therefore material-dependent, though always contained in the class of homogeneous deformations. This distinction underscores how the constitutive structure influences universality.  

A frequently encountered class of solids with internal constraints in engineering applications is that of compressible materials reinforced by inextensible fibers \citep{Pipkin1971,Pipkin1974,Pipkin1979,Erdemir2007}. This idealization captures many natural and engineered materials consisting of a soft matrix reinforced by stiff fibers. The literature on universal deformations in such solids is limited. \citet{Beskos1972} studied homogeneous compressible isotropic solids with inextensible fibers and showed that certain subsets of Families~$1$–$4$ of universal deformations remain universal for specific fiber distributions; all are homogeneous except for the shearing of a circular tube with circumferential fibers. A similar study for incompressible isotropic hyperelastic solids was presented in \citep{Beskos1973}, and universal relations for both classes were discussed in \citep{Saccomandi2002}. \citet{Beatty1978,Beatty1989} identified all fiber distributions in homogeneous compressible isotropic solids with a single family of inextensible fibers for which homogeneous deformations are universal, proving that only three such distributions exist and that in each case the fibers remain straight in both the deformed and reference configurations.
In a recent study, universal deformations in compressible isotropic Cauchy elastic solids reinforced with a single family of inextensible fibers were systematically characterized \citep{Yavari2025FibersCauchy}. This work established the first systematic classification of such deformations, thereby extending the classical results of Beskos and Beatty to the broader framework of Cauchy elasticity.

The purpose of the present work is to study universal deformations and universal material-preferred directions in anisotropic Cauchy elasticity. We show that for transverse isotropy, orthotropy, and monoclinic elasticity in both compressible and incompressible cases, the sets of universal deformations and universal preferred directions coincide exactly with those of the corresponding anisotropic hyperelasticity. This shows that, even within the more general framework of Cauchy elasticity, universality in these classes is governed entirely by material symmetry.  

This paper is organized as follows. A concise overview of nonlinear elasticity is presented in \S\ref{Nonlinear Elast}. In \S\ref{Iso-Elasticity}, the equivalence between the universality constraints in hyperelasticity and those in Cauchy elasticity for homogeneous compressible and incompressible isotropic solids is examined. The same problem is addressed for homogeneous compressible and incompressible transversely isotropic solids in \S\ref{T iso-Elasticity}, for orthotropic solids in \S\ref{Orthotropic-Elasticity}, and for monoclinic solids in \S\ref{Mono-Elasticity}. Conclusions are given in \S\ref{Sec:Conclusions}.

\section{Nonlinear elasticity} \label{Nonlinear Elast}
Within the framework of nonlinear anelasticity, an undeformed body $\mathscr{B}$ is identified with a Riemannian manifold $(\mathscr{B},\mathbf{G})$, where $\mathbf{G}$ is the material metric tensor. A deformation of this body is defined by mapping $\mathscr{B}$ to $(\mathcal{S},\mathbf{g})$, i.e., $\varphi \,: \,\mathscr{B} \to \mathcal{S}$, where $\mathcal{S}$ is also a Riemannian manifold. We assume that $(\mathcal{S},\mathbf{g})$ is the Euclidean three-dimensional space (or $\mathbb{R}^3$), and $(\mathscr{B},\mathbf{G})$ is an embedded submanifold of $\mathbb{R}^3$. The tangent map of $\varphi$ is the deformation gradient, $\mathbf{F}=T \,\varphi$, which is a linear map $\mathbf{F}(\mathbf{X})\,:\,T_{\mathbf{X}} \mathscr{B}  \to T_{\varphi(\mathbf{X})} \mathcal{S}$ at each material point $\mathbf{X} \in \mathscr{B}$. The deformation gradient tensor and its transpose $\mathbf{F}^{ \mathsf{T}}$ are expressed in components as
\begin{equation} \label{eq:deformation gradient}
F^a{}_{A}=\frac{\partial{\varphi}^a}{\partial{\mathbf{X}}^A}(\mathbf{X}) \,,
\end{equation}
and
\begin{equation} \label{eq:deformation gradient transpose}
\left(F^{\mathsf{T}}(\mathbf{X})\right)^A{}_{a}=g_{ab} \,(\mathbf{x}) \,F^b{}_{B}\,(\mathbf{X}) \,G^{AB} \,(\mathbf{X}) \,,
\end{equation}
where $\{X^A\}$ and $\{x^a\}$ are coordinate charts on $\mathscr{B}$ and $\mathcal{S}$, respectively. The right Cauchy–Green deformation tensor is defined as $\mathbf{C}(\mathbf{X})=\mathbf{F}^{\mathsf{T}}(\mathbf{X}) \,\mathbf{F}(\mathbf{X})$. The associated tensor $\mathbf{C}^\flat$, where $\flat$ denotes the flat operator induced by the metric tensor (index lowering), is the pull-back of the metric $\mathbf{g}$ by the deformation, i.e., $\mathbf{C}^\flat=\varphi^* \mathbf{g}$. In components, it reads
\begin{equation} \label{eq:Cb}
C_{AB}=g_{ab} \,F^a{}_{A} \, F^b{}_{B} \,.
\end{equation}
The Eulerian representation of $\mathbf{C}^{\flat}$ is $\mathbf{c}^\flat=\varphi_* \mathbf{G}$ which is the push-forward of the metric $\mathbf{G}$ by $\varphi$ and has components $c_{ab}=F^{-A}{}_a \,F^{-B}{}_b \,G_{AB}$, where $F^{-A}{}_a$ are components of $\mathbf{F}^{-1}$. The left Cauchy–Green deformation tensor is defined as the pull-back of the associated tensor $\mathbf{g}^\sharp$ to the reference configuration, i.e., $\mathbf{B}^\sharp = \varphi^*(\mathbf{g}^\sharp)$, where $\sharp$ is the sharp operator induced from the metric tensor (index raising). In components, it is written as $B^{AB}=F^{-A}{}_a \,F^{-B}{}_b \,g^{ab}$. Equivalently, the spatial representation of $\mathbf{B}^\sharp$ is $\mathbf{b}^\sharp=\varphi_* \,(\mathbf{G}^\sharp)$, which is called the Finger deformation tensor and is expressed in components as
\begin{equation} \label{eq:Finger deformation tensor}
b^{ab}=F^a{}_A \,F^b{}_B \,G^{AB}\,.
\end{equation}
Note that $\mathbf{c}=\mathbf{b}^{-1}$. The principal invariants of $\mathbf{b}$ or $\mathbf{C}$ (the two tensors have the same principal invariants), denoted by $I_1$, $I_2$, and $I_3$, are defined as
\begin{equation} \label{eq:principal invariants}
I_1=\operatorname{tr}\mathbf{b}=b^{ab} \,g_{ab} \,,\qquad
I_2=\frac{1}{2} \,\left(I_1^2-\operatorname{tr} \mathbf{b}^2\right)=\frac{1}{2} \,\left(I_1^2-b^{ab} \,b^{cd} \,g_{ac} \,g_{bd}\right) \,,\qquad
I_3=\operatorname{det} \mathbf{b}\,.
\end{equation}

The first and second Piola--Kirchhoff stresses are defined by
\begin{equation}
	\mathbf{P} = J\,\boldsymbol{\sigma}\,\mathbf{F}^{-\star}\,, \qquad
	\mathbf{S} = \mathbf{F}^{-1}\mathbf{P}
	= J\,\mathbf{F}^{-1}\,\boldsymbol{\sigma}\,\mathbf{F}^{-\star}\,,
\end{equation}
where $\boldsymbol{\sigma}$ is the Cauchy stress, $\mathbf{P}$ is the first Piola--Kirchhoff stress, and $\mathbf{S}$ is the second Piola--Kirchhoff stress. The volume elements $dv$ and $dV$ in the deformed and undeformed configurations, respectively, are related as $dv=J\,dV$, where $J$ is the Jacobian of deformation and is defined as
\begin{equation} \label{eq:Jacobian}
	J=\sqrt{\frac{\det\mathbf{g}}{\det\mathbf{G}}}\,\det\mathbf{F}\,.
\end{equation}
In components these read
\begin{equation} \label{eq:First piola stress}
	P^{aA} = J\,\sigma^{ab}\,F^{-A}{}_b\,,
\end{equation}
and
\begin{equation} \label{eq:Second piola stress}
	S^{AB} = F^{-A}{}_a\,P^{aB}
	= J\,F^{-A}{}_a\,\sigma^{ab}\,F^{-B}{}_b\,.
\end{equation}
In the absence of body forces, the equilibrium equations in the current configuration are expressed in terms of the Cauchy stress as
\begin{equation} \label{eq:equilibrium}
{\sigma^{ab}}_{|b}=0\,,
\end{equation}
where $(\cdot)_{|a}$ denotes covariant derivative with respect to the Levi-Civita connection of the ambient space metric $\mathbf{g}$.

An anisotropic hyperelastic solid is characterized by an energy function (per unit undeformed volume) that takes the following functional form
\begin{equation} \label{eq:Energy func}
W=\hat{W} \,(\mathbf{C}^\flat,\mathbf{G},\zeta_1,\cdots,\zeta_n) \,,
\end{equation}
where $\zeta_i, \,i=1,\cdots,n$ are the structural tensors that describe the material symmetry group of the solid. By Hilbert’s theorem, one can write $W=W \,(X,I_1,\cdots,I_m)$, where $I_1,\cdots,I_m$ are basic invariants that form an integrity basis for the set of tensors given in \eqref{eq:Energy func}. However, a Cauchy elastic solid does not necessarily have a strain-energy function. For Cauchy elastic solids, the stress at any given material point depends explicitly on the strain at that point  \citep{Cauchy1828,Truesdell1952,TruesdellNoll2004}. Material-frame indifference in Cauchy elasticity implies that the second Piola–Kirchhoff stress is expressed in the following functional form \citep{TruesdellNoll2004} 
\begin{equation} \label{eq:Second piola-general-Cauchy elast1}
\mathbf{S}=\hat{\mathbf{S}} \,(X,\mathbf{C}^\flat,\mathbf{G},\zeta_1,\cdots,\zeta_n) \,,
\end{equation}
or
\begin{equation} \label{eq:Second piola-general-Cauchy elast2}
\mathbf{S}=\mathbf{S} \,(X,I_1,\cdots,I_m) \,.
\end{equation}

\section{Universality Constraints in Isotropic Elasticity} \label{Iso-Elasticity}

\subsection{Compressible isotropic solids} \label{Comp-Iso-Elasticity}
The Cauchy stress for a compressible isotropic Cauchy elastic solid is represented as \citep{Yavari2024}
\begin{equation} \label{eq:cauchy-stress-comp-iso}
\sigma ^{ab}=\alpha_1 \,(I_1,I_2,I_3) \,g^{ab}+\alpha_2 \,(I_1,I_2,I_3) \,b^{ab}+\alpha_3 \,(I_1,I_2,I_3) \,c^{ab} \,,
\end{equation}
where $\alpha_1$, $\alpha_2$ and $\alpha_3$ are response functions. Substituting \eqref{eq:cauchy-stress-comp-iso} into the equilibrium equations \eqref{eq:equilibrium} and using metric compatibility ($g_{ab|c}=0$) gives
\begin{equation} \label{eq:equilibrium-comp-iso-1}
\begin{split}
&\alpha_2 \,{b^{ab}}_{|b}+\alpha_3 \,{c^{ab}}_{|b}+\alpha_{1,1} \,I_{1,b} \,g^{ab}+\alpha_{1,2} \,I_{2,b} \,g^{ab}+\alpha_{1,3} \,I_{3,b} \,g^{ab}+\alpha_{2,1} \,I_{1,b} \,b^{ab}+\alpha_{2,2} \,I_{2,b} \,b^{ab} \\
&+\alpha_{2,3} \,I_{3,b} \,b^{ab}+\alpha_{3,1} \,I_{1,b} \,c^{ab}+\alpha_{3,2} \,I_{2,b} \,c^{ab}+\alpha_{3,3} \,I_{3,b} \,c^{ab}=0 \,,
\end{split}
\end{equation}
in which the following relations have been used
\begin{equation} \label{eq:derivatives-comp-iso}
\alpha_{i,b} = \alpha_{i,j}\, I_{j,b}\,, \qquad (i=1,2,3;\; j=1,2,3) \,,
\end{equation}
and $\alpha_{i,j} = \dfrac{\partial{\alpha_i}}{\partial{I_j}}$ and $I_{j,b} = \dfrac{\partial{I_j}}{\partial{x_b}}$. Since $\alpha_i$ are arbitrary functions, \eqref{eq:equilibrium-comp-iso-1} holds only if the coefficients of $\alpha_i$ and $\alpha_{i,j}$ vanish, and hence,
\begin{equation} \label{eq:constraints-comp-iso}
\begin{split}
&{b^{ab}}_{|b}={c^{ab}}_{|b}=0 \,, \\
&I_{1,b} \,g^{ab}=I_{2,b} \,g^{ab}=I_{3,b} \,g^{ab}=0\,, \\
&I_{1,b} \,b^{ab}=I_{2,b} \,b^{ab}=I_{3,b} \,b^{ab}=0\,, \\
&I_{1,b} \,c^{ab}=I_{2,b} \,c^{ab}=I_{3,b} \,c^{ab}=0\,.
\end{split}
\end{equation}
From \eqref{eq:constraints-comp-iso}$_2$, it follows that $I_1$, $I_2$, and $I_3$ are constant. Consequently, the constraints~\eqref{eq:constraints-comp-iso}$_3$ and \eqref{eq:constraints-comp-iso}$_4$ hold trivially. Taken together with \eqref{eq:constraints-comp-iso}$_1$ and the compatibility equations, this result implies that the universal deformations are homogeneous. In conclusion, the universality constraints for homogeneous compressible isotropic Cauchy elastic solids are the same as those in hyperelasticity as was shown in \citep{Yavari2024}.

\subsection{Incompressible isotropic solids} \label{Incomp-Iso-Elasticity}

As a prelude to our discussion of anisotropic solids, this section examines the equivalence of the universality constraints in Cauchy elasticity and hyperelasticity for homogeneous incompressible isotropic solids.
In \citep{Yavari2024}, it was shown that the universal deformations and inhomogeneities of compressible and incompressible isotropic Cauchy elasticity are identical to those of hyperelasticity. Following the same notation defined in \citep{Yavari2024,Yavari2022Universal}, we aim to provide an alternative proof of this result in this section.

For incompressible isotropic hyperelastic and Cauchy elastic solids, the Cauchy stress tensor $\boldsymbol{\sigma}$ has the following representations \citep{Yavari2024}
\begin{equation} \label{eq:cauchy-stress-incomp-iso}
\begin{split}
&\boldsymbol{\sigma }^{\sharp}=-p \,\mathbf{g}^{\sharp}+2\,{W}_{1} \,(I_1,I_2) \,\mathbf{b}^{\sharp}-2\,{W}_{2} \,(I_1,I_2) \,\mathbf{c}^{\sharp} \,, \\
&\boldsymbol{\sigma }^{\sharp}=-p \,\mathbf{g}^{\sharp}+\alpha_1 \,(I_1,I_2) \,\mathbf{b}^{\sharp}+\alpha_2 \,(I_1,I_2) \,\mathbf{c}^{\sharp} \,, 
\end{split}
\end{equation}
where $p$ is the Lagrange multiplier corresponding to the incompressibility constraint ($I_3=1$), $W_{i} = \dfrac{\partial{{W}}}{\partial {{I}_{i}}}$, and $\alpha_{1}$ and $\alpha_{2}$ are arbitrary response functions in Cauchy elasticity. 

The process of deriving the universality constraints and material preferred directions in hyperelasticity as well as in Cauchy elasticity may be briefly explained as follows. We first substitute the corresponding Cauchy stress into the equilibrium equations~\eqref{eq:equilibrium} to obtain $p_{|a}$ as
\begin{equation} \label{eq:pa-isotropic-hyperelastic}
p_{|b} \,g^{ab}=2 \,\left[W_1 \,b^{ab}-W_2 \,c^{ab}\right]_{|b} \,,
\end{equation}
in hyperelasticity, and
\begin{equation} \label{eq:pa-isotropic-Cauchy elast}
p_{|b} \,g^{ab}=\left[\alpha_1 \,b^{ab}+\alpha_2 \,c^{ab}\right]_{|b} \,,
\end{equation}
in Cauchy elasticity. 
The integrability conditions for the existence of $p$ are $p_{|ab}=p_{|ba}$. The resulting expression in hyperelasticity is written as
\begin{equation} \label{eq:p_{|ab}-isotropic-hyperelastic}
p_{|ab}=\sum_{\kappa}{\mathcal{A}_{ab}^{\kappa} \,W_{\kappa }}\,,
\end{equation}
while in Cauchy elasticity it takes the following form
\begin{equation} \label{eq:p_{|ab}-isotropic-Cauchy elastic}
p_{|ab}=\sum_{\kappa}\left({\mathcal{B}_{ab}^{1\kappa} \,\alpha_{1\kappa}}+{\mathcal{B}_{ab}^{2\kappa} \,\alpha_{2\kappa}}\right)\,,
\end{equation}
where $W_{\kappa}=\dfrac{\partial{W}}{\partial{I_{\kappa}}}$ and $\alpha_{i\kappa} =\dfrac{\partial \alpha_i}{\partial I_{\kappa}}$, where $\kappa$ is a multi-index. The symmetries of the matrices of the coefficients of $W_{\kappa}$ and $\alpha_{i\kappa}$, namely $\mathcal{A}_{ab}^{\kappa}$ and $\mathcal{B}_{ab}^{i\kappa}$ are called universality constraints of hyperelasticity and Cauchy elasticity, respectively. In fact, to ensure the symmetry of $p_{|ab}$, it is necessary and sufficient that $\mathcal{A}_{ab}^{\kappa}=\mathcal{A}_{ba}^{\kappa}$ in hyperelasticity, and $\mathcal{B}_{ab}^{i\kappa}=\mathcal{B}_{ba}^{i\kappa}$ in Cauchy elasticity (for more details, see \citep{Yavari2024,Yavari2022Universal,YavariGoriely2021}). 

For isotropic hyperelastic solids \citep{Yavari2024,Yavari2022Universal,YavariGoriely2021}
\begin{equation} \label{eq:Universality terms-incomp-iso-hyper}
\begin{split}
&\mathcal{A}_{ab}^{1}={b^n_a}_{|bn} \,, \\
&\mathcal{A}_{ab}^{2}={c^n_a}_{|bn} \,, \\
&\mathcal{A}_{ab}^{11}={b^n_a}_{|n} \,I_{1,b}+\left(b^n_a \,I_{1,n}\right)_{|b} \,, \\
&\mathcal{A}_{ab}^{22}=-{c^{n}_{a}}_{|n} \,I_{2,b}-\left(c^{n}_{a} \,I_{2,n}\right)_{|b} \,, \\
&\mathcal{A}_{ab}^{12}={b^{n}_{a}}_{|n} \,I_{2,b}+\left(b^{n}_{a} \,I_{2,n}\right)_{|b}-{c^{n}_{a}}_{|n} \,I_{1,b}-\left(c^{n}_{a} \,I_{1,n}\right)_{|b} \,,\\
&\mathcal{A}_{ab}^{111}=b^{n}_{a} \,I_{1,n} \,I_{1,b}\,,\\
&\mathcal{A}_{ab}^{222}=-c^{n}_{a} \,I_{2,n} \,I_{2,b}\,,\\
&\mathcal{A}_{ab}^{112}=b^{n}_{a} \,(I_{1,b} \,I_{2,n}+I_{1,n} \,I_{2,b})-c^{n}_{a} \,I_{1,n} \,I_{1,b}\,,\\
&\mathcal{A}_{ab}^{122}=b^{n}_{a} \,I_{2,n} \,I_{2,b}-c^{n}_{a} \,(I_{1,b} \,I_{2,n}+I_{1,n} \,I_{2,b})\,,
\end{split}
\end{equation}
where $f_{,a}=f_{|a}=\partial {f}/\partial x^{a}$ when $f$ is a scalar field.

We know that $\mathcal{B}_{ab}^{1\kappa}$ is the matrix of the coefficient of $\alpha_{1\kappa}$, and $\mathcal{B}_{ab}^{2\kappa}$ is the matrix of the coefficient of $\alpha_{2\kappa}$. We omit $\kappa$ when it is zero, so that $\mathcal{B}_{ab}^{i}$ corresponds to the coefficients of $\alpha_{i}$. A total of twelve universality constraints for Cauchy elasticity are obtained from the symmetry conditions of the following terms \citep{Yavari2024}
\begin{equation} \label{eq:Universality terms-incomp-iso-Cauchy elast-1}
\begin{split}
&\mathcal{B}_{ab}^{1}={b^{n}_{a}}_{|bn} \,, \\
&\mathcal{B}_{ab}^{2}={c^{n}_{a}}_{|bn} \,, \\
&\mathcal{B}_{ab}^{11}={b^{n}_{a}}_{|n} \,I_{1,b}+\left(b^{n}_{a} \,I_{1,n}\right)_{|b} \,, \\
&\mathcal{B}_{ab}^{22}=-{c^{n}_{a}}_{|n} \,I_{2,b}-\left(c^{n}_{a} \,I_{2,n}\right)_{|b} \,, \\
&\mathcal{B}_{ab}^{12}={b^{n}_{a}}_{|n} \,I_{2,b}+\left(b^{n}_{a} \,I_{2,n}\right)_{|b} \,,\\
&\mathcal{B}_{ab}^{21}=-{c^{n}_{a}}_{|n} \,I_{1,b}-\left(c^{n}_{a} \,I_{1,n}\right)_{|b} \,, 
\end{split}
\end{equation}
and
\begin{equation} \label{eq:Universality terms-incomp-iso-Cauchy elast-2}
\begin{split}
&\mathcal{B}_{ab}^{111}=b^{n}_{a} \,I_{1,n} \,I_{1,b}\,,\\
&\mathcal{B}_{ab}^{222}=-c^{n}_{a} \,I_{2,n} \,I_{2,b}\,,\\
&\mathcal{B}_{ab}^{112}=b^{n}_{a} \,(I_{1,b} \,I_{2,n}+I_{1,n} \,I_{2,b})\,,\\
&\mathcal{B}_{ab}^{211}=-c^{n}_{a} \,I_{1,n} \,I_{1,b}\,,\\
&\mathcal{B}_{ab}^{122}=b^{n}_{a} \,I_{2,n} \,I_{2,b}\,,\\
&\mathcal{B}_{ab}^{221}=-c^{n}_{a} \,(I_{1,b} \,I_{2,n}+I_{1,n} \,I_{2,b})\,.
\end{split}
\end{equation}
Concerning the nine terms in hyperelasticity, we clearly have the following relations 
\begin{equation} \label{eq:hyper-Cauchy terms-relation-incomp-iso}
\begin{split}
&\mathcal{A}_{ab}^{1}=\mathcal{B}_{ab}^{1} \,,\\
&\mathcal{A}_{ab}^{2}=\mathcal{B}_{ab}^{2} \,,\\
&\mathcal{A}_{ab}^{11}=\mathcal{B}_{ab}^{11} \,,\\
&\mathcal{A}_{ab}^{22}=\mathcal{B}_{ab}^{22} \,,\\
&\mathcal{A}_{ab}^{12}=\mathcal{B}_{ab}^{12}+\mathcal{B}_{ab}^{21} \,,\\
&\mathcal{A}_{ab}^{111}=\mathcal{B}_{ab}^{111} \,,\\
&\mathcal{A}_{ab}^{222}=\mathcal{B}_{ab}^{222} \,,\\
&\mathcal{A}_{ab}^{112}=\mathcal{B}_{ab}^{112}+\mathcal{B}_{ab}^{211} \,,\\
&\mathcal{A}_{ab}^{122}=\mathcal{B}_{ab}^{122}+\mathcal{B}_{ab}^{221} \,.
\end{split}
\end{equation}

\citet{Ericksen1954} showed that if $I_1$ and $I_2$ are not constant, the terms $\mathcal{A}_{ab}^{111}$, $\mathcal{A}_{ab}^{222}$, $\mathcal{A}_{ab}^{112}$ and $\mathcal{A}_{ab}^{122}$ are symmetric only when $\nabla I_1$ and $\nabla I_2$ are parallel, with both being eigenvectors of $\mathbf{b}$ as well as $\mathbf{c}$, or
\begin{equation} \label{eq:I_1 - I_2 relation}
\begin{split}
&I_{1,a}=c_{12} \,I_{2,a} \,,\\
&b^{n}_{a} \,I_{1,n}=\lambda_{1} \,I_{1,a} \,,\\
&b^{n}_{a} \,I_{2,n}=\lambda_{1} \,I_{2,a} \,,\\
&c^{n}_{a} \,I_{1,n}=\frac{1}{\lambda_{1}}I_{1,a} \,,\\
&c^{n}_{a}I_{2,n}=\frac{1}{\lambda_{1}}I_{2,a} \,,
\end{split}
\end{equation}
where $c_{12}$ and $\lambda_{1}$ are scalar functions.

Now consider $\mathcal{B}_{ab}^{111}$. We know that if $I_1$ is not constant, this term is symmetric only if $b_{a}^{n} \,I_{1,n}=\lambda_{1} \,I_{1,a}$. If $I_2$ is not constant, the term $\mathcal{B}_{ab}^{122}$ is likewise symmetric only when $b_{a}^{n} \,I_{2,n}=\lambda_{2} \,I_{2,a}$ ($\lambda_{2}$ is a scalar function). Using these two results, $\mathcal{B}_{ab}^{112}$ is written as
\begin{equation} \label{eq:B_{ab}^{112}-incomp-iso}
\mathcal{B}_{ab}^{112}=\lambda_{2} \,I_{2,a} \,I_{1,b}+\lambda_{1} \,I_{1,a} \,I_{2,b} \,.
\end{equation}
The right-hand side of \eqref{eq:B_{ab}^{112}-incomp-iso} is symmetric either when $\lambda_{1}=\lambda_{2}$, or when $\nabla I_1$ and $\nabla I_2$ are parallel, which are equivalent. Since the eigenvectors of $\mathbf{b}$ and $\mathbf{c}$ are the same, the symmetries of the terms 
$\mathcal{B}_{ab}^{222}$, $\mathcal{B}_{ab}^{211}$ and $\mathcal{B}_{ab}^{221}$ lead to the same result. Therefore, the symmetries of $\big\{\mathcal{A}_{ab}^{111},\mathcal{A}_{ab}^{222},\mathcal{A}_{ab}^{112},\mathcal{A}_{ab}^{122}\big\}$ are equivalent to those of $\big\{\mathcal{B}_{ab}^{111},\mathcal{B}_{ab}^{222},\mathcal{B}_{ab}^{112},\mathcal{B}_{ab}^{211},\mathcal{B}_{ab}^{122},\mathcal{B}_{ab}^{221}\big\}$, since the symmetries of both sets are described by the same condition. This condition indicates that $\nabla I_1$ and $\nabla I_2$ are parallel and are eigenvectors of $\mathbf{b}$ and $\mathbf{c}$.
Let us introduce the notation $\mathcal{A}_{[ab]}^{\kappa}= \mathcal{A}_{ab}^{\kappa}-\mathcal{A}_{ba}^{\kappa}$ and $\mathcal{B}_{[ab]}^{\kappa}= \mathcal{B}_{ab}^{\kappa}-\mathcal{B}_{ba}^{\kappa}$. The corresponding universality constraints are then $\mathcal{A}_{[ab]}=0$ and $\mathcal{B}_{[ab]}=0$, which are equivalent to the symmetries of the terms $\mathcal{A}_{ab}^{\kappa}$ and $\mathcal{B}_{ab}^{\kappa}$, respectively. From this definition, we have
\begin{equation} \label{eq:term-equiv-incomp-iso-1}
\left\{\mathcal{A}_{[ab]}^{111}=\mathcal{A}_{[ab]}^{222}=\mathcal{A}_{[ab]}^{112}=\mathcal{A}_{[ab]}^{122}=0\right\} ~\text{is~equivalent~to}~
\left\{\mathcal{B}_{[ab]}^{111}=\mathcal{B}_{[ab]}^{222}=\mathcal{B}_{[ab]}^{112}=\mathcal{B}_{[ab]}^{211}=\mathcal{B}_{[ab]}^{122}=\mathcal{B}_{[ab]}^{221}=0\right\} \,,
\end{equation}
or
\begin{equation} \label{eq:Constr-equiv-incomp-iso-1}
\left\{\mathcal{A}_{ab}^{111},\mathcal{A}_{ab}^{222},\mathcal{A}_{ab}^{112},\mathcal{A}_{ab}^{122}\right\} \equiv \left\{\mathcal{B}_{ab}^{111},\mathcal{B}_{ab}^{222},\mathcal{B}_{ab}^{112},\mathcal{B}_{ab}^{211},\mathcal{B}_{ab}^{122},\mathcal{B}_{ab}^{221}\right\} \,,
\end{equation}
where $\equiv$ indicates the symmetry equivalence between the two terms which is defined as follows: 

\begin{defi}[Symmetry equivalence] 
Two sets of symmetry constraints are equivalent if they impose exactly the same conditions on the admissible deformations---that is, an admissible deformation satisfies the symmetries of one set if and only if it satisfies those of the other set.
\end{defi}

With respect to \eqref{eq:I_1 - I_2 relation}$_{1}$ and \eqref{eq:I_1 - I_2 relation}$_{3}$, the term $\mathcal{B}_{ab}^{12}$ can be rewritten as
\begin{equation} \label{eq:B_{ab}^{12}-incomp-iso-1}
\mathcal{B}_{ab}^{12}={c_{12}}\,{b^{n}_{a}}_{|n} \,I_{1,b}+\lambda_{1,b} \,I_{2,a}+\lambda_{1} \,{I_{2}}_{|ab} \,.
\end{equation}
Since $I_{2}$ is a scalar field, ${I_{2}}_{|ab}$ is symmetric and the symmetry of $\mathcal{B}_{ab}^{12}$ becomes equivalent to
\begin{equation} \label{eq:B_{ab}^{12}-incomp-iso-2}
\mathcal{B}_{ab}^{12} \equiv {c_{12}} \,\big({b^{n}_{a}}_{|n} \,I_{1,b}+\lambda_{1,b} \,I_{1,a}\big)\,.
\end{equation}
Using the same procedure, one can represent the symmetry of $\mathcal{B}_{ab}^{11}$ as
\begin{equation} \label{eq:B_{ab}^{11}-incomp-iso}
\mathcal{B}_{ab}^{11} \equiv {b^{n}_{a}}_{|n} \,I_{1,b}+\lambda_{1,b} \,I_{1,a}\,.
\end{equation}
Thus,
\begin{equation} \label{eq:Constr-equiv-incomp-iso-2}
\mathcal{B}_{ab}^{12} \equiv \mathcal{B}_{ab}^{11}\,,
\end{equation}
which means that at least one of the six symmetry constraints of the terms \eqref{eq:Universality terms-incomp-iso-Cauchy elast-1} depends on the others. Hence, we have at most five independent symmetry constraints in Cauchy elasticity ($\mathcal{B}_{[ab]}^{1}=0$, $\mathcal{B}_{[ab]}^{2}=0$, $\mathcal{B}_{[ab]}^{11}=0$, $\mathcal{B}_{[ab]}^{22}=0$ and $\mathcal{B}_{[ab]}^{21}=0$) and at most five independent symmetry constraints in hyperelasticity ($\mathcal{A}_{[ab]}^{1}=0$, $\mathcal{A}_{[ab]}^{2}=0$, $\mathcal{A}_{[ab]}^{11}=0$, $\mathcal{A}_{[ab]}^{22}=0$ and $\mathcal{A}_{[ab]}^{12}=0$) which are related as (see \eqref{eq:hyper-Cauchy terms-relation-incomp-iso}$_{1}$--\eqref{eq:hyper-Cauchy terms-relation-incomp-iso}$_{5}$)
\begin{equation} \label{eq:hyper-Cauchy constr-relation-incomp-iso}
\begin{split}
&\mathcal{A}_{[ab]}^{1}=\mathcal{B}_{[ab]}^{1}=0 \,,\\
&\mathcal{A}_{[ab]}^{2}=\mathcal{B}_{[ab]}^{2}=0 \,,\\
&\mathcal{A}_{[ab]}^{11}=\mathcal{B}_{[ab]}^{11}=0 \,,\\
&\mathcal{A}_{[ab]}^{22}=\mathcal{B}_{[ab]}^{22}=0 \,,\\
&\mathcal{A}_{[ab]}^{12}=\mathcal{B}_{[ab]}^{12}+\mathcal{B}_{[ab]}^{21}=0 \,.
\end{split}
\end{equation}
Consequently, these two sets of five universality constraints are equivalent. In conclusion, the universality constraints for homogeneous incompressible isotropic Cauchy elastic solids are the same as those in hyperelasticity as was shown in \citep{Yavari2024}.

It is worth noting that the relations~\eqref{eq:hyper-Cauchy constr-relation-incomp-iso} follow directly from \eqref{eq:hyper-Cauchy terms-relation-incomp-iso}$_{1}$--\eqref{eq:hyper-Cauchy terms-relation-incomp-iso}$_{5}$. This is simply because if $\mathcal{A}_{ab}^{i}=\mathcal{B}_{ab}^{j}+\mathcal{B}_{ab}^{k}$ ($i$, $j$ and $k$ are multi-indices), then $\mathcal{A}_{ba}^{i}=\mathcal{B}_{ba}^{j}+\mathcal{B}_{ba}^{k}$, and thus $\mathcal{A}_{ab}^{i}-\mathcal{A}_{ba}^{i}=(\mathcal{B}_{ab}^{j}-\mathcal{B}_{ba}^{j})+(\mathcal{B}_{ab}^{k}-\mathcal{B}_{ba}^{k})$.
Hence, $\mathcal{A}_{[ab]}^{i}=\mathcal{B}_{[ab]}^{j}+\mathcal{B}_{[ab]}^{k}=0$.
In other words, any relation that holds for a set of terms must also hold for the corresponding symmetry constraints.

\section{Universality Constraints in Transversely Isotropic Elasticity} \label{T iso-Elasticity}

A transversely isotropic solid is characterized at each point by a single material preferred direction, oriented normal to the local plane of isotropy. The material preferred direction is defined by a unit vector $\mathbf{N} \,(\mathbf{X})$. The strain energy function in hyperelasticity and the stress in Cauchy elasticity are then described by five independent invariants $I_1,\cdots,I_5$. The additional invariants $I_4$ and $I_5$ are defined as
\begin{equation} \label{eq:I_4,I_5-incomp-T iso} 
I_4=\mathbf{N} \cdot \mathbf{C} \cdot \mathbf{N} \,,\qquad
I_5=\mathbf{N} \cdot \mathbf{C^2} \cdot \mathbf{N} \,.
\end{equation}
For homogeneous transversely isotropic hyperelastic solids, the second Piola-Kirchhoff stress is given by \citep{Yavari2022Universal}
\begin{equation} \label{eq:second piola-T iso-hyper} 
\begin{split}
\mathbf{S} = &2\,W_1 \, \mathbf{G}^{\sharp}+2\,W_2 \,(I_2 \,\mathbf{C}^{-1}-I_3 \,\mathbf{C}^{-2})+2\,W_3 \,I_3 \,\mathbf{C}^{-1}
+2 \, W_4 \,(\mathbf{N} \otimes \mathbf{N})\\
&+2\,W_5 \,[\mathbf{N} \otimes (\mathbf{C} \cdot \mathbf{N})+(\mathbf{C} \cdot \mathbf{N}) \otimes \mathbf{N}]\,. 
\end{split}
\end{equation}
The Cauchy stress is written as \citep{Yavari2022Universal,YavariGoriely2021}
\begin{equation} \label{eq:Cauchy stress-T iso-hyper} 
\begin{split}
\boldsymbol{\sigma}=&\frac{2}{\sqrt{I_3}} \,W_1 \,\mathbf{b}^\sharp+\frac{2}{\sqrt{I_3}} \,(I_2 \,W_2+I_3 \,W_3) \,\mathbf{g}^\sharp -2 \,\sqrt{I_3} \,W_2 \, \mathbf{c}^\sharp
+\frac{2}{\sqrt{I_3}} \,W_4 \,(\mathbf{n} \otimes \mathbf{n})\\
&
+\frac{2}{\sqrt{I_3}} \,W_5 \,[\mathbf{n} \otimes (\mathbf{b} \cdot \mathbf{n})+(\mathbf{b} \cdot \mathbf{n}) \otimes \mathbf{n}] \,,
\end{split}
\end{equation}
where $W_i= \dfrac{\partial W}{\partial I_i} \,(i=1,\cdots,5)$, $\mathbf{n}=\mathbf{F} \cdot \mathbf{N}$. Thus, the components of the Cauchy stress tensor are  
\begin{equation} \label{eq:Cauchy stress-T iso-hyper-components} 
\sigma^{ab} =\frac{2}{\sqrt{I_3}}\,[W_1 \,b^{ab}+(I_2 \,W_2+I_3 \,W_3)g^{ab}-I_3 \,W_2 \, c^{ab}+W_4 \,n^{a} \,n^{b}+W_5 \,\ell^{ab}] \,,
\end{equation}
where $n^a=F^a{}_A \,N^A$ and $\ell^{ab}=n^a \,b^{bc} \,n_c+n^{b} \,b^{ac} \,n_c$.

In transversely isotropic Cauchy elasticity, the second Piola–Kirchhoff stress is represented by \citep{Spencer1970,Boehler1979,Boehler1987,YavariGoriely2025}
\begin{equation} \label{eq:second piola-T iso-Cauchy elast} 
\begin{split}
\mathbf{S} = &a_0 \,\mathbf{G}^{\sharp}+ a_1 \,\mathbf{C}^{\sharp}+a_2 \, \mathbf{C}^{2\sharp}+a_3 \,(\mathbf{N}\otimes \mathbf{N})+a_4 \,[ \mathbf{N}\otimes (\mathbf{C}\cdot \mathbf{N})+( \mathbf{C}\cdot \mathbf{N})\otimes \mathbf{N}] \\
&+a_5 \,[\mathbf{N}\otimes (\mathbf{C}^{2} \cdot \mathbf{N})+(\mathbf{C}^2 \cdot \mathbf{N} )\otimes \mathbf{N}]\,,
\end{split}
\end{equation}
and thus the Cauchy stress is written as
\begin{equation} \label{eq:Cauchy stress-T iso-Cauchy elast} 
\begin{split}
\boldsymbol{\sigma}=&\tilde{a}_0 \,\mathbf{g}^{\sharp}+\tilde{a}_1 \,\mathbf{b}^{\sharp}+\tilde{a}_2 \,\mathbf{c}^{\sharp} +\tilde{a}_3 \,(\mathbf{n} \otimes \mathbf{n})+\tilde{a}_4 \,[\mathbf{n} \otimes (\mathbf{b} \cdot \mathbf{n})+(\mathbf{b} \cdot \mathbf{n}) \otimes \mathbf{n}] \\
&+\tilde{a}_5 \,[\mathbf{n} \otimes (\mathbf{c} \cdot \mathbf{n})+(\mathbf{c} \cdot \mathbf{n}) \otimes \mathbf{n}] \,,
\end{split}
\end{equation}
where $a_i(I_1,\cdots,I_5)$ and $\tilde{a}_i(I_1,\cdots,I_5)$ , $i=0,\cdots,5$ are the response functions.

For homogeneous incompressible transversely isotropic solids $I_3=1$, and therefore the second Piola-Kirchhoff stress in hyperelasticity is represented by
\begin{equation} \label{eq:second piola-incomp-T iso-hyper} 
\begin{split}
\mathbf{S} =& -p \, \mathbf{C}^{-1}+2\,W_1 \, \mathbf{G}^{\sharp}+2\,W_2 \,(I_2 \,\mathbf{C}^{-1}-\mathbf{C}^{-2})+2 \, W_4 \,(\mathbf{N} \otimes \mathbf{N})\\
&+2\,W_5 \,[\mathbf{N} \otimes (\mathbf{C} \cdot \mathbf{N})+(\mathbf{C} \cdot \mathbf{N}) \otimes \mathbf{N}]\,, 
\end{split}
\end{equation}
where $W=W \,(I_1,I_2,I_4,I_5)$ and $p$ is the Lagrange multiplier corresponding to the incompressibility constraint $I_3=1$. Hence, the Cauchy stress reads
\begin{equation} \label{eq:Cauchy stress-incomp-T iso-hyper}
\begin{split}
\boldsymbol{\sigma} =& -p \,\mathbf{g}^\sharp+ 2 \,W_1 \, \mathbf{b}^\sharp- 2 \,W_2 \, \mathbf{c}^\sharp+ 2 \,W_4 \,(\mathbf{n} \otimes \mathbf{n}) 
+ 2 \,W_5 \, [\mathbf{n} \otimes (\mathbf{b} \cdot \mathbf{n})+(\mathbf{b} \cdot \mathbf{n}) \otimes \mathbf{n}] \,,
\end{split}
\end{equation}
which has components
\begin{equation} \label{eq:Cauchy stress-incomp-T iso-hyper-components}
\sigma^{ab} = -p \,g^{ab}+ 2 \,W_1 \, b^{ab}- 2 \,W_2 \, c^{ab}+ 2 \,W_4 \,n^{a} \,n^{b}+ 2 \,W_5 \,(n^a \,b^{bc} \,n^d \,g_{cd} + n^b \,b^{ac} \,n^d \,g_{cd}) \,.
\end{equation}
Similarly, by taking $I_3=1$ and using the Cayley-Hamilton theorem, the second Piola-Kirchhoff stress for incompressible transversely isotropic Cauchy elastic solids can be derived from \eqref{eq:second piola-T iso-Cauchy elast} as
\begin{equation} \label{eq:second piola-incomp-T iso-Cauchy elast} 
\begin{split}
\mathbf{S}= &-p \,\mathbf{C}^{-1}+\bar{a}_0 \,\mathbf{G}^{\sharp}+ \bar{a}_1 \,\mathbf{C}^{\sharp}+\bar{a}_2 \,(\mathbf{N}\otimes \mathbf{N})+\bar{a}_4 \,[\mathbf{N}\otimes (\mathbf{C}\cdot \mathbf{N})+( \mathbf{C}\cdot \mathbf{N})\otimes \mathbf{N}] \\
&+\bar{a}_5 \,[\mathbf{N}\otimes (\mathbf{C}^{-1} \cdot \mathbf{N})+(\mathbf{C}^{-1} \cdot \mathbf{N} )\otimes \mathbf{N}]\,,
\end{split}
\end{equation}
where $\bar{a}_i \,(I_1,I_2,I_4,I_5)$, $i=0,1,2,4,5$ are the response functions. We can use the Cayley-Hamilton theorem again to write the Cauchy stress for incompressible solids as
\begin{equation} \label{eq:Cauchy stress-incomp-T iso-Cauchy elast} 
\begin{split}
\boldsymbol{\sigma}=&-p \,\mathbf{g}^{\sharp}+\alpha_1 \,\mathbf{b}^{\sharp}+\alpha_2 \,\mathbf{c}^{\sharp} +\alpha_4 \,(\mathbf{n} \otimes \mathbf{n})+\alpha_5 \,[\mathbf{n} \otimes (\mathbf{b} \cdot \mathbf{n})+(\mathbf{b} \cdot \mathbf{n}) \otimes \mathbf{n}] \\
&+\alpha_6 \,[\mathbf{n} \otimes (\mathbf{c} \cdot \mathbf{n})+(\mathbf{c} \cdot \mathbf{n}) \otimes \mathbf{n}] \,,
\end{split}
\end{equation}
where $\alpha_{i}=\alpha_{i} \,(I_1,I_2,I_4,I_5)$, $i=1,2,4,5,6$ are arbitrary response functions. In the following sections, the equivalence of the universality constraints in hyperelasticity and those in Cauchy elasticity is investigated separately for compressible and incompressible cases.

\subsection{Compressible transversely isotropic solids} \label{Comp-T iso-Elasticity}

For homogeneous compressible transversely isotropic Cauchy elastic solids, the Cauchy stress~\eqref{eq:Cauchy stress-T iso-Cauchy elast} in components reads
\begin{equation} \label{eq:Cauchy stress-T iso-Cauchy elast-components} 
\sigma^{ab}=\alpha_1 \,g^{ab}+\alpha_2 \,b^{ab}+\alpha_3 \,c^{ab} +\alpha_4 \,n^a \,n^b+\alpha_5 \,\ell^{ab}+\alpha_6 \,\bar{\ell}^{ab} \,,
\end{equation}
where $\bar{\ell}^{ab}=n^a \,c^{bc} \,n_c+n^b \,c^{ac} \,n_c$, and $\alpha_{i}=\alpha_{i} \,(I_1,I_2,I_3,I_4,I_5)$, $i=1,\cdots,6$ are arbitrary response functions. Substituting \eqref{eq:Cauchy stress-T iso-Cauchy elast-components} into the equilibrium equations~\eqref{eq:equilibrium} and using metric compatibility, one obtains
\begin{equation} \label{eq:equilibrium-comp-T iso-1}
\begin{split}
&\alpha_2 \,{b^{ab}}_{|b}+\alpha_3 \,{c^{ab}}_{|b}
+\alpha_4 \,\left(n^a \,n^b\right)_{|b}+\alpha_5 \,{\ell^{ab}}_{|b}+\alpha_6 \,{\bar{\ell}^{ab}}_{|b}+
\alpha_{1,1} \,I_{1,b} \,g^{ab}+\alpha_{1,2} \,I_{2,b} \,g^{ab}+\alpha_{1,3} \,I_{3,b} \,g^{ab} \\
&+\alpha_{1,4} \,I_{4,b} \,g^{ab}+\alpha_{1,5} \,I_{5,b} \,g^{ab}+\alpha_{2,1} \,I_{1,b} \,b^{ab}+\alpha_{2,2} \,I_{2,b} \,b^{ab} +\alpha_{2,3} \,I_{3,b} \,b^{ab}+\alpha_{2,4} \,I_{4,b} \,b^{ab}+\alpha_{2,5} \,I_{5,b} \,b^{ab}\\
&+\alpha_{3,1} \,I_{1,b} \,c^{ab}+\alpha_{3,2} \,I_{2,b} \,c^{ab}+\alpha_{3,3} \,I_{3,b} \,c^{ab}+\alpha_{3,4} \,I_{4,b} \,c^{ab}+\alpha_{3,5} \,I_{5,b} \,c^{ab}+\alpha_{4,1} \,I_{1,b} \,(n^a \,n^b)+\alpha_{4,2} \,I_{2,b} \,(n^a \,n^b) \\
&+\alpha_{4,3} \,I_{3,b} \,(n^a \,n^b)+\alpha_{4,4} \,I_{4,b} \,(n^a \,n^b)+\alpha_{4,5} \,I_{5,b} \,(n^a \,n^b)+\alpha_{5,1} \,I_{1,b} \,\ell^{ab}+\alpha_{5,2} \,I_{2,b} \,\ell^{ab}+\alpha_{5,3} \,I_{3,b} \,\ell^{ab} \\
&+\alpha_{5,4} \,I_{4,b} \,\ell^{ab}+\alpha_{5,5} \,I_{5,b} \,\ell^{ab}+\alpha_{6,1} \,I_{1,b} \,\bar{\ell}^{ab}+\alpha_{6,2} \,I_{2,b} \,\bar{\ell}^{ab}+\alpha_{6,3} \,I_{3,b} \,\bar{\ell}^{ab}+\alpha_{6,4} \,I_{4,b} \,\bar{\ell}^{ab}+\alpha_{6,5} \,I_{5,b} \,\bar{\ell}^{ab}=0 \,.
\end{split}
\end{equation}
Since $\alpha_i$ and its derivatives are independent functions, \eqref{eq:equilibrium-comp-T iso-1} can be satisfied only if the coefficients of $\alpha_i$ and $\alpha_{i,j}$ vanish. This leads to the following universality constraints
\begin{equation} \label{eq:constraints-comp-T iso}
\begin{split}
&{b^{ab}}_{|b}={c^{ab}}_{|b}=0 \,, \\
&\left(n^a \,n^b\right)_{|b}=0 \,, \\
&{\ell^{ab}}_{|b}=0 \,, \\
&{\bar{\ell}^{ab}}_{|b}=0 \,, \\
&I_{i,b} \,g^{ab}=0\,, \\
&I_{i,b} \,b^{ab}=0\,, \\
&I_{i,b} \,c^{ab}=0\,, \\
&I_{i,b} \,n^a \,n^b=0\,, \\
&I_{i,b} \,\ell^{ab}=0\,, \\
&I_{i,b} \,\bar{\ell}^{ab}=0\,, 
\end{split}
\end{equation}
where $i=1,\cdots,5$. Except for \eqref{eq:constraints-comp-T iso}$_4$ and \eqref{eq:constraints-comp-T iso}$_{10}$, the remaining constraints in \eqref{eq:constraints-comp-T iso} are identical to those of compressible transversely isotropic hyperelastic solids (see \citep{Yavari2022Universal,YavariGoriely2021}). Therefore, the following results are obtained. First, by comparing with \eqref{eq:constraints-comp-iso}, one finds that \eqref{eq:constraints-comp-T iso}$_1$, \eqref{eq:constraints-comp-T iso}$_5$, \eqref{eq:constraints-comp-T iso}$_6$, and \eqref{eq:constraints-comp-T iso}$_7$ , for $i=1,2,3$, are the universality constraints for compressible isotropic solids. Thus, the universal deformations for transversely isotropic solids must be homogeneous. Second, the constraints~\eqref{eq:constraints-comp-T iso}$_5$ imply that $I_i$ $(i = 1, \ldots, 5)$ are constant (note that since $I_{i,a} = F^{-A}{}_a \,I_{i,A} = 0$, then $I_{i,A} = 0$). With this result, \eqref{eq:constraints-comp-T iso}$_6$, \eqref{eq:constraints-comp-T iso}$_7$, \eqref{eq:constraints-comp-T iso}$_8$, and \eqref{eq:constraints-comp-T iso}$_9$ are trivially satisfied. Third, from \eqref{eq:constraints-comp-T iso}$_2$ and \eqref{eq:constraints-comp-T iso}$_3$, it follows that $\mathbf{N}$ is a constant unit vector \citep{Yavari2022Universal,YavariGoriely2021}.  

It can be readily shown that when the deformation is homogeneous, $I_i$ are constant, and $\mathbf{N}$ is a constant unit vector, the additional universality constraints in Cauchy elasticity, i.e., \eqref{eq:constraints-comp-T iso}$_4$ and \eqref{eq:constraints-comp-T iso}$_{10}$, are trivially satisfied. Therefore, the universality constraints in Cauchy elasticity are equivalent to those of hyperelasticity. In summary, we have proved the following result.
\begin{prop}
The universal deformations and material preferred directions of compressible transversely isotropic Cauchy elasticity are identical to those of compressible transversely isotropic hyperelasticity.
\end{prop}

\subsection{Incompressible transversely isotropic solids} \label{Incomp-T iso-Elasticity}

The method used to obtain the universality constraints for incompressible transversely isotropic solids follows exactly the same steps as those described in \S\ref{Incomp-Iso-Elasticity} for incompressible isotropic solids: we first substitute the two expressions for the Cauchy stress, given by \eqref{eq:Cauchy stress-incomp-T iso-hyper} and \eqref{eq:Cauchy stress-incomp-T iso-Cauchy elast} for hyperelastic and Cauchy elastic solids, respectively, into the equilibrium equations~\eqref{eq:equilibrium} to determine $p_{|a}$ as
\begin{equation} \label{eq:pa-T iso-hyperelastic}
p_{|b} \,g^{ab}=2 \,\left[W_1 \,b^{ab}-W_2 \,c^{ab} +W_4 \,n^a \,n^b+W_5 \,\ell^{ab}\right]_{|b} \,,
\end{equation}
in hyperelasticity, and
\begin{equation} \label{eq:pa-T iso-Cauchy elast}
p_{|b} \,g^{ab}=\left[\alpha_1 \,b^{ab}+\alpha_2 \,c^{ab}+\alpha_4 \,n^a \,n^b+\alpha_5 \,\ell^{ab}+\alpha_6 \,\bar{\ell}^{ab}\right]_{|b} \,,
\end{equation}
in Cauchy elasticity. Recall that in both cases, the integrability condition for the existence of $p$ requires $p_{|ab}$ to be symmetric, that is, $p_{|ab}=p_{|ba}$. 

In hyperelasticity, $p_{|ab}$ is written as $p_{|ab}=\sum_{\kappa}{\mathcal{A}_{ab}^{\kappa} \,W_{\kappa }}$, where $W_{\kappa }$ are independent functions. The symmetry condition $p_{|ab}=p_{|ba}$ is then identical to the symmetries of the terms $\mathcal{A}_{ab}^{\kappa}$, i.e., $\mathcal{A}_{ab}^{\kappa}=\mathcal{A}_{ba}^{\kappa}$. \citet{YavariGoriely2023Universal} demonstrated that there are $34$ universality constraints in transversely isotropic hyperelasticity. The first nine constraints are the same as those in isotropic solids, i.e., the symmetries of the terms represented by \eqref{eq:Universality terms-incomp-iso-hyper}. They showed that the remaining $25$ constraints are the symmetries of the following terms 
\begin{equation} \label{eq:universality terms-incom-T iso-hyper-1} 
\begin{split}
&\mathcal{A}_{ab}^{4}=\left(n_{a} \,n^n\right)_{|nb} \,,\\
&\mathcal{A}_{ab}^{5}=\ell_{a|nb}^{n} \,, \\
&\mathcal{A}_{ab}^{44}=\left(n_{a} \,n^n\right)_{|n} \,I_{4,b} +\left(n_a \, n^n \,I_{4,n}\right)_{|b} \,, \\
&\mathcal{A}_{ab}^{55}=\ell_{a|n}^{n} \,I_{5,b}+\left(\ell_{a}^{n} \,I_{5,n}\right)_{|b}\,,\\
&\mathcal{A}_{ab}^{14}=b_{a|n}^n \,I_{4,b}+\left(b_{a}^{n} \,I_{4,n}\right)_{|b}+\left(n_{a} \,n^n \right)_{|n} \,I_{1,b}+\left(n_a \, n^n \,I_{1,n}\right)_{|b}\,,\\
&\mathcal{A}_{ab}^{15}=b_{a|n}^{n} \,I_{5,b}+\left(b_{a}^{n} \,I_{5,n}\right)_{|b}+\ell_{a|n}^{n}\,I_{1,b}+\left(\ell_{a}^{n} \,I_{1,n}\right)_{|b}\,,\\
&\mathcal{A}_{ab}^{24}=-c_{a|n}^{n} \,I_{4,b}-\left(c_{a}^{n} \,I_{4,n}\right)_{|b}+\left(n_{a} \,n^n\right)_{|n} \,I_{2,b}+\left(n_a \, n^n \,I_{2,n}\right)_{|b}\,,\\
&\mathcal{A}_{ab}^{25}=-c_{a|n}^{n} \,I_{5,b}-\left(c_{a}^{n} \,I_{5,n}\right)_{|b}+\ell_{a|n}^{n} \,I_{2,b}+\left(\ell_{a}^{n} \,I_{2,n}\right)_{|b}\,,\\
&\mathcal{A}_{ab}^{45}=\left(n_{a} \,n^n \right)_{|n} \,I_{5,b} +\left(n_a \, n^n \,I_{5,n}\right)_{|b} +\ell_{a|n}^{n} \,I_{4,b}+ \left(\ell_{a}^{n} \,I_{4,n}\right)_{|b} \,,
\end{split}
\end{equation}
and
\begin{equation} \label{eq:universality terms-incom-T iso-hyper-2} 
\begin{split}
&\mathcal{A}_{ab}^{444}=n_{a} \,n^{n} \,I_{4,n} \,I_{4,b}\,,\\
&\mathcal{A}_{ab}^{555}=\ell_{a}^{n}\,I_{5,n} \,I_{5,b}\,,\\
&\mathcal{A}_{ab}^{114}=b_{a}^{n} \,(I_{4,n} \,I_{1,b}+I_{4,b} \,I_{1,n})+n_{a} \,n^{n}\,I_{1,n}\,I_{1,b} \,,\\
&\mathcal{A}_{ab}^{115}=b_{a}^{n} \,(I_{5,n} \,I_{1,b}+I_{5,b} \,I_{1,n})+\ell_{a}^{n} \,I_{1,n} \,I_{1,b} \,,\\
&\mathcal{A}_{ab}^{124}=b_{a}^{n} \,(I_{4,n} \,I_{2,b}+I_{4,b} \,I_{2,n})-c_{a}^{n} \,(I_{4,n} \,I_{1,b}+I_{4,b} \,I_{1,n})+n_{a} \,n^{n}(I_{1,n} \,I_{2,b}+I_{1,b} \,I_{2,n}) \,,\\
&\mathcal{A}_{ab}^{125}=b_{a}^{n} \,(I_{5,n} \,I_{2,b}+I_{5,b} \,I_{2,n})-c_{a}^{n} \,(I_{5,n} \,I_{1,b}+I_{5,b} \,I_{1,n})+\ell_{a}^{n} \,(I_{1,n} \,I_{2,b}+I_{1,b} \,I_{2,n}) \,,\\
&\mathcal{A}_{ab}^{144}=b_{a}^{n} \,I_{4,n} \,I_{4,b}+n_{a} \,n^{n} \,(I_{4,n} \,I_{1,b}+I_{4,b} \,I_{1,n})\,,\\
&\mathcal{A}_{ab}^{145}=b_{a}^{n} \,(I_{4,n} \,I_{5,b}+I_{4,b} \,I_{5,n})+n_{a} \,n^{n} \,(I_{1,n} \,I_{5,b}+I_{1,b} \,I_{5,n})+\ell_{a}^{n} \,(I_{1,n} \,I_{4,b}+I_{1,b} \,I_{4,n}) \,,\\
&\mathcal{A}_{ab}^{155}=b_{a}^{n} \,I_{5,n} \,I_{5,b}+\ell_{a}^{n} \,(I_{1,n} \,I_{5,b}+I_{1,b} \,I_{5,n})\ \,,\\
&\mathcal{A}_{ab}^{224}=-c_{a}^{n} \,(I_{4,n} \,I_{2,b}+I_{4,b} \,I_{2,n})+n_{a} \,n^{n} \,I_{2,n} \,I_{2,b} \,,\\
&\mathcal{A}_{ab}^{225}=-c_{a}^{n} \,(I_{5,n} \,I_{2,b}+I_{5,b} \,I_{2,n})+\ell_{a}^{n} \,I_{2,n} \,I_{2,b} \,,\\
&\mathcal{A}_{ab}^{244}=-c_{a}^{n} \,I_{4,n} \,I_{4,b}+n_{a} \,n^{n} \,(I_{4,n} \,I_{2,b}+I_{4,b} \,I_{2,n}) \,,\\
&\mathcal{A}_{ab}^{245}=-c_{a}^{n} \,(I_{5,n} \,I_{4,b}+I_{5,b} \,I_{4,n})+n_{a} \,n^{n} \,(I_{2,n} \,I_{5,b}+I_{2,b} \,I_{5,n})+\ell_{a}^{n} \,(I_{2,n} \,I_{4,b}+I_{2,b} \,I_{4,n}) \,,\\
&\mathcal{A}_{ab}^{255}=-c_{a}^{n} \,I_{5,n} \,I_{5,b}+\ell_{a}^{n} \,(I_{2,n} \,I_{5,b}+I_{2,b} \,I_{5,n}) \,,\\
&\mathcal{A}_{ab}^{445}=n_{a} \,n^{n} \,(I_{4,n} \,I_{5,b}+I_{4,b} \,I_{5,n})+\ell_{a}^{n} \,I_{4,n} \,I_{4,b}\,,\\
&\mathcal{A}_{ab}^{455}=n_{a} \,n^{n} \,I_{5,n} \,I_{5,b}+\ell_{a}^{n} \,(I_{4,n} \,I_{5,b}+I_{4,b} \,I_{5,n})\,.
\end{split}
\end{equation}

In Cauchy elasticity, $p_{|ab}$ can be computed as
\begin{equation} \label{eq:p_{|ab}-T iso-Cauchy elastic}
p_{|ab}=\sum_{\kappa}({\mathcal{B}_{ab}^{1\kappa} \,\alpha_{1\kappa}}+{\mathcal{B}_{ab}^{2\kappa} \,\alpha_{2\kappa}}+{\mathcal{B}_{ab}^{4\kappa} \,\alpha_{4\kappa}}+{\mathcal{B}_{ab}^{5\kappa} \,\alpha_{5\kappa}}+{\mathcal{B}_{ab}^{6\kappa} \,\alpha_{6\kappa}})\,,
\end{equation}
where $\mathcal{B}^{1\kappa}_{ab}$, $\mathcal{B}^{2\kappa}_{ab}$, $\mathcal{B}^{4\kappa}_{ab}$, $\mathcal{B}^{5\kappa}_{ab}$ and $\mathcal{B}^{6\kappa}_{ab}$ are the matrices of the coefficients of $\alpha_{1\kappa}$, $\alpha_{2\kappa}$, $\alpha_{4\kappa}$, $\alpha_{5\kappa}$ and $\alpha_{6\kappa}$, respectively and $\alpha_{i\kappa} = \partial \alpha_i/\partial I_\kappa$ ($\kappa$ is a multi-index) are independent functions. Therefore, $p_{|ab}=p_{|ba}$ implies that all terms $\mathcal{B}^{i\kappa}_{ab}$ must be symmetric, i.e., $\mathcal{B}^{i\kappa}_{ab}=\mathcal{B}^{i\kappa}_{ba}$. The first twelve Cauchy elasticity terms $\mathcal{B}_{ab}^{\kappa}$ are the same as those of isotropic solids given in \eqref{eq:Universality terms-incomp-iso-Cauchy elast-1} and \eqref{eq:Universality terms-incomp-iso-Cauchy elast-2}. It can be shown that in transversely isotropic Cauchy elasticity the remaining terms are obtained as follows (note that when $\kappa=0$, we ignore this index in $\mathcal{B}_{ab}^{i\kappa}$, and hence, $\mathcal{B}_{ab}^{i}$ corresponds to the matrix of the coefficient of $\alpha_i$)
\begin{equation} \label{eq:universal terms-incom-T iso-Cauchy elast-1} 
\begin{split}
&\mathcal{B}_{ab}^{4}=\left(n_{a} \,n^n \right)_{|nb} \,,\\
&\mathcal{B}_{ab}^{5}=\ell_{a|nb}^{n} \,, \\
&\mathcal{B}_{ab}^{44}=\left(n_{a} \,n^n \right)_{|n} \,I_{4,b} +\left(n_a \, n^n \,I_{4,n}\right)_{|b} \,, \\
&\mathcal{B}_{ab}^{55}=\ell_{a|n}^{n} \,I_{5,b}+\left(\ell_{a}^{n} \,I_{5,n}\right)_{|b}\,,\\
&\mathcal{B}_{ab}^{14}=b_{a|n}^{n} \,I_{4,b}+\left(b_{a}^{n} \,I_{4,n}\right)_{|b} \,,\\
&\mathcal{B}_{ab}^{15}=b_{a|n}^{n} \,I_{5,b}+\left(b_{a}^{n} \,I_{5,n}\right)_{|b} \,,\\
&\mathcal{B}_{ab}^{24}=-c_{a|n}^{n} \,I_{4,b}-\left(c_{a}^{n} \,I_{4,n}\right)_{|b} \,,\\
&\mathcal{B}_{ab}^{25}=-c_{a|n}^{n} \,I_{5,b}-\left(c_{a}^{n} \,I_{5,n} \right)_{|b} \,,\\
&\mathcal{B}_{ab}^{41}=\left(n_{a} \,n^n \right)_{|n} \,I_{1,b}+\left(n_a \, n^n \,I_{1,n}\right)_{|b}\,,\\
&\mathcal{B}_{ab}^{42}=\left(n_{a} \,n^n \right)_{|n} \,I_{2,b}+\left(n_a \, n^n \,I_{2,n} \right)_{|b}\,,\\
&\mathcal{B}_{ab}^{45}=\left(n_{a} \,n^n \right)_{|n} \,I_{5,b} +\left(n_a \, n^n \,I_{5,n} \right)_{|b}\,,\\
&\mathcal{B}_{ab}^{51}=\ell_{a|n}^{n} \,I_{1,b}+\left(\ell_{a}^{n} \,I_{1,n} \right)_{|b}\,,\\
&\mathcal{B}_{ab}^{52}=\ell_{a|n}^{n} \,I_{2,b}+\left(\ell_{a}^{n} \,I_{2,n} \right)_{|b}\,,\\
&\mathcal{B}_{ab}^{54}=\ell_{a|n}^{n} \,I_{4,b}+\left(\ell_{a}^{n} \,I_{4,n}\right)_{|b} \,,
\end{split}
\end{equation}
and
\begin{equation} \label{eq:universal terms-incom-T iso-Cauchy elast-2} 
\begin{split}
&\mathcal{B}_{ab}^{444}=n_{a} \,n^{n} \,I_{4,n} \,I_{4,b}\,,\\
&\mathcal{B}_{ab}^{555}=\ell_{a}^{n} \,I_{5,n} \,I_{5,b}\,,\\
&\mathcal{B}_{ab}^{114}=b_{a}^{n} \,(I_{4,n} \,I_{1,b}+I_{4,b} \,I_{1,n}) \,,\\
&\mathcal{B}_{ab}^{115}=b_{a}^{n} \,(I_{5,n} \,I_{1,b}+I_{5,b} \,I_{1,n}) \,,\\
&\mathcal{B}_{ab}^{124}=b_{a}^{n} \,(I_{4,n} \,I_{2,b}+I_{4,b} \,I_{2,n}) \,,\\
&\mathcal{B}_{ab}^{125}=b_{a}^{n} \,(I_{5,n} \,I_{2,b}+I_{5,b} \,I_{2,n}) \,,\\
&\mathcal{B}_{ab}^{144}=b_{a}^{n} \,I_{4,n} \,I_{4,b} \,,\\
&\mathcal{B}_{ab}^{145}=b_{a}^{n} \,(I_{4,n} \,I_{5,b}+I_{4,b} \,I_{5,n}) \,,\\
&\mathcal{B}_{ab}^{155}=b_{a}^{n} \,I_{5,n} \,I_{5,b} \,,\\
&\mathcal{B}_{ab}^{214}=-c_{a}^{n} \,(I_{4,n} \,I_{1,b}+I_{4,b} \,I_{1,n}) \,,\\
&\mathcal{B}_{ab}^{215}=-c_{a}^{n} \,(I_{5,n} \,I_{1,b}+I_{5,b} \,I_{1,n}) \,,\\
&\mathcal{B}_{ab}^{224}=-c_{a}^{n} \,(I_{4,n} \,I_{2,b}+I_{4,b} \,I_{2,n}) \,,\\
&\mathcal{B}_{ab}^{225}=-c_{a}^{n} \,(I_{5,n} \,I_{2,b}+I_{5,b} \,I_{2,n}) \,,\\
&\mathcal{B}_{ab}^{244}=-c_{a}^{n} \,I_{4,n} \,I_{4,b} \,,\\
&\mathcal{B}_{ab}^{245}=-c_{a}^{n} \,(I_{5,n} \,I_{4,b}+I_{5,b} \,I_{4,n}) \,,\\
&\mathcal{B}_{ab}^{255}=-c_{a}^{n} \,I_{5,n} \,I_{5,b} \,,\\
&\mathcal{B}_{ab}^{411}=n_{a} \,n^{n} \,I_{1,n} \,I_{1,b}\,,\\
&\mathcal{B}_{ab}^{412}=n_{a} \,n^{n} \,(I_{1,n} \,I_{2,b}+I_{1,b} \,I_{2,n})\,,\\
&\mathcal{B}_{ab}^{415}=n_{a} \,n^{n} \,(I_{1,n} \,I_{5,b}+I_{1,b} \,I_{5,n})\,,\\
&\mathcal{B}_{ab}^{422}=n_{a} \,n^{n} \,I_{2,n} \,I_{2,b}\,,\\
&\mathcal{B}_{ab}^{425}=n_{a} \,n^{n} \,(I_{2,n} \,I_{5,b}+I_{2,b} \,I_{5,n})\,,\\
&\mathcal{B}_{ab}^{441}=n_{a} \,n^{n} \,(I_{4,n} \,I_{1,b}+I_{4,b} \,I_{1,n})\,,\\
&\mathcal{B}_{ab}^{442}=n_{a} \,n^{n} \,(I_{4,n} \,I_{2,b}+I_{4,b} \,I_{2,n})\,,\\
&\mathcal{B}_{ab}^{445}=n_{a} \,n^{n} \,(I_{4,n} \,I_{5,b}+I_{4,b} \,I_{5,n})\,,\\
&\mathcal{B}_{ab}^{455}=n_{a} \,n^{n} \,I_{5,n} \,I_{5,b}\,,\\
&\mathcal{B}_{ab}^{511}=\ell_{a}^{n} \,I_{1,n} \,I_{1,b}\,,\\
&\mathcal{B}_{ab}^{512}=\ell_{a}^{n} \,(I_{1,n} \,I_{2,b}+I_{1,b} \,I_{2,n})\,,\\
&\mathcal{B}_{ab}^{514}=\ell_{a}^{n} \,(I_{1,n} \,I_{4,b}+I_{1,b} \,I_{4,n})\,,\\
&\mathcal{B}_{ab}^{515}=\ell_{a}^{n} \,(I_{1,n} \,I_{5,b}+I_{1,b} \,I_{5,n})\,,\\
&\mathcal{B}_{ab}^{522}=\ell_{a}^{n} \,I_{2,n} \,I_{2,b}\,,\\
&\mathcal{B}_{ab}^{524}=\ell_{a}^{n} \,(I_{2,n} \,I_{4,b}+I_{2,b} \,I_{4,n})\,,\\
&\mathcal{B}_{ab}^{525}=\ell_{a}^{n} \,(I_{2,n} \,I_{5,b}+I_{2,b} \,I_{5,n})\,,\\
&\mathcal{B}_{ab}^{544}=\ell_{a}^{n} \,I_{4,n} \,I_{4,b}\,,\\
&\mathcal{B}_{ab}^{554}=\ell_{a}^{n} \,(I_{4,n} \,I_{5,b}+I_{4,b} \,I_{5,n})\,.
\end{split}
\end{equation}
Moreover, there are $15$ additional terms in Cauchy elasticity which are associated with the coefficients of $\alpha _{6\kappa}$: 
\begin{equation}  \label{eq:universal terms-incom-T iso-Cauchy elast-3}
\begin{split} 
&\mathcal{B}_{ab}^{6}=\bar{\ell}_{a|nb}^{n}\,, \\
&\mathcal{B}_{ab}^{61}=\left(\bar{\ell}_{a}^{n} \,I_{1,n}\right)_{|b}+\bar{\ell}_{a|n}^{n} \,I_{1,b}\,, \\
&\mathcal{B}_{ab}^{62}=\left(\bar{\ell}_{a}^{n} \,I_{2,n}\right)_{|b}+\bar{\ell}_{a|n}^{n} \,I_{2,b}\,, \\
&\mathcal{B}_{ab}^{64}=\left(\bar{\ell}_{a}^{n} \,I_{4,n}\right)_{|b}+\bar{\ell}_{a|n}^{n} \,I_{4,b}\,, \\
&\mathcal{B}_{ab}^{65}=\left(\bar{\ell}_{a}^{n} \,I_{5,n}\right)_{|b}+\bar{\ell}_{a|n}^{n} \,I_{5,b}\,, \\
&\mathcal{B}_{ab}^{611}=\bar{\ell}_{a}^{n} \, I_{1,n} \,I_{1,b}\,, \\
&\mathcal{B}_{ab}^{622}=\bar{\ell}_{a}^{n} \, I_{2,n} \,I_{2,b}\,, \\
&\mathcal{B}_{ab}^{644}=\bar{\ell}_{a}^{n} \, I_{4,n} \,I_{4,b}\,, \\
&\mathcal{B}_{ab}^{655}=\bar{\ell}_{a}^{n} \, I_{5,n} \,I_{5,b}\,, \\
&\mathcal{B}_{ab}^{612}=\bar{\ell}_{a}^{n} \,(I_{2,n} \,I_{1,b}+I_{2,b} \,I_{1,n})\,, \\
&\mathcal{B}_{ab}^{614}=\bar{\ell}_{a}^{n} \,(I_{4,n} \,I_{1,b}+I_{4,b} \,I_{1,n})\,, \\
&\mathcal{B}_{ab}^{615}=\bar{\ell}_{a}^{n} \,(I_{5,n} \,I_{1,b}+I_{5,b} \,I_{1,n})\,, \\
&\mathcal{B}_{ab}^{624}=\bar{\ell}_{a}^{n} \,(I_{4,n} \,I_{2,b}+I_{4,b} \,I_{2,n})\,, \\
&\mathcal{B}_{ab}^{625}=\bar{\ell}_{a}^{n} \,(I_{5,n} \,I_{2,b}+I_{5,b} \,I_{2,n})\,, \\
&\mathcal{B}_{ab}^{645}=\bar{\ell}_{a}^{n} \,(I_{5,n} \,I_{4,b}+I_{5,b} \,I_{4,n})\,.
\end{split}
\end{equation}

Let us write the relations between the symmetry constraints $\mathcal{B}_{[ab]}^{\kappa}=0$ and $\mathcal{A}_{[ab]}^{\kappa}=0$ for transversely isotropic solids. The first nine relations linking the symmetry constraints in hyperelasticity with those in Cauchy elasticity are identical to those derived directly from \eqref{eq:hyper-Cauchy terms-relation-incomp-iso} in isotropic solids. According to \eqref{eq:universality terms-incom-T iso-hyper-1},~  \eqref{eq:universality terms-incom-T iso-hyper-2},~\eqref{eq:universal terms-incom-T iso-Cauchy elast-1} and \eqref{eq:universal terms-incom-T iso-Cauchy elast-2}, the remaining constraints are given by 
\begin{equation}  \label{eq:hyper-Cauchy constr-relation-incomp-T iso-1}
\begin{split}
&\mathcal{A}_{[ab]}^{4}=\mathcal{B}_{[ab]}^{4}=0\,, \\
&\mathcal{A}_{[ab]}^{5}=\mathcal{B}_{[ab]}^{5}=0\,, \\
&\mathcal{A}_{[ab]}^{44}=\mathcal{B}_{[ab]}^{44}=0\,, \\
&\mathcal{A}_{[ab]}^{55}=\mathcal{B}_{[ab]}^{55}=0\,, \\
&\mathcal{A}_{[ab]}^{14}=\mathcal{B}_{[ab]}^{14}+\mathcal{B}_{[ab]}^{41}=0\,, \\
&\mathcal{A}_{[ab]}^{15}=\mathcal{B}_{[ab]}^{15}+\mathcal{B}_{[ab]}^{51}=0\,, \\
&\mathcal{A}_{[ab]}^{24}=\mathcal{B}_{[ab]}^{24}+\mathcal{B}_{[ab]}^{42}=0\,, \\
&\mathcal{A}_{[ab]}^{25}=\mathcal{B}_{[ab]}^{25}+\mathcal{B}_{[ab]}^{52}=0\,, \\
&\mathcal{A}_{[ab]}^{45}=\mathcal{B}_{[ab]}^{45}+\mathcal{B}_{[ab]}^{54}=0\,,
\end{split}
\end{equation}
and
\begin{equation}  \label{eq:hyper-Cauchy constr-relation-incomp-T iso-2}
\begin{split}
&\mathcal{A}_{[ab]}^{444}=\mathcal{B}_{[ab]}^{444}=0\,,\\
&\mathcal{A}_{[ab]}^{555}=\mathcal{B}_{[ab]}^{555}=0\,,\\
&\mathcal{A}_{[ab]}^{114}=\mathcal{B}_{[ab]}^{114}+\mathcal{B}_{[ab]}^{411}=0\,, \\
&\mathcal{A}_{[ab]}^{115}=\mathcal{B}_{[ab]}^{115}+\mathcal{B}_{[ab]}^{511}=0\,, \\
&\mathcal{A}_{[ab]}^{124}=\mathcal{B}_{[ab]}^{124}+\mathcal{B}_{[ab]}^{214}+\mathcal{B}_{[ab]}^{412}=0\,, \\
&\mathcal{A}_{[ab]}^{125}=\mathcal{B}_{[ab]}^{125}+\mathcal{B}_{[ab]}^{215}+\mathcal{B}_{[ab]}^{512}=0\,, \\
&\mathcal{A}_{[ab]}^{144}=\mathcal{B}_{[ab]}^{144}+\mathcal{B}_{[ab]}^{414}=0\,, \\
&\mathcal{A}_{[ab]}^{145}=\mathcal{B}_{[ab]}^{145}+\mathcal{B}_{[ab]}^{415}+\mathcal{B}_{[ab]}^{514}=0\,, \\
&\mathcal{A}_{[ab]}^{155}=\mathcal{B}_{[ab]}^{155}+\mathcal{B}_{[ab]}^{515}=0\,, \\
&\mathcal{A}_{[ab]}^{224}=\mathcal{B}_{[ab]}^{224}+\mathcal{B}_{[ab]}^{422}=0\,, \\
&\mathcal{A}_{[ab]}^{225}=\mathcal{B}_{[ab]}^{225}+\mathcal{B}_{[ab]}^{522}=0\,, \\
&\mathcal{A}_{[ab]}^{244}=\mathcal{B}_{[ab]}^{244}+\mathcal{B}_{[ab]}^{424}=0\,, \\
&\mathcal{A}_{[ab]}^{245}=\mathcal{B}_{[ab]}^{245}+\mathcal{B}_{[ab]}^{425}+\mathcal{B}_{[ab]}^{524}=0\,, \\
&\mathcal{A}_{[ab]}^{255}=\mathcal{B}_{[ab]}^{255}+\mathcal{B}_{[ab]}^{525}=0\,, \\
&\mathcal{A}_{[ab]}^{445}=\mathcal{B}_{[ab]}^{445}+\mathcal{B}_{[ab]}^{544}=0\,, \\
&\mathcal{A}_{[ab]}^{455}=\mathcal{B}_{[ab]}^{455}+\mathcal{B}_{[ab]}^{545}=0\,.
\end{split}
\end{equation}
Therefore, there are a total of $75$ universality constraints in Cauchy elasticity, compared to $34$ in hyperelasticity. In what follows, we will prove that the universality constraints in transversely isotropic Cauchy elasticity and hyperelasticity are equivalent. Note that $\mathbf{n}, I_{1}, I_{2}, I_{4}$ and $I_{5}$ are assumed to be non-constant, although the result remains unchanged even if they are constant. 
Moreover, note that the twelve constraints of Cauchy elasticity and the nine of hyperelasticity for isotropic solids coincide, so the proof is not repeated here.

One of the common constraints is $\mathcal{A}_{[ab]}^{555}=\mathcal{B}_{[ab]}^{555}=0$ which implies that $\nabla I_5$ is an eigenvector of $\ell_{a}^{n}$, i.e., 
\begin{equation}  \label{eq:I_5-eigenvect-ell}
\ell_{a}^{n} \,I_{5,n}=\bar{\lambda}_{5} \,I_{5,a}\,,
\end{equation}
where $\bar{\lambda}_{5}$ is the corresponding eigenvalue. The second common constraint is $\mathcal{A}_{[ab]}^{444}=\mathcal{B}_{[ab]}^{444}=0$ which indicates that either $n_{a} \,I_{4,b}$ is symmetric or $n^{n} \,I_{4,n}=0$. The former entails that $\nabla I_4$ and $\mathbf{n}$ are parallel, that is,
\begin{equation}  \label{eq:n and I_4-parallel}
n_a=c_4 \,I_{4,a}\,,
\end{equation}
where $c_{4}$ is a scalar function. However, the latter is written as
\begin{equation}  \label{eq:I_4 and n-inner pro}
    I_{4,n} \,n^n
    = \llangle \nabla I_4,\mathbf{n} \rrangle_{\mathbf{g}}=0 \,,
\end{equation}
in which $\llangle .,. \rrangle_{\mathbf{g}}$ designates the inner product with respect to the metric tensor $\mathbf{g}$, implying that $\nabla I_4$ and $\mathbf{n}$ are orthogonal. As a result, the symmetry constraints in both hyperelasticity and Cauchy elasticity are satisfied if $\nabla I_4$ and $\mathbf{n}$ are either parallel or orthogonal. These cases are discussed separately.

\subsubsection{Case 1: $\mathbf{n \,(x)}$ and $\nabla I_4$ are parallel} \label{Incomp-T iso- n and I_4 are parallel}

Let us assume that $\nabla I_4$ and $\mathbf{n}$ are parallel. We first consider the terms $\mathcal{A}_{ab}^{\kappa}$ and $\mathcal{B}_{ab}^{i\kappa}$, where $\kappa$ is a three-component multi-index in hyperelasticity, or equivalently, a double index in Cauchy elasticity.

\vspace{1em}
\noindent\textbf{\small Symmetry equivalence of $\mathcal{A}_{ab}^{\kappa}$ and $\mathcal{B}_{ab}^{\kappa}$ with $\kappa$ a triple index.}
Consider the terms $\mathcal{A}_{ab}^{445}$ and $\mathcal{A}_{ab}^{455}$ in hyperelasticity. From \eqref{eq:I_5-eigenvect-ell} and \eqref{eq:n and I_4-parallel}, these terms can be rewritten as 
\begin{equation}  \label{eq:A445 and A455-incomp-T iso-n I_4 parallel}
\begin{split}
&\mathcal{A}_{ab}^{445}=c_{4} \,I_{4,a} \,I_{4,b} \,n^{n} \,I_{5,n}+c_{4} \,I_{4,a} \,I_{5,b} \,n^{n} \,I_{4,n}+\ell^{n}_{a} \,I_{4,n} \,I_{4,b} \,,\\
&\mathcal{A}_{ab}^{455}=\bar{\lambda}_{5} \,I_{5,a} \,I_{4,b}+ \ell^n_{a} \,I_{4,n} \,I_{5,b}+c_{4} \,I_{4,a} \,I_{5,b} \,n^{n} \,I_{5,n}\,.
\end{split}
\end{equation}
If the symmetric term $I_{4,a} \,I_{4,b}$ on the right-hand side of \eqref{eq:A445 and A455-incomp-T iso-n I_4 parallel}$_1$ is neglected, one can write the following expression representing the symmetry equivalence of $\mathcal{A}_{ab}^{445}$ 
\begin{equation}  \label{eq:A445-incomp-T iso-n I_4 parallel}
\mathcal{A}_{ab}^{445} \equiv c_{4} \,I_{4,a} \,I_{5,b} \,n^{n} \,I_{4,n}+\ell^{n}_{a} \,I_{4,n} \,I_{4,b} \,.
\end{equation}
The symmetry of \eqref{eq:A445-incomp-T iso-n I_4 parallel} implies that 
\begin{equation}  \label{eq:A445 symmetry-incomp-T iso-1}
c_{4} \,I_{4,a} \,I_{5,b} \,n^{n} \,I_{4,n}+\ell^{n}_{a} \,I_{4,n} \,I_{4,b}=c_{4} \,I_{4,b} \,I_{5,a} \,n^{n} \,I_{4,n}+\ell^{n}_{b} \,I_{4,n} \,I_{4,a}\,,
\end{equation}
which leads to
\begin{equation}  \label{eq:A445 symmetry-incomp-T iso-2}
\left(c_{4} \,n^{n} \,I_{4,n} \,I_{5,b}-\ell^{n}_{b} \,I_{4,n}\right) \,I_{4,a}=\left(c_{4} \,n^{n} \,I_{4,n} \,I_{5,a}-\ell^{n}_{a}\,I_{4,n}\right) \,I_{4,b}\,.
\end{equation}
Eq. \eqref{eq:A445 symmetry-incomp-T iso-2} holds if either
\begin{equation}  \label{eq:A445 symmetry-incomp-T iso-3}
c_{4} \,n^{n} \,I_{4,n} \,I_{5,a}-\ell^{n}_{a} \,I_{4,n}=\bar{c}_1 \,I_{4,a}\,,
\end{equation}
or
\begin{equation}  \label{eq:A445 symmetry-incomp-T iso-4}
\ell^{n}_{a} \,I_{4,n}=c_{4} \,n^{n} \,I_{4,n} \,I_{5,a}-\bar{c_1} \,I_{4,a}\,,
\end{equation}
where $\bar{c}_1$ is a scalar function. Substituting \eqref{eq:A445 symmetry-incomp-T iso-4} into \eqref{eq:A445 and A455-incomp-T iso-n I_4 parallel}$_2$ gives us
\begin{equation}  \label{eq:A455-incomp-T iso-1}
\mathcal{A}_{ab}^{455}=\bar{\lambda}_{5} \,I_{5,a} \,I_{4,b}+\left(c_{4} \,n^{n} \,I_{4,n} \,I_{5,a}-\bar{c}_1\,I_{4,a}\right) \,I_{5,b}+c_{4} \,I_{4,a} \,I_{5,b} \,n^{n} \,I_{5,n}\,.
\end{equation}
Again, omitting the symmetric term $I_{5,a} \,I_{5,b}$ in \eqref{eq:A455-incomp-T iso-1} yields
\begin{equation}  \label{eq:A455-incomp-T iso-2}
\mathcal{A}_{ab}^{455}\equiv \bar{\lambda}_{5} \,I_{5,a} \,I_{4,b}+(c_{4} \,n^{n} \,I_{5,n}-\bar{c_1}) \,I_{4,a} \,I_{5,b}\,.
\end{equation}
Therefore, if $\bar{\lambda}_{5}\ne c_{4} \,n^{n}\,I_{5,n}-\bar{c_{1}}$, the term $\mathcal{A}_{ab}^{455}$ is symmetric if and only if $\nabla I_4$ and $\nabla I_5$ are parallel, i.e.,
\begin{equation}  \label{eq:I_4,I_5 parallel}
I_{4,a}=c_{45} \,I_{5,a} \,,
\end{equation}
where $c_{45}$ is a scalar function. It should be noted that, as given by \eqref{eq:A455-incomp-T iso-2}, the other condition preserving the symmetry of $\mathcal{A}_{ab}^{455}$ is $\bar{\lambda}_{5} = c_{4} \,n^{n} \,I_{5,n} - \bar{c}_1$. However, this is a highly specific case that is not satisfied by any class of the universal deformations in hyperelasticity (this can be verified by the results in \citep{YavariGoriely2021}). Thus, \eqref{eq:I_4,I_5 parallel} is the only solution preserving the symmetries of $\mathcal{A}_{ab}^{445}$ and $\mathcal{A}_{ab}^{455}$ when $\mathbf{n}$ and $\nabla I_4$ are parallel. We can apply a similar approach to examine $\mathcal{A}_{ab}^{114}$ and $\mathcal{A}_{ab}^{144}$. From \eqref{eq:n and I_4-parallel} we have
\begin{equation}  \label{eq:A144 and A114-incomp-T iso-n I_4 parallel}
\begin{split}
&\mathcal{A}_{ab}^{144}=b^n_{a} \,I_{4,n} \,I_{4,b}+c_{4} \,n^{n} \,I_{4,n} \,I_{4,a} \,I_{1,b}+c_{4} \,n^{n} \,I_{1,n} \,I_{4,a} \,I_{4,b}\,,\\
&\mathcal{A}_{ab}^{114}=b^n_{a} \,I_{4,n} \,I_{1,b}+b^n_{a} \,I_{1,n} \,I_{4,b}+c_{4} \,n^{n} \,I_{1,n} \,I_{4,a} \,I_{1,b}\,.
\end{split}
\end{equation}
After ignoring the symmetric term, the symmetry of \eqref{eq:A144 and A114-incomp-T iso-n I_4 parallel}$_1$ gives the following relation
\begin{equation}  \label{eq:A144 symmetry-incomp-T iso-Case 1}
b^n_{a} \,I_{4,n}=c_{4} \,n^{n} \,I_{4,n} \,I_{1,a}-\bar{c}_{2} \,I_{4,a}\,,
\end{equation}
where $\bar{c}_{2}$ is a scalar function. Substituting \eqref{eq:A144 symmetry-incomp-T iso-Case 1} into \eqref{eq:A144 and A114-incomp-T iso-n I_4 parallel}$_{2}$ and employing \eqref{eq:I_1 - I_2 relation}$_2$, $\mathcal{A}_{ab}^{114}$ reduces to
\begin{equation}  \label{eq:A114 symmetry-incomp-T iso-Case 1-1}
\mathcal{A}_{ab}^{114}=c_{4} \,n^{n} \,I_{4,n} \,I_{1,a} \,I_{1,b}-\bar{c}_{2} \,I_{4,a} \,I_{1,b}+\lambda_{1} \,I_{1,a} \,I_{4,b}+c_{4} \,n^{n} \,I_{1,n} \,I_{4,a} \,I_{1,b}\,.
\end{equation}
Eq. \eqref{eq:A114 symmetry-incomp-T iso-Case 1-1} can be rewritten in a simplified form as follows
\begin{equation}  \label{eq:A114 symmetry-incomp-T iso-Case 1-2}
\mathcal{A}_{ab}^{114}=(c_{4} \,n^{n} \,I_{1,n}-\bar{c}_{2}) \,I_{4,a} \,I_{1,b}+\lambda_{1} \,I_{1,a} \,I_{4,b}\,.
\end{equation}
Thus, the symmetry of $\mathcal{A}_{ab}^{114}$ indicates that $\nabla I_1$ and $\nabla I_4$ are parallel. This remains the sole solution as the other possibility $\lambda_{1}=c_{4} \,n^{n} \,I_{1,n}-\bar{c}_{2}$, which is a very specific case, and is not satisfied by any of the universal deformations.

Similarly, the symmetries of $\left\{\mathcal{A}_{ab}^{115},\mathcal{A}_{ab}^{155}\right\}$, $\left\{\mathcal{A}_{ab}^{224},\mathcal{A}_{ab}^{244}\right\}$, and $\left\{\mathcal{A}_{ab}^{225},\mathcal{A}_{ab}^{255}\right\}$ imply, respectively, that $I_{1,a}=c_{15} \,I_{5,a}$, $I_{2,a}=c_{24} \,I_{4,a}$, and $I_{2,a}=c_{25} \,I_{5,a}$, where $c_{15}$, $c_{24}$ and $c_{25}$ are scalar functions. Given these results, the symmetries of the remaining terms $\left\{\mathcal{A}_{ab}^{124},\mathcal{A}_{ab}^{125},\mathcal{A}_{ab}^{145},\mathcal{A}_{ab}^{245}\right\}$ hold identically. Therefore, the symmetries of $\mathcal{A}_{ab}^{\kappa}$, where $\kappa$ is a triple index, hold if and only if $\mathbf{n}$, $\nabla I_1$, $\nabla I_2$, $\nabla I_4$ and $\nabla I_5$ are parallel. To prove the symmetry equivalence of $\mathcal{A}_{ab}^{\kappa}$ and $\mathcal{B}_{ab}^{i\kappa}$, it suffices to show that this condition is a unique solution that satisfies the symmetries of the Cauchy elasticity constraints as well. In what follows, this is discussed in detail.

Recall that $\mathbf{n}$ and $\nabla I_4$ are parallel, and that $\nabla I_5$ is an eigenvector of $\ell^{n}_{a}$, which together preserve the symmetries of $\mathcal{B}_{ab}^{444}$ and $\mathcal{B}_{ab}^{555}$ in Cauchy elasticity. Based on these relations, one can express $\mathcal{B}_{ab}^{445}$ as
\begin{equation}  \label{eq:B445 incomp-T iso-Case 1-1}
\mathcal{B}_{ab}^{445} =c_{4} \,I_{4,a} \,I_{4,b} \,n^{n} \,I_{5,n}+c_{4} \,I_{4,a} \,I_{5,b} \,n^{n} \,I_{4,n}\,,
\end{equation}
which after neglecting the symmetric term simplifies to
\begin{equation}  \label{eq:B445 incomp-T iso-Case 1-2}
\mathcal{B}_{ab}^{445} \equiv c_{4} \,I_{4,a} \,I_{5,b} \,n^{n} \,I_{4,n}\,. 
\end{equation}
Since $n^{n} \,I_{4,n}\ne 0$, from the symmetry of \eqref{eq:B445 incomp-T iso-Case 1-2} it follows that $\nabla I_4$ and $\nabla I_5$ are parallel. Thus, $\mathcal{B}_{ab}^{544}$, $\mathcal{B}_{ab}^{455}$ and $\mathcal{B}_{ab}^{554}$ are also symmetric. Equivalently, this may be written as
\begin{equation}  \label{eq:hyper-Cauchy elast-equiv-incomp-T iso-Case1-1}
\left\{\mathcal{B}_{ab}^{445},\mathcal{B}_{ab}^{544},\mathcal{B}_{ab}^{455},\mathcal{B}_{ab}^{554} \right\} \equiv \left\{\mathcal{A}_{ab}^{445},\mathcal{A}_{ab}^{455} \right\} \,. 
\end{equation}
Similarly, we again consider \eqref{eq:n and I_4-parallel} to rewrite $\mathcal{B}_{ab}^{441}$ as follows
\begin{equation}  \label{eq:B441-incopm-T iso-Case1-1}
\mathcal{B}_{ab}^{441} =c_{4} \,I_{4,a} \,I_{4,b} \,n^{n} \,I_{1,n}+c_{4} \,I_{4,a} \,I_{1,b} \,n^{n} \,I_{4,n}\,.
\end{equation}
Since the first term on the right-hand side of \eqref{eq:B441-incopm-T iso-Case1-1} is symmetric and $n^{n} \,I_{4,n} \ne 0$, the symmetry of $\mathcal{B}_{ab}^{441}$ reduces to that of $c_4 \,I_{4,a} \,I_{1,b}$, implying that $\nabla I_4$ and $\nabla I_1$ are parallel. Consequently, $\mathcal{B}_{[ab]}^{144}=0$, $\mathcal{B}_{[ab]}^{411}=0$ and $\mathcal{B}_{[ab]}^{114}=0$ hold identically, and hence 
\begin{equation}  \label{eq:hyper-Cauchy elast-equiv-incomp-T iso-Case1-2}
\left\{\mathcal{B}_{ab}^{441},\mathcal{B}_{ab}^{144},\mathcal{B}_{ab}^{411},\mathcal{B}_{ab}^{114}\right\} \equiv \left\{\mathcal{A}_{ab}^{114},\mathcal{A}_{ab}^{144} \right\} \,. 
\end{equation}
Proceeding with the same approach leads to the following results
\begin{equation}  \label{eq:hyper-Cauchy elast-equiv-incomp-T iso-Case1-3}
\begin{split}
&\left\{\mathcal{B}_{ab}^{115},\mathcal{B}_{ab}^{511},\mathcal{B}_{ab}^{155},\mathcal{B}_{ab}^{551} \right\} \equiv \left\{\mathcal{A}_{ab}^{115},\mathcal{A}_{ab}^{155}\right\} \,, \\
&\left\{\mathcal{B}_{ab}^{224},\mathcal{B}_{ab}^{422},\mathcal{B}_{ab}^{244},\mathcal{B}_{ab}^{442} \right\} \equiv \left\{\mathcal{A}_{ab}^{224},\mathcal{A}_{ab}^{244}\right\} \,, \\
&\left\{\mathcal{B}_{ab}^{225},\mathcal{B}_{ab}^{522},\mathcal{B}_{ab}^{255},\mathcal{B}_{ab}^{552} \right\} \equiv \left\{\mathcal{A}_{ab}^{225},\mathcal{A}_{ab}^{255} \right\} \,, 
\end{split}
\end{equation}
which is equivalent to saying that $\nabla I_i \,(i=1,2,4,5)$ are parallel. The remaining terms $\mathcal{B}_{ab}^{124}$, $\mathcal{B}_{ab}^{214}$, $\mathcal{B}_{ab}^{412}$, $\mathcal{B}_{ab}^{125}$, $\mathcal{B}_{ab}^{215}$, $\mathcal{B}_{ab}^{512}$, $\mathcal{B}_{ab}^{145}$, $\mathcal{B}_{ab}^{415}$, $\mathcal{B}_{ab}^{514}$, $\mathcal{B}_{ab}^{245}$, $\mathcal{B}_{ab}^{425}$,  and $\mathcal{B}_{ab}^{524}$ can then be shown to be trivially symmetric.

Since $\mathbf{n}$ and $\nabla I_1$ are parallel, $\mathbf{n}$ is an eigenvector of $b^{n}_{a}$, that is, $b_{a}^{n} \,n_{n} = \lambda_{1} \,n_{a}$. As a result, we have
\begin{align}  
&\ell_{a}^{n}=2 \,\lambda_{1} \,n_{a} \,n^{n}\,, \label{eq:ell and n-incomp-T iso-Case1}\\
&\bar{\ell}_{a}^{n}=\frac{2}{\lambda_1} \,n_{a} \,n^n \label{eq:ellbar and n-incomp-T iso-Case1}\,.
\end{align}
With the given relations, it is straightforward to show that
\begin{equation}  \label{eq:cauchy term-additional-equiv-incomp-T iso-Case1}
\begin{split}
&\mathcal{B}_{ab}^{611}\equiv\mathcal{B}_{ab}^{511}\,, \\
&\mathcal{B}_{ab}^{622}\equiv\mathcal{B}_{ab}^{522}\,, \\
&\mathcal{B}_{ab}^{644}\equiv\mathcal{B}_{ab}^{544}\,, \\
&\mathcal{B}_{ab}^{655}\equiv\mathcal{B}_{ab}^{555}\,, \\
&\mathcal{B}_{ab}^{612}\equiv\mathcal{B}_{ab}^{512}\,, \\
&\mathcal{B}_{ab}^{614}\equiv\mathcal{B}_{ab}^{514}\,, \\
&\mathcal{B}_{ab}^{615}\equiv\mathcal{B}_{ab}^{515}\,, \\
&\mathcal{B}_{ab}^{624}\equiv\mathcal{B}_{ab}^{524}\,, \\
&\mathcal{B}_{ab}^{625}\equiv\mathcal{B}_{ab}^{525}\,, \\
&\mathcal{B}_{ab}^{645}\equiv\mathcal{B}_{ab}^{545}\,.
\end{split}
\end{equation} 

In conclusion, the symmetry constraints associated with the terms $\mathcal{A}_{ab}^{\kappa}$, where $\kappa$ is a triple index, in hyperelasticity and those corresponding to $\mathcal{B}_{ab}^{1\kappa}$, $\mathcal{B}_{ab}^{2\kappa}$, $\mathcal{B}_{ab}^{4\kappa}$, $\mathcal{B}_{ab}^{5\kappa}$, and $\mathcal{B}_{ab}^{6\kappa}$, where $\kappa$ is a double index, in Cauchy elasticity are equivalent in Case 1. Both sets are symmetric if and only if $\mathbf{n}$, $\nabla I_1$, $\nabla I_2$, $\nabla I_4$, and $\nabla I_5$ are parallel.

\vspace{1em}
\noindent\textbf{\small Symmetry equivalence of $\mathcal{A}_{ab}^{\kappa}$ and $\mathcal{B}_{ab}^{\kappa}$ with $\kappa$ a double index.}
We next turn our attention to the terms $\mathcal{A}_{ab}^{\kappa}$ and $\mathcal{B}_{ab}^{i\kappa}$, where $\kappa$ is a two-component multi-index in hyperelasticity or a single index in Cauchy elasticity. First, $\mathcal{A}_{ab}^{44}$ and $\mathcal{B}_{ab}^{44}$ are expanded as follows 
\begin{equation}  \label{eq:A44 and B44-incomp-T iso-Case1-1}
\mathcal{A}_{ab}^{44}=\mathcal{B}_{ab}^{44}={n_a}_{|n} \,n^n \, I_{4,b}+n_{a} \,n_{|n}^{n} \,I_{4,b}+{n_a}_{|b} \,n^n \,I_{4,n}+n_{a} \, \left(n^n \,I_{4,n}\right)_{|b}\,. 
\end{equation}
Using \eqref{eq:n and I_4-parallel}, the above expression simplifies to read
\begin{equation}  \label{eq:A44 and B44-incomp-T iso-Case1-2}
\mathcal{A}_{ab}^{44}=\mathcal{B}_{ab}^{44}={n_a}_{|n} \,n^n \, I_{4,b}+c_4 \,n_{|n}^{n} \,I_{4,a} \,I_{4,b}+c_{4,b} \,I_{4,a} \,n^n \,I_{4,n}+c_4 \,I_{4|ab} \,n^n \,I_{4,n}+c_4 \,I_{4,a} \,\left(n^n \,I_{4,n}\right)_{|b}\,.
\end{equation} 
Because the terms $I_{4,a} \,I_{4,b}$ and $I_{4|ab}$ in \eqref{eq:A44 and B44-incomp-T iso-Case1-2} are symmetric, the symmetry equivalence is described by
\begin{equation}  \label{eq:A44 and B44-incomp-T iso-Case1-3} 
\mathcal{A}_{ab}^{44}=\mathcal{B}_{ab}^{44} \equiv {n_a}_{|n} \,n^n \, I_{4,b}+c_{4,b} \,I_{4,a}\,n^n \,I_{4,n} +c_4 \,I_{4,a}\,\left(n^n \,I_{4,n}\right)_{|b}\,.
\end{equation}
Hence
\begin{equation}  \label{eq:A44 and B44-incomp-T iso-Case1-4} 
\mathcal{A}_{ab}^{44}=\mathcal{B}_{ab}^{44} \equiv {n_a}_{|n} \,n^n \, I_{4,b}+\left(c_4 \,I_{4,n} \,n^n\right)_{|b} \,I_{4,a}\,,  
\end{equation}
or
\begin{equation}  \label{eq:A44 and B44-incomp-T iso-Case1-5} 
\mathcal{A}_{ab}^{44}=\mathcal{B}_{ab}^{44} \equiv {n_a}_{|n} \,n^n \, I_{4,b}+\left(n_n \,n^n\right)_{|b} \,I_{4,a}\,.
\end{equation}
The symmetry of \eqref{eq:A44 and B44-incomp-T iso-Case1-5} is represented as follows
\begin{equation}  \label{eq:A44 and B44 symmetry-incomp-T iso-Case1-1} 
{n_a}_{|n} \,n^n \, I_{4,b}+\left(n_n \,n^n\right)_{|b} \,I_{4,a}={n_b}_{|n} \,n^n \, I_{4,a}+\left(n_n \,n^n\right)_{|a} \,I_{4,b} \,,
\end{equation} 
which can be written in a more simplified form as
\begin{equation}  \label{eq:A44 and B44 symmetry-incomp-T iso-Case1-2} 
\left[{n_a}_{|n} \,n^n-\left(n_n \,n^n\right)_{|a}\right] \,I_{4,b}=\left[{n_b}_{|n} \,n^n-\left(n_n \,n^n\right)_{|b}\right] \,I_{4,a}\,.
\end{equation}
Eq.~\eqref{eq:A44 and B44 symmetry-incomp-T iso-Case1-2} suggests that   
\begin{equation}  \label{eq:A44 and B44 symmetry-incomp-T iso-Case1-Final} 
{n_a}_{|n} \,n^n-\left(n_n \,n^n\right)_{|a}=\bar{c}_3 \,I_{4,a}\,,
\end{equation}
where $\bar{c}_{3}$ is a scalar function. Now, consider $\mathcal{B}_{ab}^{41}$. We proceed as in the case of $\mathcal{B}_{ab}^{44}$, taking $n_{a}=c_{1}\,I_{1,a}$, where $c_{1}$ is a scalar function, and neglecting the symmetric terms. This yields the following expression
\begin{equation}  \label{eq:B41 symmetry-incomp-T iso-Case1-1} 
\mathcal{B}_{ab}^{41} \equiv {n_a}_{|n} \,n^n \, I_{1,b}+(n_n \,n^n)_{|b} \,I_{1,a}\,.
\end{equation}
From \eqref{eq:A44 and B44 symmetry-incomp-T iso-Case1-Final}, we have
\begin{equation}  \label{eq:na|n-incomp-T iso-Case1}
{n_a}_{|n} \,n^n=\bar{c}_3 \,I_{4,a}+\left(n_n \,n^n\right)_{|a}\,.
\end{equation}
Substituting \eqref{eq:na|n-incomp-T iso-Case1} into \eqref{eq:B41 symmetry-incomp-T iso-Case1-1} and taking $I_{4,a}=c_{41} \,I_{1,a}$ yields
\begin{equation}  \label{eq:B41 symmetry-incomp-T iso-Case1-2}
\mathcal{B}_{ab}^{41} \equiv \bar{c}_3 \, \,c_{41} \,I_{1,a} \,I_{1,b}+\left(n_n \,n^n\right)_{|a} \,I_{1,b}+\left(n_n \,n^n\right)_{|b} \,I_{1,a}\,,
\end{equation}
which is clearly symmetric. This means that if $\mathcal{B}_{ab}^{44}$ is symmetric, then $\mathcal{B}_{ab}^{41}$ will also be symmetric. Therefore,
\begin{equation}  \label{eq:B41-equiv-B44-incomp-T iso-Case1}
\mathcal{B}_{ab}^{41} \equiv \mathcal{B}_{ab}^{44}\,.
\end{equation}
Similarly
\begin{equation}  \label{eq:B42,B45 equiv B44-incomp-T iso-Case1}
\mathcal{B}_{ab}^{42} \equiv \mathcal{B}_{ab}^{44}\,, \qquad
\mathcal{B}_{ab}^{45} \equiv \mathcal{B}_{ab}^{44}\,.
\end{equation}
Moreover, we may use \eqref{eq:ell and n-incomp-T iso-Case1} to rewrite $\mathcal{A}_{ab}^{55} \,(=\mathcal{B}_{ab}^{55})$ as follows
\begin{equation}  \label{eq:A55 B55-incomp-T iso-Case1-1} 
\mathcal{A}_{ab}^{55}=\mathcal{B}_{ab}^{55}=\left(2 \,\lambda_{1} \,n^n \,n_{a}\right)_{|n} \,I_{5,b}+\left(2 \,\lambda_{1} \,n_{a} \,n^n \,I_{5,n}\right)_{|b}\,,
\end{equation}
which is simplified as
\begin{equation}  \label{eq:A55 B55-incomp-T iso-Case1-2}
\mathcal{A}_{ab}^{55}=\mathcal{B}_{ab}^{55}={n_a}_{|n} \,(2\, \lambda_{1} \,n^n)\, I_{5,b}+n_{a} \,{\left(2\,\lambda_{1}\,n^n \right)}_{|n} \,I_{5,b}+{n_a}_{|b} \,(2\, \lambda_{1} \,n^n \,I_{5,n})+n_{a}\, \left(2\,\lambda_{1} \,n^n \,I_{5,n}\right)_{|b}\,. 
\end{equation}
Following an approach similar to that used in \eqref{eq:A44 and B44-incomp-T iso-Case1-2}--\eqref{eq:A44 and B44 symmetry-incomp-T iso-Case1-2}, one can show that the symmetry of $\mathcal{A}_{ab}^{55}$ or $\mathcal{B}_{ab}^{55}$ results in 
\begin{equation}  \label{eq:A55 B55-incomp-T iso-Case1-Final}
{n_a}_{|n} \, (2\, \lambda_{1} \,n^n)=\bar{c}_4 \, I_{5,a}+\left(2\, \lambda_{1} \,n_n \,n^n\right)_{|a}\,,
\end{equation}
and that the following symmetry equivalence also holds 
\begin{equation}  \label{eq:B51 symmetry-incomp-T iso-Case1-1} 
\mathcal{B}_{ab}^{51} \equiv {n_a}_{|n} \,(2 \,\lambda_{1} \,n^n) \,I_{1,b}+\left(2 \,\lambda_{1} \,n_n \,n^n\right)_{|b} \,I_{1,a}\,,
\end{equation}
where $\bar{c}_4$ is a scalar function. We substitute \eqref{eq:A55 B55-incomp-T iso-Case1-Final} into \eqref{eq:B51 symmetry-incomp-T iso-Case1-1} to get
\begin{equation}  \label{eq:B51 symmetry-incomp-T iso-Case1-2} 
\mathcal{B}_{ab}^{51} \equiv \bar{c}_4 \, I_{5,a} \,I_{1,b}+\left(2\, \lambda_{1} \,n_n \,n^n\right)_{|a} \,I_{1,b}+\left(2 \,\lambda_{1} \,n_n \,n^n\right)_{|b} \,I_{1,a}\,. 
\end{equation}
Due to the functional dependence of $I_1$ and $I_5$, the right-hand side of \eqref{eq:B51 symmetry-incomp-T iso-Case1-2} is symmetric. As a result, the symmetry of $\mathcal{B}_{ab}^{51}$ is equivalent to the symmetry of $\mathcal{B}_{ab}^{55}$, i.e.,
\begin{equation}  \label{eq:B51 B55 symmetry-equiv-incomp-T iso-Case1} 
\mathcal{B}_{ab}^{51} \equiv \mathcal{B}_{ab}^{55}\,.
\end{equation}
The following results can be obtained in a similar manner
\begin{equation}  \label{eq:B52,B54- B55 symmetry-equiv-incomp-T iso-Case1}
\mathcal{B}_{ab}^{52} \equiv \mathcal{B}_{ab}^{55}\,, \qquad
\mathcal{B}_{ab}^{54} \equiv \mathcal{B}_{ab}^{55}\,.
\end{equation}

With respect to \eqref{eq:ell and n-incomp-T iso-Case1}, $\mathcal{B}_{ab}^{51}$ takes the following form
\begin{equation}  \label{eq: B51-second form-incomp-T iso-case1-1}
\mathcal{B}_{ab}^{51}= \left(2 \,\lambda_{1} \,n^n \,n_{a}\right)_{|n} \,I_{1,b}+\left(2 \,\lambda_{1} \,n_{a} \,n^n \,I_{1,n}\right)_{|b}\,,
\end{equation}
which can be further simplified to
\begin{equation}  \label{eq: B51-second form-incomp-T iso-case1-2}
\mathcal{B}_{ab}^{51}= 2 \,\lambda_{1} \,\left[ \left(\,n^n \,n_{a}\right)_{|n} \,I_{1,b}+\left(n_{a} \,n^n \,I_{1,n}\right)_{|b}\right]+2 \,\lambda_{1,n} \,n^n \,n_{a} \,I_{1,b}+2\,\lambda_{1,b} \,n_{a} \,n^n \,I_{1,n}\,,
\end{equation}
or
\begin{equation}  \label{eq: B51-second form-incomp-T iso-case1-3}
\mathcal{B}_{ab}^{51}= 2 \,\lambda_{1} \,\mathcal{B}_{ab}^{41}+2 \,\lambda_{1,n} \,n^n \,n_{a} \,I_{1,b}+2\,\lambda_{1,b} \,n_{a} \,n^n \,I_{1,n}\,.
\end{equation}
Given that $\mathcal{B}_{ab}^{41}$ is symmetric and $\mathbf{n}$ and $\nabla I_1$ are parallel, the following relation is implied
\begin{equation}  \label{eq: B51-second form-equiv-incomp-T iso-case1}
\mathcal{B}_{ab}^{51} \equiv 2\,\lambda_{1,b} \,n_{a} \,n^n \,I_{1,n}\,.
\end{equation}
Now, let us use the relation~\eqref{eq:ellbar and n-incomp-T iso-Case1} to write $\mathcal{B}_{ab}^{61}$ as
\begin{equation}  \label{eq: B61-incomp-T iso-case1-1}
\mathcal{B}_{ab}^{61}= \left(\frac{2}{\lambda_{1}} \,n^n \,n_{a}\right)_{|n} \,I_{1,b}+\left(\frac{2}{\lambda_{1}} \,n_{a} \,n^n \,I_{1,n}\right)_{|b}\,.
\end{equation}

We can perform a similar manipulation for $\mathcal{B}_{ab}^{61}$ to obtain
\begin{equation}  \label{eq: B61-equiv-incomp-T iso-case1}
\mathcal{B}_{ab}^{61} \equiv -\frac{2\,\lambda_{1,b}}{\lambda_{1}^{2}} \,n_{a} \,n^n \,I_{1,n}\,.
\end{equation}
Hence,
\begin{equation}  \label{eq: B61 B51-equiv-incomp-T iso-case1}
\mathcal{B}_{ab}^{61} \equiv \mathcal{B}_{ab}^{51}\,,
\end{equation}
and similarly
\begin{equation}  \label{eq:B6i-B5i-equiv-incomp-T iso-Case1}
\mathcal{B}_{ab}^{62} \equiv \mathcal{B}_{ab}^{52}\,, \qquad
\mathcal{B}_{ab}^{64} \equiv \mathcal{B}_{ab}^{54}\,, \qquad
\mathcal{B}_{ab}^{65} \equiv \mathcal{B}_{ab}^{55}\,.
\end{equation}
Therefore, although this part involves $16$ universality constraints in Cauchy elasticity, only $6$ of them, namely those associated with the symmetries of the terms $\mathcal{B}_{ab}^{44}$, $\mathcal{B}_{ab}^{55}$, $\mathcal{B}_{ab}^{15}$, $\mathcal{B}_{ab}^{25}$, $\mathcal{B}_{ab}^{14}$ and $\mathcal{B}_{ab}^{24}$ are independent, and these correspond to $6$ independent universality constraints in hyperelasticity, as given by \eqref{eq:hyper-Cauchy constr-relation-incomp-T iso-1}$_3$--\eqref{eq:hyper-Cauchy constr-relation-incomp-T iso-1}$_8$. Note that the symmetry of $\mathcal{A}_{ab}^{45}$ is not an independent constraint in this case because with respect to \eqref{eq:hyper-Cauchy constr-relation-incomp-T iso-1}$_3$, \eqref{eq:hyper-Cauchy constr-relation-incomp-T iso-1}$_4$, \eqref{eq:hyper-Cauchy constr-relation-incomp-T iso-1}$_9$, \eqref{eq:B42,B45 equiv B44-incomp-T iso-Case1}$_2$ and \eqref{eq:B52,B54- B55 symmetry-equiv-incomp-T iso-Case1}$_2$, one can show that
\begin{equation}  \label{eq:A45-equiv-incomp-T iso-Case1}
\mathcal{A}_{ab}^{45}\equiv \mathcal{A}_{ab}^{44}+\mathcal{A}_{ab}^{55} \,,
\end{equation}
and hence there are $6$ independent symmetry constraints in hyperelasticity in this part. 

In conclusion, in Case 1 the symmetries of the terms $\mathcal{A}_{ab}^{\kappa}$, where $\kappa$ is a double index, in hyperelasticity are equivalent to the symmetries of $\mathcal{B}_{ab}^{1\kappa}$, $\mathcal{B}_{ab}^{2\kappa}$, $\mathcal{B}_{ab}^{4\kappa}$, $\mathcal{B}_{ab}^{5\kappa}$, and $\mathcal{B}_{ab}^{6\kappa}$, where $\kappa$ is a single index, in Cauchy elasticity.

\vspace{1em}
\noindent\textbf{\small Symmetry equivalence of $\mathcal{A}_{ab}^{\kappa}$ and $\mathcal{B}_{ab}^{\kappa}$ with $\kappa$ a single index.}
Finally, we consider $\mathcal{A}_{ab}^{\kappa}$ and $\mathcal{B}_{ab}^{i\kappa}$, where $\kappa$ is a single index in hyperelasticity, or equivalently, $\kappa = 0$ in Cauchy elasticity. Since we have
\begin{equation}  \label{eq:A4,A5-B4,B5-relation-incomp-T iso-Case1}
\mathcal{A}_{ab}^{4}=\mathcal{B}_{ab}^{4}\,, \qquad
\mathcal{A}_{ab}^{5}=\mathcal{B}_{ab}^{5}\,,
\end{equation}
we only need to prove that the symmetry of the remaining term in Cauchy elasticity, i.e., $\mathcal{B}_{ab}^{6}$, does not admit an independent constraint. To this end, relation~\eqref{eq:ellbar and n-incomp-T iso-Case1} is applied to \eqref{eq:universal terms-incom-T iso-Cauchy elast-3}$_1$ to obtain
\begin{equation}  \label{eq:B6-incomp-T iso-Case1-1}
\mathcal{B}_{ab}^{6} =\left(\frac{2}{\lambda_1} \,n^n \,n_{a}\right)_{|nb}\,,
\end{equation}
which is expanded as
\begin{equation}  \label{eq:B6-incomp-T iso-Case1-2}
\mathcal{B}_{ab}^{6} =\left(\frac{2}{\lambda_1}\right)_{|nb} \,n^n \,n_{a}+\left(\frac{2}{\lambda_1}\right)_{|n} \,\left(n^n \,n_a\right)_{|b}+\left(\frac{2}{\lambda_1}\right)_{|b} \,\left(n^n \,n_a\right)_{|n}+\frac{2}{\lambda_1} \,\left(n^n \,n_a\right)_{|nb} \,.
\end{equation}
Because the term $(n^n \,n_a)_{|nb}$ is symmetric according to the symmetry of $\mathcal{B}_{ab}^{4}$, this term can be ignored. After some simplifications, \eqref{eq:B6-incomp-T iso-Case1-2} can be expressed as
\begin{equation}  \label{eq:B6-equiv-incomp-T iso-Case1-1}
\mathcal{B}_{ab}^{6} \equiv -\left(\frac{2}{\lambda _{1}^{2}}\right)_{|b} \,\lambda_{1,n} \,n^n \,n_a-\frac{1}{\lambda _{1}^{2}} \,(2 \,\lambda_{1} \,n_a \,n^n)_{|nb}\,. 
\end{equation}
Taking \eqref{eq:ell and n-incomp-T iso-Case1} into account, we know that the term $(2 \,\lambda_{1} \,n_a \,n^n)_{|nb}$ is equal to ${\ell_{a}^{n}}_{|nb}$, and hence is symmetric due to the symmetry of $\mathcal{B}_{ab}^{5}$. So, the symmetry equivalence \eqref{eq:B6-equiv-incomp-T iso-Case1-1} takes the form
\begin{equation}  \label{eq:B6-equiv-incomp-T iso-Case1-2}
\mathcal{B}_{ab}^{6} \equiv \frac{4}{\lambda_{1}^{3}} \,\lambda_{1,b} \,\lambda_{1,n} \,n^n \,n_a\,.
\end{equation}
To prove the symmetry of \eqref{eq:B6-equiv-incomp-T iso-Case1-2}, it is enough to show that $\nabla \lambda_1$ and $\mathbf{n}$ are parallel. For this reason, attention is given to \eqref{eq: B51-second form-equiv-incomp-T iso-case1}. From this equation, the following term must be symmetric
\begin{equation}  \label{eq:lambda1,b-n-incomp-T iso-Case1}
\lambda_{1,b} \,n_a \,n^n \,I_{1,n}\,.
\end{equation}
Since $n^n \,I_{1,n} \ne 0$, the symmetry of \eqref{eq:lambda1,b-n-incomp-T iso-Case1} indicates that $\nabla \lambda_1$ and $\mathbf{n}$ are parallel. Accordingly, \eqref{eq:B6-equiv-incomp-T iso-Case1-2} is symmetric and the symmetry condition of the term $\mathcal{B}_{ab}^{6}$ is satisfied. Therefore, $\mathcal{B}_{[ab]}^{6}=0$ is not an independent symmetry constraint in Cauchy elasticity in this case. As a result, the symmetries of the terms $\mathcal{A}_{ab}^{\kappa}$, where $\kappa$ is a single index in hyperelasticity, and the symmetries of $\mathcal{B}_{ab}^{4}$, $\mathcal{B}_{ab}^{5}$, and $\mathcal{B}_{ab}^{6}$ in Cauchy elasticity are equivalent. 

So far, we have assumed that $\mathbf{n}, I_{1}, I_{2}, I_{4}$ and $I_{5}$ are not constant. However, it can be readily shown that if any of them or any combination thereof becomes constant (as in Family 5, where $I_i$ are constant), the equivalence still holds. For example, suppose that $I_{4}$ is constant. Then,
\begin{equation}  \label{eq:example-I4 const-incomp-T iso-Case1-1}
\begin{split}
&\mathcal{A}_{ab}^{44}=\mathcal{B}_{ab}^{44}=0\,, \\
&\mathcal{B}_{ab}^{14}=\mathcal{B}_{ab}^{24}=\mathcal{B}_{ab}^{54}=\mathcal{B}_{ab}^{64}=0\,,\\
&\mathcal{A}_{ab}^{444}=\mathcal{B}_{ab}^{444}=0\,,\\
&\mathcal{A}_{ab}^{144}=\mathcal{B}_{ab}^{144}=\mathcal{B}_{ab}^{414}=0\,,\\
&\mathcal{A}_{ab}^{244}=\mathcal{B}_{ab}^{244}=\mathcal{B}_{ab}^{424}=0\,,\\
&\mathcal{A}_{ab}^{445}=\mathcal{B}_{ab}^{544}=\mathcal{B}_{ab}^{644}=\mathcal{B}_{ab}^{445}=0\,,\\
&\mathcal{B}_{ab}^{114}=\mathcal{B}_{ab}^{214}=\mathcal{B}_{ab}^{154}=\mathcal{B}_{ab}^{514}=\mathcal{B}_{ab}^{614}=\mathcal{B}_{ab}^{224}=\mathcal{B}_{ab}^{524}=\mathcal{B}_{ab}^{624}=\mathcal{B}_{ab}^{545}=\mathcal{B}_{ab}^{645}=0\,,
\end{split}
\end{equation}
and accordingly,
\begin{equation}  \label{eq:example-I4 const-incomp-T iso-Case1-2}
\begin{split}
&\mathcal{A}_{ab}^{14}\equiv \mathcal{B}_{ab}^{41}\,,\\
&\mathcal{A}_{ab}^{24}\equiv \mathcal{B}_{ab}^{42}\,,\\
&\mathcal{A}_{ab}^{45}\equiv \mathcal{B}_{ab}^{45}\,,\\
&\mathcal{A}_{ab}^{114}\equiv \mathcal{B}_{ab}^{411}\,,\\
&\mathcal{A}_{ab}^{124}\equiv \mathcal{B}_{ab}^{412}\,,\\
&\mathcal{A}_{ab}^{145}\equiv \mathcal{B}_{ab}^{415}\,,\\
&\mathcal{A}_{ab}^{224}\equiv \mathcal{B}_{ab}^{422}\,,\\
&\mathcal{A}_{ab}^{245}\equiv \mathcal{B}_{ab}^{425}\,,
\end{split}
\end{equation}
while the remaining symmetry relations remain valid, thereby preserving the equivalence.

In summary, for incompressible transversely isotropic solids in Case 1 (when $\mathbf{n \,(x)}$ and $\nabla I_4$ are parallel), the universality constraints in Cauchy elasticity and those in hyperelasticity are equivalent, and their material preferred directions are identical.

\subsubsection{Case 2: $\mathbf{n \,(x)}$ and $\nabla I_4$ are orthogonal} \label{Incomp-T iso- n and I_4 are orthogonal}
As discussed earlier, the symmetries of the terms $\mathcal{A}_{ab}^{444}$ and $\mathcal{B}_{ab}^{444}$ are also maintained when $\mathbf{n}$ and $\nabla I_4$ are orthogonal. In hyperelasticity, the term $\mathcal{A}_{ab}^{445}$ is then written as 
\begin{equation}  \label{eq:A445-incomp-T iso-Case2-1}
\mathcal{A}_{ab}^{445}=(n_a \,n^n \,I_{5,n}+\ell^{n}_{a} \,I_{4,n}) \,I_{4,b} \,,
\end{equation}
which is symmetric only when
\begin{equation}  \label{eq:A445-symmetry cond-incomp-T iso-Case2-1}
n_a \,n^n \,I_{5,n}+\ell^{n}_{a} \,I_{4,n}=\bar{c}_5 \,I_{4,a} \,,
\end{equation}
or
\begin{equation}  \label{eq:A445-symmetry cond-incomp-T iso-Case2-2}
\ell^{n}_{a} \,I_{4,n}=\bar{c}_5 \,I_{4,a}-n_a \,n^n \,I_{5,n} \,,
\end{equation}
where $\bar{c}_5$ is a scalar function. We first assume that $n^n \,I_{5,n}\ne 0$ and substitute \eqref{eq:A445-symmetry cond-incomp-T iso-Case2-2} and \eqref{eq:I_5-eigenvect-ell} into \eqref{eq:universality terms-incom-T iso-hyper-2}$_{16}$ to represent $\mathcal{A}_{ab}^{455}$ as follows 
\begin{equation}  \label{eq:A455-incomp-T iso-Case2-1}
\mathcal{A}_{ab}^{455}=\bar{\lambda}_5 \,I_{5,b} \,I_{4,a}+\bar{c}_5 \,I_{4,b} \,I_{5,a}  \,,
\end{equation}
which is symmetric when $\nabla I_4$ and $\nabla I_5$ are parallel (note that the very special solution in which $\bar{\lambda}_5=\bar{c}_5$ is excluded because it is not valid for any family of universal deformations). Another solution to the symmetry of \eqref{eq:A445-incomp-T iso-Case2-1} is obtained when $n^n \,I_{5,n}=0$ and $\ell^{n}_{a} \,I_{4,n}=\bar{\lambda}_4 \,I_{4,a}$, where $\bar{\lambda}_{4}$ is the associated eigenvalue. In this case, the term $\mathcal{A}_{ab}^{455}$ reads
\begin{equation}  \label{eq:A455-incomp-T iso-Case2-2}
\mathcal{A}_{ab}^{455}=\bar{\lambda}_5 \,I_{5,a} \,I_{4,b}+\bar{\lambda}_4 \,I_{4,a} \,I_{5,b}  \,,
\end{equation}
which tells us that either $\bar{\lambda}_4=\bar{\lambda}_5$, or $\nabla I_4$ and $\nabla I_5$ are parallel. As a result, in hyperelasticity when $\mathbf{n} \perp \nabla I_4$, the symmetries of $\mathcal{A}_{ab}^{445}$ and $\mathcal{A}_{ab}^{455}$ hold for only two scenarios: 1) when $\nabla I_4$ and $\nabla I_5$ are parallel, and 2) when $\mathbf{n}$ is orthogonal to $\nabla I_5$, with $\nabla I_4$ and $\nabla I_5$ both being independent eigenvectors of $\ell^{n}_{a}$ associated with the same eigenvalue. 

Now consider $\mathcal{B}_{ab}^{445}$, $\mathcal{B}_{ab}^{544}$ and $\mathcal{B}_{ab}^{554}$ in Cauchy elasticity. Since $\mathbf{n}$ and $\nabla I_4$ are not parallel, $\mathcal{B}_{ab}^{445}=n_a \,I_{4,b} \,n^n \,I_{5,n}$ is symmetric if and only if $\mathbf{n}$ is orthogonal to $\nabla I_5$ ($n^n \,I_{5,n}=0$). Moreover, the symmetry of $\mathcal{B}_{ab}^{544}$ implies that $\nabla I_4$ is also an eigenvector of $\ell^{n}_{a}$. Consequently, $\mathcal{B}_{ab}^{554}$ becomes
\begin{equation}  \label{eq:B554-incomp-T iso-Case2}
\mathcal{B}_{ab}^{554}=\bar{\lambda}_5 \,I_{5,a} \,I_{4,b}+\bar{\lambda}_4 \,I_{4,a} \,I_{5,b} \,.  
\end{equation}
The symmetry constraint associated with \eqref{eq:B554-incomp-T iso-Case2} has two solutions: 1) $\nabla I_4$ and $\nabla I_5$ are parallel, and 2) $\bar{\lambda}_{4}=\bar{\lambda}_{5}$ and $\nabla I_4$ and $\nabla I_5$ are not parallel while both are orthogonal to $\mathbf{n}$. These two solutions are clearly the same as those obtained for the symmetry conditions of $\mathcal{A}_{ab}^{445}$ and $\mathcal{A}_{ab}^{455}$. Let us first address the equivalence for the first scenario in which $\nabla I_4$ and $\nabla I_5$ are parallel.  

\vspace{1em}
\noindent\textbf{\small Equivalence of $\mathcal{A}_{ab}^{\kappa}$ and $\mathcal{B}_{ab}^{\kappa}$ with $\kappa$ a triple index.} 
We know that $\mathbf{n}$ is orthogonal to $\nabla I_4$, and that $\nabla I_4$ and $\nabla I_5$ are parallel. It immediately follows that $\mathbf{n} \perp \nabla I_5$ and $\ell^{n}_{a} \,I_{4,n}=\bar{\lambda}_{5} \,I_{4,a}$. Thus $\mathcal{B}_{ab}^{445}=\mathcal{B}_{ab}^{455}=0$, and the terms
\begin{equation}  \label{eq:A445,B445-incomp-T iso-Case2}
\begin{split}
&\mathcal{A}_{ab}^{445}=\mathcal{B}_{ab}^{544}=\ell_{a}^{n} \,I_{4,n} \,I_{4,b}=\bar{\lambda}_5 \,I_{4,a} \,I_{4,b} \,,\\
&\mathcal{A}_{ab}^{455}=\mathcal{B}_{ab}^{554}=\ell_{a}^{n} \,(I_{5,n} \,I_{4,b}+I_{5,b} \,I_{4,n})=\bar{\lambda}_5 \,(I_{5,a} \,I_{4,b}+I_{4,a} \,I_{5,b})\,, 
\end{split}
\end{equation}
are symmetric. Therefore, we have
\begin{equation}  \label{eq:hyper Cauchy elast-equiv-incomp-T iso-Case2-1}
\left\{\mathcal{B}_{ab}^{445},\mathcal{B}_{ab}^{544},\mathcal{B}_{ab}^{455},\mathcal{B}_{ab}^{554}\right\} \equiv \left\{\mathcal{A}_{ab}^{445},\mathcal{A}_{ab}^{455}\right\} \,. 
\end{equation}

Proceeding further and with reference to \eqref{eq:I_1 - I_2 relation}$_2$, the terms $\mathcal{A}_{ab}^{144}$ and $\mathcal{A}_{ab}^{114}$ take the following form
\begin{equation}  \label{eq:A144,A114-incomp-T iso-Case2}
\begin{split}
&\mathcal{A}_{ab}^{144}=b_{a}^{n} \,I_{4,n} \,I_{4,b}+n_a \,n^n \,I_{1,n} \,I_{4,b} \,, \\
&\mathcal{A}_{ab}^{114}=b_{a}^{n} \,I_{4,n} \,I_{1,b}+\lambda_1 \,I_{1,a} \,I_{4,b}+n_a \,n^n \,I_{1,n} \,I_{1,b} \,.
\end{split}
\end{equation}
Regarding \eqref{eq:A144,A114-incomp-T iso-Case2}$_1$, the constraint $\mathcal{A}_{[ab]}^{144}=0$ gives $b_{a}^{n} \,I_{4,n}=\bar{c}_6 \,I_{4,a}-n_a \,n^n \,I_{1,n}$ which can be substituted into \eqref{eq:A144,A114-incomp-T iso-Case2}$_2$ to yield
\begin{equation}  \label{eq:A114-incomp-T iso-Case2}
\mathcal{A}_{ab}^{114}=\bar{c}_6 \,I_{4,a} \,I_{1,b}+\lambda_1 \,I_{1,a} \,I_{4,b} \,, 
\end{equation} 
where $\bar{c}_6$ is a scalar function. Enforcing the symmetry of \eqref{eq:A114-incomp-T iso-Case2} requires $\nabla I_1$ and $\nabla I_4$ to be parallel, which is the only possible solution. On the other hand, the symmetry of $\mathcal{B}_{ab}^{441}=n_a \,I_{4,b} \,n^n \,I_{1,n}$ is preserved if and only if $n^n \,I_{1,n}=0$, since $\mathbf{n}$ and $\nabla I_4$ are not parallel. $\mathcal{B}_{ab}^{144}$ is also symmetric only when $b^{n}_{a} \,I_{4,n}$ and $I_{4,a}$ are parallel. We therefore conclude that a functional dependence between $I_{1}$ and $I_{4}$ constitutes a unique condition ensuring the symmetries of $\mathcal{B}_{ab}^{144}$ and $\mathcal{B}_{ab}^{441}$. Given this solution, the terms $\mathcal{B}_{ab}^{114}$ and $\mathcal{B}_{ab}^{411}$ are symmetric as well. Since the symmetries in both sets, $\left\{\mathcal{B}_{ab}^{441}, \mathcal{B}_{ab}^{144}, \mathcal{B}_{ab}^{114},\mathcal{B}_{ab}^{411}\right\}$ and $\left\{\mathcal{A}_{ab}^{114}, \mathcal{A}_{ab}^{144}\right\}$, are preserved if and only if $\nabla I_1$ and $\nabla I_4$ are parallel, the corresponding symmetry constraints are equivalent, i.e.,
\begin{equation}  \label{eq:hyper Cauchy elast-equiv-incomp-T iso-Case2-2}
\left\{\mathcal{B}_{ab}^{441}, \mathcal{B}_{ab}^{144}, \mathcal{B}_{ab}^{114},\mathcal{B}_{ab}^{411}\right\} \equiv \left\{\mathcal{A}_{ab}^{114}, \mathcal{A}_{ab}^{144}\right\} \,.
\end{equation}
Following the same approach, one can show that
\begin{equation}  \label{eq:hyper Cauchy elast-equiv-incomp-T iso-Case2-3}
\begin{split}
&\left\{\mathcal{B}_{ab}^{224}, \mathcal{B}_{ab}^{422}, \mathcal{B}_{ab}^{244},\mathcal{B}_{ab}^{442}\right\} \equiv \left\{\mathcal{A}_{ab}^{224}, \mathcal{A}_{ab}^{244}\right\} \,, \\
&\left\{\mathcal{B}_{ab}^{115}, \mathcal{B}_{ab}^{511}, \mathcal{B}_{ab}^{155},\mathcal{B}_{ab}^{551}\right\} \equiv \left\{\mathcal{A}_{ab}^{115}, \mathcal{A}_{ab}^{155}\right\} \,,\\
&\left\{\mathcal{B}_{ab}^{225}, \mathcal{B}_{ab}^{522}, \mathcal{B}_{ab}^{255},\mathcal{B}_{ab}^{552}\right\} \equiv \left\{\mathcal{A}_{ab}^{225}, \mathcal{A}_{ab}^{255}\right\} \,,
\end{split}
\end{equation}
which are equivalent to the functional dependence of the pairs $\left\{I_{2},I_{4}\right\}$, $\left\{I_{1},I_{5}\right\}$ and $\left\{I_{2},I_{5}\right\}$, respectively. Thus, $\nabla I_i \,(i=1,2,4,5)$ are mutually parallel and orthogonal to $\mathbf{n}$. The symmetry constraints corresponding to the remaining terms in hyperelasticity, $\left\{\mathcal{A}_{ab}^{124},\mathcal{A}_{ab}^{125},\mathcal{A}_{ab}^{145},\mathcal{A}_{ab}^{245}\right\}$, as well as those corresponding to the similar ones in Cauchy elasticity including $\mathcal{B}_{ab}^{124}$, $\mathcal{B}_{ab}^{214}$, $\mathcal{B}_{ab}^{412}$, $\mathcal{B}_{ab}^{125}$,$\mathcal{B}_{ab}^{215}$, $\mathcal{B}_{ab}^{512}$, $\mathcal{B}_{ab}^{145}$, $\mathcal{B}_{ab}^{415}$, $\mathcal{B}_{ab}^{514}$, $\mathcal{B}_{ab}^{245}$, $\mathcal{B}_{ab}^{425}$, and $\mathcal{B}_{ab}^{524}$ are then identically satisfied. 

To complete this discussion, we need to investigate the additional terms in Cauchy elasticity, i.e., $\mathcal{B}_{ab}^{611}$, $\mathcal{B}_{ab}^{622}$, $\mathcal{B}_{ab}^{644}$, $\mathcal{B}_{ab}^{655}$, $\mathcal{B}_{ab}^{612}$, $\mathcal{B}_{ab}^{614}$, $\mathcal{B}_{ab}^{615}$, $\mathcal{B}_{ab}^{624}$, $\mathcal{B}_{ab}^{625}$ and $\mathcal{B}_{ab}^{645}$. Let us use \eqref{eq:I_1 - I_2 relation}$_4$ to simplify $\bar{\ell}^{n}_{a} \,I_{i,n} \,(i=1,2,4,5)$ as follows
\begin{equation}  \label{eq:ellbar-Ii-incomp-T iso-Case2-1}
\bar{\ell}^{n}_{a} \,I_{i,n}=n^n \,c^{c}_{a} \, n_c \,I_{i,n}+n_a \,c^{n}_{c} \,n^c \,I_{i,n}=n^n \,I_{i,n} \,c^{c}_{a} \,n_c + n_a \,\frac{1}{\lambda_1} \,I_{i,c} \,n^c\,,
\end{equation}
which according to the orthogonality of $\mathbf{n}$ and $\nabla I_i$ leads to
\begin{equation}  \label{eq:ellbar-Ii-incomp_T iso-Case2-2}
\bar{\ell}^{n}_{a} \,I_{i,n}=0\,.
\end{equation}
Consequently, with reference to \eqref {eq:universal terms-incom-T iso-Cauchy elast-3}$_{6}$--\eqref {eq:universal terms-incom-T iso-Cauchy elast-3}$_{15}$, we see that the additional symmetry constraint terms vanish and do not impose further constraints beyond the existing ones. Therefore, $\mathcal{A}_{ab}^{\kappa}$ and $\mathcal{B}_{ab}^{\kappa}$, where $\kappa$ is a triple index, are equivalent.

Finally, it is worth noting that by following the same calculation as used in \eqref{eq:ellbar-Ii-incomp_T iso-Case2-2}, one also obtains $\ell^{n}_{a} \,I_{i,n}=0$. As a consequence, the symmetry equivalences in \eqref{eq:hyper Cauchy elast-equiv-incomp-T iso-Case2-1}, \eqref{eq:hyper Cauchy elast-equiv-incomp-T iso-Case2-2} and \eqref{eq:hyper Cauchy elast-equiv-incomp-T iso-Case2-3} can be written more precisely as 
\begin{equation}  \label{eq:hyper Cauchy elast-equiv-incomp-T iso-Case2-final}
\begin{split}
&\mathcal{A}_{ab}^{114}=\mathcal{B}_{ab}^{114} \,, \\
&\mathcal{A}_{ab}^{115}=\mathcal{B}_{ab}^{115} \,, \\
&\mathcal{A}_{ab}^{144}=\mathcal{B}_{ab}^{144} \,, \\
&\mathcal{A}_{ab}^{155}=\mathcal{B}_{ab}^{155} \,, \\
&\mathcal{A}_{ab}^{224}=\mathcal{B}_{ab}^{224} \,, \\
&\mathcal{A}_{ab}^{225}=\mathcal{B}_{ab}^{225} \,, \\
&\mathcal{A}_{ab}^{244}=\mathcal{B}_{ab}^{244} \,, \\
&\mathcal{A}_{ab}^{255}=\mathcal{B}_{ab}^{255} \,, \\
&\mathcal{A}_{ab}^{445}=\mathcal{B}_{ab}^{445}=0 \,, \\
&\mathcal{A}_{ab}^{455}=\mathcal{B}_{ab}^{455}=0 \,. 
\end{split}
\end{equation}

\vspace{1em}
\noindent\textbf{\small Equivalence of $\mathcal{A}_{ab}^{\kappa}$ and $\mathcal{B}_{ab}^{\kappa}$ with $\kappa$ a double index.}
In Case 2, we know that
\begin{equation}  \label{eq:A44,B44,B55,A55-incomp-T iso-Case2}
\begin{split}
&\mathcal{B}_{ab}^{44}=\mathcal{A}_{ab}^{44}=\left(n_a \,n^n\right)_{|n} \,I_{4,b}\,, \\
&\mathcal{B}_{ab}^{55}=\mathcal{A}_{ab}^{55}={\ell^{n}_{a}}_{|n} \,I_{5,b}\,.
\end{split}
\end{equation}
Since $\nabla I_1$ and $\nabla I_4$ are parallel, the symmetry of $\mathcal{B}_{ab}^{41}=(n_a \,n^n)_{|n} \,I_{1,b}$ is equivalent to that of $\mathcal{B}_{ab}^{44}$. Likewise, we have
\begin{equation}  \label{eq:Cauchy elast-equiv-incomp-T iso-Case2-1}
\begin{split}
&\mathcal{B}_{ab}^{51} \equiv \mathcal{B}_{ab}^{55} \,, \\
&\mathcal{B}_{ab}^{42} \equiv \mathcal{B}_{ab}^{44}\,,\\
&\mathcal{B}_{ab}^{52} \equiv \mathcal{B}_{ab}^{55}\,,\\
&\mathcal{B}_{ab}^{45} \equiv \mathcal{B}_{ab}^{44} \,, \\
&\mathcal{B}_{ab}^{54} \equiv \mathcal{B}_{ab}^{55} \,.
\end{split}
\end{equation}
Subsequently, with reference to \eqref{eq:hyper-Cauchy constr-relation-incomp-T iso-1}$_5$--\eqref{eq:hyper-Cauchy constr-relation-incomp-T iso-1}$_9$, the following relations hold between the hyperelasticity and the Cauchy elasticity constraints
\begin{equation}  \label{eq:hyper-Cauchy elast-equiv-incomp-T iso-Case2-part2}
\begin{split}
&\mathcal{A}_{ab}^{14} \equiv \mathcal{B}_{ab}^{14} +\mathcal{B}_{ab}^{44} \,, \\
&\mathcal{A}_{ab}^{15} \equiv \mathcal{B}_{ab}^{15} +\mathcal{B}_{ab}^{55}\,,\\
&\mathcal{A}_{ab}^{24} \equiv \mathcal{B}_{ab}^{24} +\mathcal{B}_{ab}^{44}\,,\\
&\mathcal{A}_{ab}^{25} \equiv \mathcal{B}_{ab}^{25} +\mathcal{B}_{ab}^{55} \,, \\
&\mathcal{A}_{ab}^{45} \equiv \mathcal{A}_{ab}^{44} +\mathcal{A}_{ab}^{55}\equiv \mathcal{B}_{ab}^{44} +\mathcal{B}_{ab}^{55} \,.
\end{split}
\end{equation}
Thus, in hyperelasticity there are six independent symmetry constraints namely $\mathcal{A}_{[ab]}^{44}=0$, $\mathcal{A}_{[ab]}^{55}=0$, $\mathcal{A}_{[ab]}^{14}=0$, $\mathcal{A}_{[ab]}^{15}=0$, $\mathcal{A}_{[ab]}^{24}=0$ and $\mathcal{A}_{[ab]}^{25}=0$. These correspond to six independent symmetry constraints in Cauchy elasticity, $\mathcal{B}_{[ab]}^{44}=0$, $\mathcal{B}_{[ab]}^{55}=0$, $\mathcal{B}_{[ab]}^{14}=0$, $\mathcal{B}_{[ab]}^{15}=0$, $\mathcal{B}_{[ab]}^{24}=0$, and $\mathcal{B}_{[ab]}^{25}=0$, indicating that the two sets are equivalent.

Because $\mathcal{A}_{ab}^{4}=\mathcal{B}_{ab}^{4}$ and $\mathcal{A}_{ab}^{5}=\mathcal{B}_{ab}^{5}$, we conclude that in Case 2, all symmetry constraints in hyperelasticity are equivalent to the corresponding ones in Cauchy elasticity. In other words, the universal deformations and universal material preferred directions in Cauchy elasticity are a subset of those in hyperelasticity. In order to complete the proof, it is necessary to show that the symmetries of the extra terms $\mathcal{B}_{ab}^{6}$, $\mathcal{B}_{ab}^{61}$, $\mathcal{B}_{ab}^{62}$, $\mathcal{B}_{ab}^{64}$ and $\mathcal{B}_{ab}^{65}$ hold trivially. These terms in Cauchy elasticity, if not trivially symmetric, can only impose further constraints on the existing universal deformations and material preferred directions in hyperelasticity given in \citep{YavariGoriely2021}. Following \citep{YavariGoriely2021,Yavari2022Universal}, we find that for all families of universal deformations of incompressible transversely isotropic hyperelastic solids in which $\mathbf{n} \perp \nabla I_4$, the following relations hold
\begin{equation}  \label{eq:elln and ellbarn=0-incomp-T iso-Case2}
{\bar{\ell}^{n}_{a}}_{|n}={\ell^{n}_{a}}_{|n}=0 \,. 
\end{equation}
Concerning \eqref{eq:universal terms-incom-T iso-Cauchy elast-3}$_1$--\eqref{eq:universal terms-incom-T iso-Cauchy elast-3}$_5$, it follows that $\mathcal{B}_{ab}^{6}=\mathcal{B}_{ab}^{61}=\mathcal{B}_{ab}^{62}=\mathcal{B}_{ab}^{64}=\mathcal{B}_{ab}^{65}=0$, and therefore the universality constraints in hyperelasticity and Cauchy elasticity in Case 2 are equivalent.

As outlined previously, in addition to Case 1 and Case 2, there remains one other possibility to be addressed. This possibility indicates that $\nabla I_4$ and $\nabla I_5$ are perpendicular to $\mathbf{n}$, and that both are eigenvectors of $\ell^{n}_{a}$ associated with the same eigenvalue while they are not parallel. With reference to \citep{YavariGoriely2021}, one observes that $I_{i} \,(i=1,2,4,5)$ are functionally dependent for all families of universal deformations, and thus this case is not an admissible solution in hyperelasticity. Thus, let us investigate it for Cauchy elasticity. If a solution satisfies the symmetries of the Cauchy elasticity terms, then the Cauchy elasticity constraints of \eqref{eq:hyper-Cauchy constr-relation-incomp-T iso-1} and \eqref{eq:hyper-Cauchy constr-relation-incomp-T iso-2} are all satisfied, implying that the hyperelasticity symmetry constraints also hold. Therefore, this solution must also be valid for hyperelasticity, which leads to a contradiction. As a result, this case is inadmissible in both hyperelasticity and Cauchy elasticity.

In summary, we have proved the following result.
\begin{prop}
The universal deformations and material preferred directions of incompressible transversely isotropic Cauchy elasticity are identical to those of incompressible transversely isotropic hyperelasticity.
\end{prop}

\section{Universality Constraints in Orthotropic Elasticity} \label{Orthotropic-Elasticity}

At each point in the reference configuration, an orthotropic solid exhibits reflection symmetry with respect to three mutually perpendicular planes. Accordingly, the orthotropic directions at a point $\mathbf{X}$ are defined by a set of three vectors $\mathbf{N}_1 \,(\mathbf{X})$, $\mathbf{N}_2 \,(\mathbf{X})$ and $\mathbf{N}_3 \,(\mathbf{X})$ that are orthonormal with respect to the metric tensor $\mathbf{G}$. In hyperelasticity, the energy function of an orthotropic solid is described by seven independent invariants, denoted as $I_{1}$, $I_{2}$, $I_{3}$, $I_{4}$, $I_{5}$, $I_{6}$ and $I_{7}$. 
The first three invariants are defined in \eqref{eq:principal invariants}.
The remaining invariants are introduced as follows 
\begin{equation}  \label{eq:invariants-Ortho}
I_4=\mathbf{N}_1 \cdot \mathbf{C}\cdot \mathbf{N}_1 \,, \qquad
I_5=\mathbf{N}_1 \cdot \mathbf{C}^{2}\cdot \mathbf{N}_1\,,\qquad
I_6=\mathbf{N}_2 \cdot \mathbf{C} \cdot \mathbf{N}_2 \,, \qquad
I_7=\mathbf{N}_2 \cdot \mathbf{C}^{2}\cdot \mathbf{N}_2\,.
\end{equation}

For orthotropic hyperelastic solids, the second Piola–Kirchhoff stress and the Cauchy stress are given, respectively, by \citep{YavariGoriely2021,Yavari2022Universal}
\begin{equation} \label{eq:second piola-Ortho-hyper} 
\begin{split}
\mathbf{S} = &2\,W_1 \, \mathbf{G}^{\sharp}+2\,W_2 \,(I_2 \,\mathbf{C}^{-1}-I_3 \,\mathbf{C}^{-2})+2\,W_3 \,I_3 \,\mathbf{C}^{-1}\\
&+2 \, W_4 \,(\mathbf{N}_1 \otimes \mathbf{N}_1)+2\,W_5 \,[\mathbf{N}_1 \otimes (\mathbf{C} \cdot \mathbf{N}_1)+(\mathbf{C} \cdot \mathbf{N}_1) \otimes \mathbf{N}_1] \\
&+2 \, W_6 \,(\mathbf{N}_2 \otimes \mathbf{N}_2)+2\,W_7 \,[\mathbf{N}_2 \otimes (\mathbf{C} \cdot \mathbf{N}_2)+(\mathbf{C} \cdot \mathbf{N}_2) \otimes \mathbf{N}_2] \,,
\end{split}
\end{equation}
and
\begin{equation} \label{eq:Cauchy stress-Ortho-hyper} 
\begin{split}
\boldsymbol{\sigma}=&\frac{2}{\sqrt{I_3}} \,W_1 \,\mathbf{b}^\sharp+\frac{2}{\sqrt{I_3}} \,(I_2 \,W_2+I_3 \,W_3) \,\mathbf{g}^\sharp -2 \,\sqrt{I_3} \,W_2 \, \mathbf{c}^\sharp\\
&+\frac{2}{\sqrt{I_3}} \,W_4 \,(\mathbf{n}_1 \otimes \mathbf{n}_1)+\frac{2}{\sqrt{I_3}} \,W_5 \,[\mathbf{n}_1 \otimes (\mathbf{b} \cdot \mathbf{n}_1)+(\mathbf{b} \cdot \mathbf{n}_1) \otimes \mathbf{n}_1] \\
&+\frac{2}{\sqrt{I_3}} \,W_6 \,(\mathbf{n}_2 \otimes \mathbf{n}_2)+\frac{2}{\sqrt{I_3}} \,W_7 \,[\mathbf{n}_2 \otimes (\mathbf{b} \cdot \mathbf{n}_2)+(\mathbf{b} \cdot \mathbf{n}_2) \otimes \mathbf{n}_2] \,,
\end{split}
\end{equation}
where $W_i= \dfrac{\partial W}{\partial I_i} \,(i=1,\cdots,7)$, $\mathbf{n}_1=\mathbf{F} \cdot \mathbf{N}_1$ and $\mathbf{n}_2=\mathbf{F} \cdot \mathbf{N}_2$. In components, the Cauchy stress reads
\begin{equation} \label{eq:Cauchy stress-Ortho-hyper-components} 
\begin{split}
\sigma^{ab} =&\frac{2}{\sqrt{I_3}}\,[W_1 \,b^{ab}+(I_2 \,W_2+I_3 \,W_3) \,g^{ab}-I_3 \,W_2 \, c^{ab}\\
&+W_4 \,n_{1}^{a} \,n_{1}^{b}+W_5 \,(n_{1}^{a} \,b^{bc} \,n_{1}^{d} \,g_{cd} + n_{1}^{b} \,b^{ac} \,n_{1}^{d} \,g_{cd}) \\
&+W_6 \,n_{2}^{a} \,n_{2}^{b}+W_7 \,(n_{2}^{a} \,b^{bc} \,n_{2}^{d} \,g_{cd} + n_{2}^{b} \,b^{ac} \,n_{2}^{d} \,g_{cd})] \,,
\end{split}
\end{equation}
where $n_{1}^{a}=F^{a}{}_A \,N_{1}^{A}$ and $n_{2}^{a}=F^{a}{}_{A} \,N_{2}^{A}$.

For orthotropic Cauchy elastic solids, we have the following representation for the second Piola-Kirchhoff stress tensor \citep{YavariGoriely2025}
\begin{equation} \label{eq:second piola-Ortho-Cauchy elast} 
\begin{split}
\mathbf{S} = &a_0 \,\mathbf{G}^{\sharp}+ a_1 \,\mathbf{C}^{\sharp}+a_2 \, \mathbf{C}^{2\sharp}+a_3 \,(\mathbf{N}_1\otimes \mathbf{N}_1)+a_4 \,[ \mathbf{N}_1\otimes (\mathbf{C}\cdot \mathbf{N}_1)+( \mathbf{C}\cdot \mathbf{N}_1)\otimes \mathbf{N}_1] \\
&+a_5 \,[\mathbf{N}_1\otimes (\mathbf{C}^{2} \cdot \mathbf{N}_1)+(\mathbf{C}^2 \cdot \mathbf{N}_1)\otimes \mathbf{N}_1] \\
&+a_6 \,(\mathbf{N}_2\otimes \mathbf{N}_2)+a_7 \,[\mathbf{N}_2 \otimes (\mathbf{C}\cdot \mathbf{N}_2)+(\mathbf{C}\cdot \mathbf{N}_2)\otimes \mathbf{N}_2] \\
&+a_8 \,[\mathbf{N}_2\otimes (\mathbf{C}^{2} \cdot \mathbf{N}_2)+(\mathbf{C}^2 \cdot \mathbf{N}_2)\otimes \mathbf{N}_2] \,,
\end{split}
\end{equation}
where $a_i \,(I_1,\cdots,I_7)$, $i=0,\cdots,8$ are the response functions. The Cauchy stress tensor is written as
\begin{equation} \label{eq:Cauchy stress-Ortho-Cauchy elast} 
\begin{split}
\boldsymbol{\sigma}=&\tilde{a}_0 \,\mathbf{g}^{\sharp}+\tilde{a}_1 \,\mathbf{b}^{\sharp}+\tilde{a}_2 \,\mathbf{c}^{\sharp} +\tilde{a}_3 \,(\mathbf{n}_1 \otimes \mathbf{n}_1)+\tilde{a}_4 \,[\mathbf{n}_1 \otimes (\mathbf{b} \cdot \mathbf{n}_1)+(\mathbf{b} \cdot \mathbf{n}_1) \otimes \mathbf{n}_1] \\
&+\tilde{a}_5 \,[\mathbf{n}_1 \otimes (\mathbf{c} \cdot \mathbf{n}_1)+(\mathbf{c} \cdot \mathbf{n}_1) \otimes \mathbf{n}_1] \\
&+\tilde{a}_6 \,(\mathbf{n}_2 \otimes \mathbf{n}_2)+\tilde{a}_7 \,[\mathbf{n}_2 \otimes (\mathbf{b} \cdot \mathbf{n}_2)+(\mathbf{b} \cdot \mathbf{n}_2) \otimes \mathbf{n}_2] \\
&+\tilde{a}_8 \,[\mathbf{n}_2 \otimes (\mathbf{c} \cdot \mathbf{n}_2)+(\mathbf{c} \cdot \mathbf{n}_2) \otimes \mathbf{n}_2] \,,
\end{split}
\end{equation}
where $\tilde{a}_i \,(I_1,\cdots,I_7) \,,i=0,\cdots,8$ are the response functions.

The second Piola–Kirchhoff stress tensor for incompressible orthotropic hyperelastic solids is represented as \citep{Yavari2022Universal,YavariGoriely2021}
\begin{equation} \label{eq:second piola-incomp-Ortho-hyper} 
\begin{split}
\mathbf{S} =& -p \, \mathbf{C}^{-1}+2\,W_1 \, \mathbf{G}^{\sharp}+2\,W_2 \,(I_2 \,\mathbf{C}^{-1}-\mathbf{C}^{-2}) \\
&+2 \, W_4 \,(\mathbf{N}_1 \otimes \mathbf{N}_1)
+2\,W_5 \,[\mathbf{N}_1 \otimes (\mathbf{C} \cdot \mathbf{N}_1)+(\mathbf{C} \cdot \mathbf{N}_1) \otimes \mathbf{N}_1] \\
&+2 \, W_6 \,(\mathbf{N}_2 \otimes \mathbf{N}_2)
+2\,W_7 \,[\mathbf{N}_2 \otimes (\mathbf{C} \cdot \mathbf{N}_2)+(\mathbf{C} \cdot \mathbf{N}_2) \otimes \mathbf{N}_2] \,,
\end{split}
\end{equation}
where $W=W \,(I_1,I_2,I_4,I_5,I_6,I_7)$. Moreover, the Cauchy stress reads
\begin{equation} \label{eq:Cauchy stress-incomp-Ortho-hyper}
\begin{split}
\boldsymbol{\sigma} =& -p \,\mathbf{g}^\sharp+ 2 \,W_1 \, \mathbf{b}^\sharp- 2 \,W_2 \, \mathbf{c}^\sharp+ 2 \,W_4 \,(\mathbf{n}_1 \otimes \mathbf{n}_1) \\
&+ 2 \,W_5 \, [\mathbf{n}_1 \otimes (\mathbf{b} \cdot \mathbf{n}_1)+(\mathbf{b} \cdot \mathbf{n}_1) \otimes \mathbf{n}_1] + 2 \,W_6 \,(\mathbf{n}_2 \otimes \mathbf{n}_2) \\
&+ 2 \,W_7 \, [\mathbf{n}_2 \otimes (\mathbf{b} \cdot \mathbf{n}_2)+(\mathbf{b} \cdot \mathbf{n}_2) \otimes \mathbf{n}_2]\,.
\end{split}
\end{equation}
In components, it is written as
\begin{equation} \label{eq:Cauchy stress-incomp-Ortho-hyper-components}
\sigma^{ab} = -p \,g^{ab}+ 2 \,W_1 \, b^{ab}- 2 \,W_2 \, c^{ab}+ 2 \,W_4 \,n_1^{a} \,n_1^{b}+ 2 \,W_5 \, \ell_1^{ab} + 2 \,W_6 \,n_2^{a} \,n_2^{b} + 2 \,W_7 \, \ell_2^{ab} \,,
\end{equation}
where $\ell_1^{ab}=n_1^a \,b^{bc} \,n_1^d \,g_{cd} + n_1^b \,b^{ac} \,n_1^d \,g_{cd}$ and $\ell_2^{ab}=n_2^a \,b^{bc} \,n_2^d \,g_{cd} + n_2^b \,b^{ac} \,n_2^d \,g_{cd}$. 
For incompressible orthotropic Cauchy elastic solids, the second Piola–Kirchhoff stress tensor is written as \citep{Spencer1970,Boehler1979,Boehler1987}
\begin{equation} \label{eq:second piola-incomp-Ortho-Cauchy elast} 
\begin{split}
\mathbf{S}= &-p \,\mathbf{C}^{-1}+\bar{a}_0 \,\mathbf{G}^{\sharp}+ \bar{a}_1 \,\mathbf{C}^{\sharp}+\bar{a}_2 \,(\mathbf{N}_1\otimes \mathbf{N}_1)+\bar{a}_4 \,[\mathbf{N}_1\otimes (\mathbf{C}\cdot \mathbf{N}_1)+( \mathbf{C}\cdot \mathbf{N}_1)\otimes \mathbf{N}_1] \\
&+\bar{a}_5 \,[\mathbf{N}_1\otimes (\mathbf{C}^{-1} \cdot \mathbf{N}_1)+(\mathbf{C}^{-1} \cdot \mathbf{N}_1)\otimes \mathbf{N}_1] \\
&+\bar{a}_6 \,(\mathbf{N}_2\otimes \mathbf{N}_2)+\bar{a}_7 \,[\mathbf{N}_2\otimes (\mathbf{C}\cdot \mathbf{N}_2)+(\mathbf{C}\cdot \mathbf{N}_2)\otimes \mathbf{N}_2] \\
&+\bar{a}_8 \,[\mathbf{N}_2\otimes (\mathbf{C}^{-1} \cdot \mathbf{N}_2)+(\mathbf{C}^{-1} \cdot \mathbf{N}_2)\otimes \mathbf{N}_2] \,,
\end{split}
\end{equation}
where $\bar{a}_i \,(I_1,I_2,I_4,I_5,I_6,I_7)$, $i=0,1,2,4,5,6,7,8$ are the response functions. Thus, one can write the Cauchy stress tensor for incompressible Cauchy elastic solids as
\begin{equation} \label{eq:Cauchy stress-incomp-Ortho-Cauchy elast} 
\begin{split}
\boldsymbol{\sigma}=&-p \,\mathbf{g}^{\sharp}+\alpha_1 \,\mathbf{b}^{\sharp}+\alpha_2 \,\mathbf{c}^{\sharp} +\alpha_4 \,(\mathbf{n}_1 \otimes \mathbf{n}_1)+\alpha_5 \,[\mathbf{n}_1 \otimes (\mathbf{b} \cdot \mathbf{n}_1)+(\mathbf{b} \cdot \mathbf{n}_1) \otimes \mathbf{n}_1] \\
&+\alpha_6 \,[\mathbf{n}_1 \otimes (\mathbf{c} \cdot \mathbf{n}_1)+(\mathbf{c} \cdot \mathbf{n}_1) \otimes \mathbf{n}_1] \\
&+\alpha_7 \,(\mathbf{n}_2 \otimes \mathbf{n}_2)+\alpha_8 \,[\mathbf{n}_2 \otimes (\mathbf{b} \cdot \mathbf{n}_2)+(\mathbf{b} \cdot \mathbf{n}_2) \otimes \mathbf{n}_2] \\
&+\alpha_9 \,[\mathbf{n}_2 \otimes (\mathbf{c} \cdot \mathbf{n}_2)+(\mathbf{c} \cdot \mathbf{n}_2) \otimes \mathbf{n}_2] \,,
\end{split}
\end{equation}
where $\alpha_{i}=\alpha_{i} \,(I_1, I_2, I_4,I_5,I_6,I_7)$, $i=1,2,4,5,6,7,8,9$ are arbitrary response functions.

\subsection{Compressible orthotropic solids}
Let us write the Cauchy stress~\eqref{eq:Cauchy stress-Ortho-Cauchy elast} in components as
\begin{equation} \label{eq:Cauchy stress-Ortho-Cauchy elast-components} 
\sigma^{ab}=\alpha_1 \,g^{ab}+\alpha_2 \,b^{ab}+\alpha_3 \,c^{ab} +\alpha_4 \,n_1^a \,n_1^b+\alpha_5 \,\ell_1^{ab}+\alpha_6 \,\bar{\ell}_1^{ab}+\alpha_7 \,n_2^a \,n_2^b+\alpha_8 \,\ell_2^{ab}+\alpha_9 \,\bar{\ell}_2^{ab} \,,
\end{equation}
where $\bar{\ell}_1^{ab}=n_1^a \,c^{bc} \,n_1^d \,g_{cd} + n_1^b \,c^{ac} \,n_1^d \,g_{cd}$ and $\bar{\ell}_2^{ab}=n_2^a \,c^{bc} \,n_2^d \,g_{cd} + n_2^b \,c^{ac} \,n_2^d \,g_{cd}$, and $\alpha_i \,(I_1,I_2,I_3,I_4,I_5,I_6,I_7) \,,i=1,\cdots,9$ are arbitrary response  functions. Thus, the equilibrium equations read
\begin{equation} \label{eq:equilibrium-comp-Ortho-1}
\begin{split}
&\alpha_2 \,{b^{ab}}_{|b}+\alpha_3 \,{c^{ab}}_{|b}
+\alpha_4 \,\left(n_1^a \,n_1^b\right)_{|b}+\alpha_5 \,{\ell_1^{ab}}_{|b}+\alpha_6 \,{\bar{\ell}_1^{ab}}_{|b}+\alpha_7 \,\left(n_2^a \,n_2^b\right)_{|b}+\alpha_8 \,{\ell_2^{ab}}_{|b}+\alpha_9 \,{\bar{\ell}_2^{ab}}_{|b} \\
&+\alpha_{1,i} \,I_{i,b} \,g^{ab}+\alpha_{2,i} \,I_{i,b} \,b^{ab}+\alpha_{3,i} \,I_{i,b} \,c^{ab}+\alpha_{4,i} \,I_{i,b} \,(n_1^a \,n_1^b)+\alpha_{5,i} \,I_{i,b} \,\ell_1^{ab}+\alpha_{6,i} \,I_{i,b} \,\bar{\ell}_1^{ab} \\
&+\alpha_{7,i} \,I_{i,b} \,(n_2^a \,n_2^b)+\alpha_{8,i} \,I_{i,b} \,\ell_2^{ab}+\alpha_{9,i} \,I_{i,b} \,\bar{\ell}_2^{ab}=0 \,,
\end{split}
\end{equation}
where $i=1,\cdots,7$. Therefore, the universality constraints are written as
\begin{equation} \label{eq:constraints-comp-Ortho}
\begin{split}
&{b^{ab}}_{|b}={c^{ab}}_{|b}=0 \,, \\
&\left(n_1^a \,n_1^b\right)_{|b}=0 \,,\\
&\left(n_2^a \,n_2^b\right)_{|b}=0 \,, \\
&{\ell_1^{ab}}_{|b}=0 \,,\\
&{\ell_2^{ab}}_{|b}=0 \,, \\
&I_{i,b} \,g^{ab}=0\,, \\
&I_{i,b} \,b^{ab}=0\,, \\
&I_{i,b} \,c^{ab}=0\,, \\
&I_{i,b} \,n_1^a \,n_1^b=0 \,,\\
&I_{i,b} \,n_2^a \,n_2^b=0\,, \\
&I_{i,b} \,\ell_1^{ab}=0 \,,\\
&I_{i,b} \,\ell_2^{ab}=0\,, \\
&{\bar{\ell}_1^{ab}}_{|b}=0 \,, \\
&{\bar{\ell}_2^{ab}}_{|b}=0 \,, \\
&I_{i,b} \,\bar{\ell}_1^{ab}=0\,, \\
&I_{i,b} \,\bar{\ell}_2^{ab}=0\,.
\end{split}
\end{equation}
The constraints~\eqref{eq:constraints-comp-Ortho}$_1$–\eqref{eq:constraints-comp-Ortho}$_{12}$ coincide with those obtained for homogeneous compressible orthotropic hyperelastic solids \citep{Yavari2022Universal,YavariGoriely2021}. Thus, to prove that the universality constraints in Cauchy elasticity and hyperelasticity are equivalent, it is sufficient to show that the extra constraints in Cauchy elasticity, i.e., \eqref{eq:constraints-comp-Ortho}$_{13}$--\eqref{eq:constraints-comp-Ortho}$_{16}$, are trivially satisfied.

Considering \eqref{eq:constraints-comp-Ortho}$_6$, one concludes that $I_i \,(i=1,\cdots,7)$ are constant. Consequently, \eqref{eq:constraints-comp-Ortho}$_7$–\eqref{eq:constraints-comp-Ortho}$_{12}$ hold identically. Moreover, the forms of the constraints~\eqref{eq:constraints-comp-Ortho}$_2$, \eqref{eq:constraints-comp-Ortho}$_3$, and \eqref{eq:constraints-comp-Ortho}$_4$, \eqref{eq:constraints-comp-Ortho}$_5$ are identical to \eqref{eq:constraints-comp-T iso}$_2$ and \eqref{eq:constraints-comp-T iso}$_3$, respectively. Therefore, the constraints~\eqref{eq:constraints-comp-Ortho}$_1$--\eqref{eq:constraints-comp-Ortho}$_5$ are the same as those for transversely isotropic solids. This shows that universal deformations are homogeneous, and $\mathbf{N}_1$ and $\mathbf{N}_2$ are constant unit vectors. By combining these results, one can show that
\begin{equation} \label{eq:constraints-comp-Ortho-extra}
{\bar{\ell}_1^{ab}}_{|b}={\bar{\ell}_2^{ab}}_{|b}=I_{i,b} \,\bar{\ell}_1^{ab}=I_{i,b} \,\bar{\ell}_2^{ab}=0 \,.
\end{equation}
Hence, the extra constraints in Cauchy elasticity are satisfied identically. In summary, we have proved the following result.
\begin{prop}
The universal deformations and material preferred directions of compressible orthotropic Cauchy elasticity are identical to those of compressible orthotropic hyperelasticity.
\end{prop}

\subsection{Incompressible orthotropic solids} \label{Incomp-Ortho-Elasticity}
The derivation of the universality constraints for incompressible orthotropic solids proceeds in precisely the same manner as that employed for incompressible isotropic solids (see \S\ref{Incomp-Iso-Elasticity}) and transversely isotropic solids (see \S\ref{Incomp-T iso-Elasticity}): the corresponding Cauchy stress tensors, given in \eqref{eq:Cauchy stress-incomp-Ortho-hyper} for hyperelastic solids and in \eqref{eq:Cauchy stress-incomp-Ortho-Cauchy elast} for Cauchy elastic solids, are substituted into the equilibrium equations~\eqref{eq:equilibrium} to determine $p_{|a}$ as 
\begin{equation} \label{eq:pa-Ortho-hyperelastic}
p_{|b} \,g^{ab}=2 \,{\left[W_1 \,b^{ab}-W_2 \,c^{ab} +W_4 \,n_1^a \,n_1^b+W_5 \,\ell_1^{ab}+W_6 \,n_2^a \,n_2^b+W_7 \,\ell_2^{ab}\right]}_{|b} \,,
\end{equation}
for hyperelastic solids, and
\begin{equation} \label{eq:pa-Ortho-Cauchy elast}
p_{|b} \,g^{ab}={\left[\alpha_1 \,b^{ab}+\alpha_2 \,c^{ab}+\alpha_4 \,n_1^a \,n_1^b+\alpha_5 \,\ell_1^{ab}+\alpha_6 \,\bar{\ell}_1^{ab}+\alpha_7 \,n_2^a \,n_2^b+\alpha_8 \,\ell_2^{ab}+\alpha_9 \,\bar{\ell}_2^{ab}\right]}_{|b} \,,
\end{equation}
for Cauchy elastic solids. Upon imposing the integrability condition $p_{|ab}=p_{|ba}$, the universality constraints are then determined. 

In hyperelasticity, we get
\begin{equation} \label{eq:p_{|ab}-Ortho-hyperelastic}
p_{|ab}=\sum_{\kappa}{\mathcal{A}_{ab}^{\kappa} \,W_{\kappa }}\,.
\end{equation}
Recall that $p_{|ab}=p_{|ba}$ if and only if all the coefficients of partial derivatives of $W$ are symmetric, that is, $\mathcal{A}_{ab}^{\kappa}=\mathcal{A}_{ba}^{\kappa}$. As discussed in \citep{YavariGoriely2023Universal} and \citep{YavariGoriely2021}, the matrix $\mathcal{A}_{ab}^{\kappa}$ contains $83$ terms that can be categorized as follows:
\begin{enumerate}[label=\roman*)]
    \item The nine terms that already appear in the isotropic hyperelastic case and are required to be symmetric:
\begin{equation}  \label{eq:nine iso terms-incomp-Ortho-hyper}
\kappa \in \mathcal{K}_{iso}=\left\{1, 2, 11, 22, 12, 111, 222,112, 122\right\}\,.
\end{equation}
    \item $25$ terms associated with $\mathbf{N}_1$:
\begin{equation}  \label{eq:N1 terms-incomp-Ortho-hyper}
\begin{split}
\kappa \in \mathcal{K}_{i}=\{&4, 5, 44, 55, 14, 15, 24, 25, 45, 444, 555, 114, 115, 124, 125, \\
&144, 145, 155, 224, 225, 244, 245, 255, 445, 455\}\,.
\end{split}
\end{equation}
    \item $25$ terms associated with $\mathbf{N}_2$:
\begin{equation}  \label{eq:N2 terms-incomp-Ortho-hyper}
\begin{split}
\kappa \in \mathcal{K}_{ii}=\{&6, 7, 66, 77, 16, 17, 26, 27, 67, 666, 777, 116, 117, 126, 127, \\
&166, 167, 177, 226, 227, 266, 267, 277, 667, 677\}\,.
\end{split}
\end{equation}
   \item 24 terms corresponding to coupling of $\mathbf{N}_1$ and $\mathbf{N}_2$:
\begin{equation}  \label{eq: N1-N2 coupling terms-incomp-Ortho-hyper}
\begin{split}
\kappa \in \mathcal{K}_{iii}=\{&46, 47, 56, 57, 146, 147, 156, 157, 246, 247, 256, 257, 446, 447,\\
&456, 457, 556, 557, 466, 467, 566, 567, 477, 577\}\,.
\end{split}
\end{equation}
\end{enumerate}
The terms $\mathcal{K}_{i}$ and $\mathcal{K}_{ii}$ are equivalent in form to \eqref{eq:universality terms-incom-T iso-hyper-1} and \eqref{eq:universality terms-incom-T iso-hyper-2} in transversely isotropic hyperelasticity. This leads to the following conclusion for orthotropic hyperelastic solids \citep{YavariGoriely2021,Yavari2022Universal} as well as orthotropic Cauchy elastic solids: the material preferred directions $\mathbf{N}_{1}$, $\mathbf{N}_{2}$ and $\mathbf{N}_{3}$ are universal for orthotropic solids if each direction is universal for transversely isotropic solids, and if the pairs $(\mathbf{N}_{1},\mathbf{N}_{2})$, $(\mathbf{N}_{1},\mathbf{N}_{3})$ and $(\mathbf{N}_{2},\mathbf{N}_{3})$ satisfy the $\mathcal{K}_{iii}$ constraints in hyperelasticity as well as the corresponding ones in Cauchy elasticity (we can extend this statement to Cauchy elasticity because we have shown that for incompressible transversely isotropic solids, the universality constraints in hyperelasticity and Cauchy elasticity are equivalent).

Let $\mathbf{n}=\mathbf{n}_1$ and $\mathbf{m}=\mathbf{n}_2$, and assume that $I_i \,(i=1,2,4,5,6,7)$ are not constant. Since $\mathbf{n}$ and $\mathbf{m}$ should satisfy the corresponding constraints of transversely isotropic solids, each of them has the same two possibilities introduced in \S\ref{Incomp-T iso-Elasticity} (recall that each set gives $63$ universality constraints in Cauchy elasticity). Together, they yield three cases including: 1) $\mathbf{n}$ and $\nabla I_i \,(i=1,2,4,5,6,7)$ are mutually parallel and orthogonal to $\mathbf{m}$, 2) $\nabla I_i \,(i=1,2,4,5,6,7)$ are mutually parallel and orthogonal to both $\mathbf{n}$ and $\mathbf{m}$, and 3) $\mathbf{n}$, $\mathbf{m}$ and $\nabla I_i \, (i=1,2,4,5,6,7)$ are all parallel. From the universal deformations and material preferred directions for orthotropic hyperelasticity reported in \citep{YavariGoriely2021,Yavari2022Universal}, it can be recognized that the third case is not valid in hyperelasticity. Therefore, in order to investigate a possible equivalence between the universality constraints in hyperelasticity and those in Cauchy elasticity, we need to study the terms corresponding to coupling $\mathbf{N}_1$ and $\mathbf{N}_2$, taking only the first two possibilities into account. Furthermore, note that in transversely isotropic solids, the symmetries of the terms related to $\mathbf{N}_1$ require that $\nabla I_1$, $\nabla I_2$, $\nabla I_4$ and $\nabla I_5$ be parallel. A similar argument can be applied to the symmetries of the terms corresponding to $\mathbf{N}_2$ as well, resulting in the functional dependence of $I_{1}$, $I_{2}$, $I_{6}$ and $I_{7}$. As a consequence, $\nabla I_i \,(i=1,2,4,5,6,7)$ in both cases are functionally dependent.

We first represent the terms $\mathcal{A}_{ab}^{\kappa}$ associated with $\mathcal{K}_{iii}$ as follows \citep{YavariGoriely2021,Yavari2022Universal}
\begin{equation} \label{eq:universality terms-incom-Ortho-hyper-1} 
\begin{split}
&\mathcal{A}_{ab}^{46}=\left(n_{a} \,n^n\right)_{|n} \,I_{6,b} +\left(n_a \, n^n \,I_{6,n}\right)_{|b}+\left(m_{a} \,m^n\right)_{|n} \,I_{4,b} +\left(m_a \, m^n \,I_{4,n}\right)_{|b} \,, \\
&\mathcal{A}_{ab}^{47}=\left(n_{a} \,n^n\right)_{|n} \,I_{7,b} +\left(n_a \, n^n \,I_{7,n}\right)_{|b}+{\mathfrak{K}^n_a}_{|n} \,I_{4,b} +\left(\mathfrak{K}^n_a \,I_{4,n}\right)_{|b} \,, \\
&\mathcal{A}_{ab}^{56}={\mathfrak{L}^n_a}_{|n} \,I_{6,b} +\left(\mathfrak{L}^n_a \,I_{6,n}\right)_{|b}+\left(m_{a} \,m^n\right)_{|n} \,I_{5,b} +\left(m_a \, m^n \,I_{5,n}\right)_{|b} \,, \\
&\mathcal{A}_{ab}^{57}={\mathfrak{L}^n_a}_{|n} \,I_{7,b} +\left(\mathfrak{L}^n_a \,I_{7,n}\right)_{|b}+{\mathfrak{K}^n_a}_{|n} \,I_{5,b} +\left(\mathfrak{K}^n_a \,I_{5,n}\right)_{|b} \,,
\end{split}
\end{equation}
and
\begin{equation} \label{eq:universality terms-incom-Ortho-hyper-2} 
\begin{split}
&\mathcal{A}_{ab}^{146}=b_{a}^{n} \,(I_{4,n} \,I_{6,b}+I_{4,b} \,I_{6,n}) \,,\\
&\mathcal{A}_{ab}^{147}=b_{a}^{n} \,(I_{4,n} \,I_{7,b}+I_{4,b} \,I_{7,n}) \,,\\
&\mathcal{A}_{ab}^{156}=b_{a}^{n} \,(I_{5,n} \,I_{6,b}+I_{5,b} \,I_{6,n}) \,,\\
&\mathcal{A}_{ab}^{157}=b_{a}^{n} \,(I_{5,n} \,I_{7,b}+I_{5,b} \,I_{7,n}) \,,\\
&\mathcal{A}_{ab}^{246}=c_{a}^{n} \,(I_{4,n} \,I_{6,b}+I_{4,b} \,I_{6,n}) \,,\\
&\mathcal{A}_{ab}^{247}=c_{a}^{n} \,(I_{4,n} \,I_{7,b}+I_{4,b} \,I_{7,n}) \,,\\
&\mathcal{A}_{ab}^{256}=c_{a}^{n} \,(I_{5,n} \,I_{6,b}+I_{5,b} \,I_{6,n}) \,,\\
&\mathcal{A}_{ab}^{257}=c_{a}^{n} \,(I_{5,n} \,I_{7,b}+I_{5,b} \,I_{7,n}) \,,\\
&\mathcal{A}_{ab}^{446}=n_a \,n^n \,(I_{4,n} \,I_{6,b}+I_{4,b} \,I_{6,n}) \,,\\
&\mathcal{A}_{ab}^{447}=n_a \,n^n \,(I_{4,n} \,I_{7,b}+I_{4,b} \,I_{7,n}) \,,\\
&\mathcal{A}_{ab}^{456}=n_a \,n^n \,(I_{5,n} \,I_{6,b}+I_{5,b} \,I_{6,n})+\mathfrak{L}^n_{a} \,(I_{4,n} \,I_{6,b}+I_{4,b} \,I_{6,n}) \,,\\
&\mathcal{A}_{ab}^{457}=n_a \,n^n \,(I_{5,n} \,I_{7,b}+I_{5,b} \,I_{7,n})+\mathfrak{L}^n_{a} \,(I_{4,n} \,I_{7,b}+I_{4,b} \,I_{7,n}) \,,\\
&\mathcal{A}_{ab}^{466}=m_a \,m^n \,(I_{4,n} \,I_{6,b}+I_{4,b} \,I_{6,n}) \,,\\
&\mathcal{A}_{ab}^{467}=m_a \,m^n \,(I_{4,n} \,I_{7,b}+I_{4,b} \,I_{7,n})+\mathfrak{K}^n_{a} \,(I_{4,n} \,I_{6,b}+I_{4,b} \,I_{6,n}) \,,\\
&\mathcal{A}_{ab}^{477}=\mathfrak{K}^n_{a} \,(I_{4,n} \,I_{7,b}+I_{4,b} \,I_{7,n}) \,,\\
&\mathcal{A}_{ab}^{556}=\mathfrak{L}^n_{a} \,(I_{5,n} \,I_{6,b}+I_{5,b} \,I_{6,n}) \,,\\
&\mathcal{A}_{ab}^{557}=\mathfrak{L}^n_{a} \,(I_{5,n} \,I_{7,b}+I_{5,b} \,I_{7,n}) \,,\\
&\mathcal{A}_{ab}^{566}=m_a \,m^n \,(I_{5,n} \,I_{6,b}+I_{5,b} \,I_{6,n}) \,\\
&\mathcal{A}_{ab}^{567}=m_a \,m^n \,(I_{5,n} \,I_{7,b}+I_{5,b} \,I_{7,n})+\mathfrak{K}^n_{a} \,(I_{5,n} \,I_{6,b}+I_{5,b} \,I_{6,n}) \,\\
&\mathcal{A}_{ab}^{577}=\mathfrak{K}^n_{a} \,(I_{5,n} \,I_{7,b}+I_{5,b} \,I_{7,n}) \,,
\end{split}
\end{equation}
where $\mathfrak{L}^{ab}=\ell_{1}^{ab}$ and $\mathfrak{K}^{ab}=\ell_{2}^{ab}$.

In Cauchy elasticity, $p_{|ab}$ is expressed as
\begin{equation} \label{eq:p_{|ab}-Ortho-Cauchy elastic}
p_{|ab}=\sum_{\kappa}\left({\mathcal{B}_{ab}^{1\kappa} \,\alpha_{1\kappa}}+{\mathcal{B}_{ab}^{2\kappa} \,\alpha_{2\kappa}}+{\mathcal{B}_{ab}^{4\kappa} \,\alpha_{4\kappa}}+{\mathcal{B}_{ab}^{5\kappa} \,\alpha_{5\kappa}}+{\mathcal{B}_{ab}^{6\kappa} \,\alpha_{6\kappa}}+{\mathcal{B}_{ab}^{7\kappa} \,\alpha_{7\kappa}}+{\mathcal{B}_{ab}^{8\kappa} \,\alpha_{8\kappa}}+{\mathcal{B}_{ab}^{9\kappa} \,\alpha_{9\kappa}}\right)\,,
\end{equation}
where $\mathcal{B}_{ab}^{1\kappa}$, $\mathcal{B}_{ab}^{2\kappa}$, $\mathcal{B}_{ab}^{4\kappa}$, $\mathcal{B}_{ab}^{5\kappa}$, $\mathcal{B}_{ab}^{6\kappa}$, $\mathcal{B}_{ab}^{7\kappa}$, $\mathcal{B}_{ab}^{8\kappa}$ and $\mathcal{B}_{ab}^{9\kappa}$ are the matrices of coefficients of $\alpha_{1\kappa}$, $\alpha_{2\kappa}$, $\alpha_{4\kappa}$, $\alpha_{5\kappa}$, $\alpha_{6\kappa}$, $\alpha_{7\kappa}$, $\alpha_{8\kappa}$ and $\alpha_{9\kappa}$, respectively ($\alpha_{i\kappa} = \partial \alpha_i/\partial I_\kappa$, where $\kappa$ is a multi-index). It can be shown that there are $44$ terms in Cauchy elasticity corresponding to coupling of $\mathbf{N}_1$ and $\mathbf{N}_2$, which are 
\begin{equation} \label{eq:universality terms-incom-Ortho-Cauchy elast-1} 
\begin{split}
&\mathcal{B}_{ab}^{46}=\left(n_{a} \,n^n\right)_{|n} \,I_{6,b} +\left(n_a \, n^n \,I_{6,n}\right)_{|b} \,,\\
&\mathcal{B}_{ab}^{47}=\left(n_{a} \,n^n\right)_{|n} \,I_{7,b} +\left(n_a \, n^n \,I_{7,n}\right)_{|b} \,,\\
&\mathcal{B}_{ab}^{56}={\mathfrak{L}^n_a}_{|n} \,I_{6,b} +\left(\mathfrak{L}^n_a \,I_{6,n}\right)_{|b} \,, \\
&\mathcal{B}_{ab}^{57}={\mathfrak{L}^n_a}_{|n} \,I_{7,b} +\left(\mathfrak{L}^n_a \,I_{7,n}\right)_{|b} \,, \\
&\mathcal{B}_{ab}^{66}={\bar{\mathfrak{L}}^n_a}_{|n} \,I_{6,b} +\left(\bar{\mathfrak{L}}^n_a \,I_{6,n}\right)_{|b} \,,\\
&\mathcal{B}_{ab}^{67}={\bar{\mathfrak{L}}^n_a}_{|n} \,I_{7,b} +\left(\bar{\mathfrak{L}}^n_a \,I_{7,n}\right)_{|b} \,,\\
&\mathcal{B}_{ab}^{74}=\left(m_{a} \,m^n\right)_{|n} \,I_{4,b} +\left(m_a \, m^n \,I_{4,n}\right)_{|b} \,, \\
&\mathcal{B}_{ab}^{75}=\left(m_{a} \,m^n\right)_{|n} \,I_{5,b} +\left(m_a \, m^n \,I_{5,n}\right)_{|b} \,, \\
&\mathcal{B}_{ab}^{84}={\mathfrak{K}^n_a}_{|n} \,I_{4,b} +\left(\mathfrak{K}^n_a \,I_{4,n}\right)_{|b} \,, \\
&\mathcal{B}_{ab}^{85}={\mathfrak{K}^n_a}_{|n} \,I_{5,b} +\left(\mathfrak{K}^n_a \,I_{5,n}\right)_{|b} \,, \\
&\mathcal{B}_{ab}^{94}={\bar{\mathfrak{K}}^n_a}_{|n} \,I_{4,b} +\left(\bar{\mathfrak{K}}^n_a \,I_{4,n}\right)_{|b} \,, \\
&\mathcal{B}_{ab}^{95}={\bar{\mathfrak{K}}^n_a}_{|n} \,I_{5,b} +\left(\bar{\mathfrak{K}}^n_a \,I_{5,n}\right)_{|b} \,, 
\end{split}
\end{equation}
and
\begin{equation} \label{eq:universality terms-incom-Ortho-Cauchy elast-2} 
\begin{split}
&\mathcal{B}_{ab}^{146}=b^{n}_{a} \,(I_{6,n} \,I_{4,b}+I_{6,b} \,I_{4,n}) \,,\\
&\mathcal{B}_{ab}^{147}=b^{n}_{a} \,(I_{4,n} \,I_{7,b}+I_{4,b} \,I_{7,n}) \,,\\
&\mathcal{B}_{ab}^{156}=b^{n}_{a} \,(I_{5,n} \,I_{6,b}+I_{5,b} \,I_{6,n}) \,,\\
&\mathcal{B}_{ab}^{157}=b^{n}_{a} \,(I_{5,n} \,I_{7,b}+I_{5,b} \,I_{7,n}) \,,\\
&\mathcal{B}_{ab}^{246}=c^{n}_{a} \,(I_{6,n} \,I_{4,b}+I_{6,b} \,I_{4,n}) \,,\\
&\mathcal{B}_{ab}^{247}=c^{n}_{a} \,(I_{4,n} \,I_{7,b}+I_{4,b} \,I_{7,n}) \,,\\
&\mathcal{B}_{ab}^{256}=c^{n}_{a} \,(I_{5,n} \,I_{6,b}+I_{5,b} \,I_{6,n}) \,,\\
&\mathcal{B}_{ab}^{257}=c^{n}_{a} \,(I_{5,n} \,I_{7,b}+I_{5,b} \,I_{7,n}) \,,\\
&\mathcal{B}_{ab}^{446}=n_a \,n^n \,(I_{4,n} \,I_{6,b}+I_{4,b} \,I_{6,n}) \,,\\
&\mathcal{B}_{ab}^{447}=n_a \,n^n \,(I_{4,n} \,I_{7,b}+I_{4,b} \,I_{7,n}) \,,\\
&\mathcal{B}_{ab}^{456}=n_a \,n^n \,(I_{5,n} \,I_{6,b}+I_{5,b} \,I_{6,n}) \,,\\
&\mathcal{B}_{ab}^{457}=n_a \,n^n \,(I_{5,n} \,I_{7,b}+I_{5,b} \,I_{7,n}) \,,\\
&\mathcal{B}_{ab}^{546}=\mathfrak{L}^{n}_{a} \,(I_{4,n} \,I_{6,b}+I_{4,b} \,I_{6,n}) \,,\\
&\mathcal{B}_{ab}^{547}=\mathfrak{L}_{a}^{n} \,(I_{4,n} \,I_{7,b}+I_{4,b} \,I_{7,n}) \,,\\
&\mathcal{B}_{ab}^{556}=\mathfrak{L}^{n}_{a} \,(I_{5,n} \,I_{6,b}+I_{5,b} \,I_{6,n}) \,,\\
&\mathcal{B}_{ab}^{557}=\mathfrak{L}^{n}_{a} \,(I_{5,n} \,I_{7,b}+I_{5,b} \,I_{7,n}) \,,\\
&\mathcal{B}_{ab}^{646}=\bar{\mathfrak{L}}^{n}_{a} \,(I_{4,n} \,I_{6,b}+I_{4,b} \,I_{6,n}) \,, \\
&\mathcal{B}_{ab}^{647}=\bar{\mathfrak{L}}_{a}^{n} \,(I_{4,n} \,I_{7,b}+I_{4,b} \,I_{7,n}) \,, \\
&\mathcal{B}_{ab}^{656}=\bar{\mathfrak{L}}^{n}_{a} \,(I_{5,n} \,I_{6,b}+I_{5,b} \,I_{6,n}) \,, \\
&\mathcal{B}_{ab}^{657}=\bar{\mathfrak{L}}^{n}_{a} \,(I_{5,n} \,I_{7,b}+I_{5,b} \,I_{7,n}) \,, \\
&\mathcal{B}_{ab}^{746}=m_a \,m^n \,(I_{4,n} \,I_{6,b}+I_{4,b} \,I_{6,n}) \,,\\
&\mathcal{B}_{ab}^{747}=m_a \,m^n \,(I_{4,n} \,I_{7,b}+I_{4,b} \,I_{7,n}) \,,\\
&\mathcal{B}_{ab}^{756}=m_a \,m^n \,(I_{5,n} \,I_{6,b}+I_{5,b} \,I_{6,n}) \,,\\
&\mathcal{B}_{ab}^{757}=m_a \,m^n \,(I_{5,n} \,I_{7,b}+I_{5,b} \,I_{7,n}) \,,\\
&\mathcal{B}_{ab}^{846}=\mathfrak{K}_{a}^{n} \,(I_{4,n} \,I_{6,b}+I_{4,b} \,I_{6,n}) \,,\\
&\mathcal{B}_{ab}^{847}=\mathfrak{K}_{a}^{n} \,(I_{4,n} \,I_{7,b}+I_{4,b} \,I_{7,n}) \,,\\
&\mathcal{B}_{ab}^{856}=\mathfrak{K}_{a}^{n} \,(I_{5,n} \,I_{6,b}+I_{5,b} \,I_{6,n}) \,,\\
&\mathcal{B}_{ab}^{857}=\mathfrak{K}_{a}^{n} \,(I_{5,n} \,I_{7,b}+I_{5,b} \,I_{7,n}) \,,\\
&\mathcal{B}_{ab}^{946}= \bar{\mathfrak{K}}_{a}^{n} \,(I_{4,n} \,I_{6,b}+I_{4,b} \,I_{6,n}) \,, \\
&\mathcal{B}_{ab}^{947}= \bar{\mathfrak{K}}_{a}^{n} \,(I_{4,n} \,I_{7,b}+I_{4,b} \,I_{7,n}) \,, \\
&\mathcal{B}_{ab}^{956}=\bar{\mathfrak{K}}_{a}^{n} \,(I_{5,n} \,I_{6,b}+I_{5,b} \,I_{6,n}) \,, \\
&\mathcal{B}_{ab}^{957}= \bar{\mathfrak{K}}_{a}^{n} \,(I_{5,n} \,I_{7,b}+I_{5,b} \,I_{7,n}) \,, 
\end{split}
\end{equation}
where $\bar{\mathfrak{L}}^{ab}=\bar{\ell}_{1}^{ab}$ and $\bar{\mathfrak{K}}^{ab}=\bar{\ell}_{2}^{ab}$. Consequently, the following relations are established between the two groups of terms
\begin{equation}  \label{eq:hyper-Cauchy elast-term-relation-incopm-ortho-1}
\begin{split}
&\mathcal{A}_{ab}^{46}=\mathcal{B}_{ab}^{46}+\mathcal{B}_{ab}^{74}\,, \\
&\mathcal{A}_{ab}^{47}=\mathcal{B}_{ab}^{47}+\mathcal{B}_{ab}^{84}\,, \\
&\mathcal{A}_{ab}^{56}=\mathcal{B}_{ab}^{56}+\mathcal{B}_{ab}^{75}\,, \\
&\mathcal{A}_{ab}^{57}=\mathcal{B}_{ab}^{57}+\mathcal{B}_{ab}^{85}\,, 
\end{split}
\end{equation}
and
\begin{equation}  \label{eq:hyper-Cauchy elast-term-relation-incopm-ortho-2}
\begin{split}
&\mathcal{A}_{ab}^{146}=\mathcal{B}_{ab}^{146}\,, \\
&\mathcal{A}_{ab}^{147}=\mathcal{B}_{ab}^{147}\,, \\
&\mathcal{A}_{ab}^{156}=\mathcal{B}_{ab}^{156}\,, \\
&\mathcal{A}_{ab}^{157}=\mathcal{B}_{ab}^{157}\,, \\
&\mathcal{A}_{ab}^{246}=\mathcal{B}_{ab}^{246}\,, \\
&\mathcal{A}_{ab}^{247}=\mathcal{B}_{ab}^{247}\,, \\
&\mathcal{A}_{ab}^{256}=\mathcal{B}_{ab}^{256}\,, \\
&\mathcal{A}_{ab}^{257}=\mathcal{B}_{ab}^{257}\,, \\
&\mathcal{A}_{ab}^{446}=\mathcal{B}_{ab}^{446}\,, \\
&\mathcal{A}_{ab}^{447}=\mathcal{B}_{ab}^{447}\,, \\
&\mathcal{A}_{ab}^{456}=\mathcal{B}_{ab}^{456}+\mathcal{B}_{ab}^{546}\,, \\
&\mathcal{A}_{ab}^{457}=\mathcal{B}_{ab}^{457}+\mathcal{B}_{ab}^{547}\,, \\
&\mathcal{A}_{ab}^{466}=\mathcal{B}_{ab}^{746}\,, \\
&\mathcal{A}_{ab}^{467}=\mathcal{B}_{ab}^{747}+\mathcal{B}_{ab}^{846}\,, \\
&\mathcal{A}_{ab}^{477}=\mathcal{B}_{ab}^{847}\,, \\
&\mathcal{A}_{ab}^{556}=\mathcal{B}_{ab}^{556}\,, \\
&\mathcal{A}_{ab}^{557}=\mathcal{B}_{ab}^{557}\,, \\
&\mathcal{A}_{ab}^{566}=\mathcal{B}_{ab}^{756}\,, \\
&\mathcal{A}_{ab}^{567}=\mathcal{B}_{ab}^{757}+\mathcal{B}_{ab}^{856}\,, \\
&\mathcal{A}_{ab}^{577}=\mathcal{B}_{ab}^{857}\,.
\end{split}
\end{equation}
In what follows, the symmetry equivalence between the terms in hyperelasticity given in \eqref{eq:universality terms-incom-Ortho-hyper-1} and \eqref{eq:universality terms-incom-Ortho-hyper-2}, and those in Cauchy elasticity represented by \eqref{eq:universality terms-incom-Ortho-Cauchy elast-1} and \eqref{eq:universality terms-incom-Ortho-Cauchy elast-2} for the two possible cases is examined separately.

\subsubsection{Case 1: $\mathbf{n} \,(\mathbf{x})$ and $\nabla I_i$ are mutually parallel and orthogonal to $\mathbf{m} \,(\mathbf{x})$ } \label{Case1-incomp-ortho}

This section discusses the symmetry equivalence between the symmetry constraint terms in hyperelasticity and those in Cauchy elasticity, when $\mathbf{n}$ and $\nabla I_i \,(i=1,2,4,5,6,7)$ are parallel to each other and orthogonal to $\mathbf{m}$ (because of the existing symmetry, this case is equivalent to that in which $\mathbf{m}$ and $\nabla I_i \,(i=1,2,4,5,6,7)$ are parallel to each other and orthogonal to $\mathbf{n}$).

We first investigate the symmetry equivalence between the terms $\mathcal{A}^\kappa_{ab}$ and $\mathcal{B}^\kappa_{ab}$ when $\kappa$ is a three-component index. Regarding the relations~\eqref{eq:hyper-Cauchy elast-term-relation-incopm-ortho-2} and \eqref{eq:universality terms-incom-Ortho-Cauchy elast-2}, only the terms appearing in \eqref{eq:hyper-Cauchy elast-term-relation-incopm-ortho-2}$_{11}$, \eqref{eq:hyper-Cauchy elast-term-relation-incopm-ortho-2}$_{12}$, \eqref{eq:hyper-Cauchy elast-term-relation-incopm-ortho-2}$_{14}$ and \eqref{eq:hyper-Cauchy elast-term-relation-incopm-ortho-2}$_{19}$ need to be examined. Since $\mathbf{n}$, $\nabla I_5$, $\nabla I_6$, and $\nabla I_7$ are parallel, the symmetry conditions for the terms $\mathcal{B}_{ab}^{456}$ and $\mathcal{B}_{ab}^{457}$ are trivially satisfied, and thus
\begin{equation}  \label{eq:A456,A457-equiv-incomp-ortho-Case1-1}
\mathcal{A}_{[ab]}^{456}=\mathcal{B}_{[ab]}^{546}=0\,,\qquad
\mathcal{A}_{[ab]}^{457}=\mathcal{B}_{[ab]}^{547}=0\,,
\end{equation}
or equivalently
\begin{equation}  \label{eq:A456,A457-equiv-incomp-ortho-Case1-2}
\mathcal{A}_{ab}^{456} \equiv \mathcal{B}_{ab}^{546}\,,\qquad
\mathcal{A}_{ab}^{457} \equiv \mathcal{B}_{ab}^{547}\,.
\end{equation}
Besides, owing to the orthogonality of $\mathbf{m}$ to $\nabla I_4$, $\nabla I_5$ and $\nabla I_7$, it follows that $\mathcal{B}_{ab}^{747}=\mathcal{B}_{ab}^{757}=0$, which leads to
\begin{equation}  \label{eq:A467,A567-equiv-incomp-ortho-Case1}
\mathcal{A}_{ab}^{467} \equiv \mathcal{B}_{ab}^{846}\,,\qquad
\mathcal{A}_{ab}^{567} \equiv \mathcal{B}_{ab}^{856}\,.
\end{equation}
The remaining terms in Cauchy elasticity are in one-to-one correspondence with those in hyperelasticity, as represented in \eqref{eq:hyper-Cauchy elast-term-relation-incopm-ortho-2}. It remains to show that the symmetries of the extra terms in Cauchy elasticity, namely $\mathcal{B}_{ab}^{646}$, $\mathcal{B}_{ab}^{647}$, $\mathcal{B}_{ab}^{656}$, $\mathcal{B}_{ab}^{657}$, $\mathcal{B}_{ab}^{946}$, $\mathcal{B}_{ab}^{947}$, $\mathcal{B}_{ab}^{956}$ and $\mathcal{B}_{ab}^{957}$ hold identically. Knowing that $\mathbf{n}$ is an eigenvector of $\mathbf{c}$, one may write $\bar{\mathfrak{L}}^n_a=\dfrac{2}{\lambda_1} \,n_a \,n^n$ (see \eqref{eq:ellbar and n-incomp-T iso-Case1}), and hence 
\begin{equation}  \label{eq:B646,B647-incomp-ortho-Case1}
\begin{split}
&\mathcal{B}_{ab}^{646}= \frac{2}{\lambda_1} \,I_{4,n} \,n^n \, n_a \,I_{6,b}+ \frac{2}{\lambda_1} \,I_{6,n} \,n^n \, n_a \,I_{4,b}\,,\\
&\mathcal{B}_{ab}^{647}= \frac{2}{\lambda_1} \,I_{4,n} \,n^n \, n_a \,I_{7,b}+ \frac{2}{\lambda_1} \,I_{7,n} \,n^n \, n_a \,I_{4,b}\,,\\
&\mathcal{B}_{ab}^{656}= \frac{2}{\lambda_1} \,I_{5,n} \,n^n \, n_a \,I_{6,b}+ \frac{2}{\lambda_1} \,I_{6,n} \,n^n \, n_a \,I_{5,b}\,,\\
&\mathcal{B}_{ab}^{657}= \frac{2}{\lambda_1} \,I_{5,n} \,n^n \, n_a \,I_{7,b}+ \frac{2}{\lambda_1} \,I_{7,n} \,n^n \, n_a \,I_{5,b}\,,
\end{split}
\end{equation}
which are symmetric because $\mathbf{n}$ and $\nabla I_i$ are parallel. In addition, the orthogonality of $\mathbf{m}$ to $\nabla I_i \,(i=1,2,4,5,6,7)$ together with $c^n_a \,I_{i,n}=\dfrac{I_{i,a}}{\lambda_1}$ gives the following relation
\begin{equation}  \label{eq:Kbar-incomp-ortho-case1-1}
\bar{\mathfrak{K}}^{n}_{a} \,I_{i,n}=m^n \,c^{c}_{a} \, m_c \,I_{i,n}+m_a \,c^{n}_{c} \,m^c \,I_{i,n}=m^n \,I_{i,n} \,c^{c}_{a} \,m_c + m_a \,\frac{1}{\lambda_1} \,I_{i,c} \,m^c=0 \,.
\end{equation}
It follows that
\begin{equation}  \label{eq:B946,B947-incomp-ortho-Case1}
\mathcal{B}_{ab}^{946}=\mathcal{B}_{ab}^{947}=\mathcal{B}_{ab}^{956}=\mathcal{B}_{ab}^{957}=0 \,,
\end{equation}
and thus the additional Cauchy elasticity terms introduce no new constraints. Consequently, in this case the symmetry of the term $\mathcal{A}_{ab}^\kappa$ is equivalent to that of $\mathcal{B}_{ab}^\kappa$, where $\kappa$ is a three-component index.

We next address the symmetry equivalence between the terms \eqref{eq:universality terms-incom-Ortho-hyper-1} and \eqref{eq:universality terms-incom-Ortho-Cauchy elast-1} (i.e., $\mathcal{A}^\kappa_{ab}$ and $\mathcal{B}^\kappa_{ab}$, where $\kappa$ is a two-component index). Given that $\mathbf{n}$ and $\nabla I_6$ are parallel, we have $n_a \,I_{6,n}=n_n \,I_{6,a}$ and the term $\mathcal{B}_{ab}^{46}$ becomes
\begin{equation}  \label{eq:B46-incomp-ortho-Case1-1}
\mathcal{B}_{ab}^{46}=\left(n_{a} \,n^n\right)_{|n} \,I_{6,b} +\left(I_{6,a} \, n^n \,n_n\right)_{|b} \,.
\end{equation}
Expanding the right-hand side of \eqref{eq:B46-incomp-ortho-Case1-1} and omitting the symmetric terms $n_a \,I_{6,b}$ and ${I_6}_{|ab}$, one has the following representation
\begin{equation}  \label{eq:B46-incomp-ortho-Case1-2}
\mathcal{B}_{ab}^{46}={n_a}_{|n} \,n^n \,I_{6,b}+I_{6,a} \,\left(n^n \,n_n\right)_{|b} \,.
\end{equation}
Referring to \eqref{eq:A44 and B44-incomp-T iso-Case1-5}, which provides an equivalent expression for $\mathcal{B}_{ab}^{44}$ when $\mathbf{n}$ and $\nabla I_4$ are parallel in transversely isotropic solids, and with respect to the functional dependence of $I_4$ and $I_6$, it is inferred that
\begin{equation}  \label{eq:B46-B44-equiv-incomp-ortho-Case1}
\mathcal{B}_{ab}^{46} \equiv \mathcal{B}_{ab}^{44}  \,.
\end{equation}
Similarly, we have
\begin{equation}  \label{eq:B47-B44-equiv-incomp-ortho-Case1}
\mathcal{B}_{ab}^{47} \equiv \mathcal{B}_{ab}^{44}  \,.
\end{equation}
We can also use \eqref{eq:ell and n-incomp-T iso-Case1} and $n_a \,I_{6,n}=n_n \,I_{6,a}$ to rewrite $\mathcal{B}_{ab}^{56}$ as
\begin{equation}  \label{eq:B56-incomp-ortho-Case1-1}
\mathcal{B}_{ab}^{56}=\left(2 \,\lambda_1 \,n_a \,n^n\right)_{|n} \,I_{6,b}+\left(2 \,\lambda_1 \,n_n \,n^n \,I_{6,a}\right)_{|b} \,,
\end{equation}
which is further simplified to read
\begin{equation}  \label{eq:B56-incomp-ortho-Case1-2}
\mathcal{B}_{ab}^{56} \equiv {n_a}_{|n} \,(2 \,\lambda_1 \,n^n) \,I_{6,b}+\left(2 \,\lambda_1 \,n_n \,n^n \right)_{|b} \,I_{6,a} \,.
\end{equation}
By applying \eqref{eq:A55 B55-incomp-T iso-Case1-Final} which is obtained from the symmetry of $\mathcal{B}_{ab}^{55}$ for transversely isotropic solids, \eqref{eq:B56-incomp-ortho-Case1-2} can be expressed in the following form
\begin{equation}  \label{eq:B56-incomp-ortho-Case1-3}
\mathcal{B}_{ab}^{56} \equiv \bar{c}_4 \,I_{5,a} \,I_{6,b}+\left(2 \,\lambda_1 \,n_n \,n^n\right)_{|b} \,I_{6,a}+\left(2 \,\lambda_1 \,n_n \,n^n\right)_{|a} \,I_{6,b} \,,
\end{equation}
which is evidently symmetric. Thus
\begin{equation}  \label{eq:B56-B55-equiv-incomp-ortho-Case1}
\mathcal{B}_{ab}^{56} \equiv \mathcal{B}_{ab}^{55}  \,.
\end{equation}
Employing the same approach, the following symmetry equivalences are likewise established
\begin{equation}  \label{eq:B57,B66,B67-B55-equiv-incomp-ortho-Case1}
\mathcal{B}_{ab}^{57} \equiv \mathcal{B}_{ab}^{55}  \,, \qquad
\mathcal{B}_{ab}^{66} \equiv \mathcal{B}_{ab}^{55}  \,, \qquad
\mathcal{B}_{ab}^{67} \equiv \mathcal{B}_{ab}^{55}  \,.
\end{equation}
From \eqref{eq:hyper-Cauchy elast-term-relation-incopm-ortho-1}, \eqref{eq:B46-B44-equiv-incomp-ortho-Case1}, \eqref{eq:B47-B44-equiv-incomp-ortho-Case1}, \eqref{eq:B56-B55-equiv-incomp-ortho-Case1} and \eqref{eq:B57,B66,B67-B55-equiv-incomp-ortho-Case1}$_1$, one deduces that
\begin{equation}  \label{eq:hyper-Cauchy elast-term-equiv-double index-incopm-ortho-1}
\begin{split}
&\mathcal{A}_{ab}^{46} \equiv \mathcal{B}_{ab}^{44}+\mathcal{B}_{ab}^{74}\,, \\
&\mathcal{A}_{ab}^{47} \equiv \mathcal{B}_{ab}^{44}+\mathcal{B}_{ab}^{84}\,, \\
&\mathcal{A}_{ab}^{56} \equiv \mathcal{B}_{ab}^{55}+\mathcal{B}_{ab}^{75}\,, \\
&\mathcal{A}_{ab}^{57} \equiv \mathcal{B}_{ab}^{55}+\mathcal{B}_{ab}^{85}\,.
\end{split}
\end{equation}
We know that the terms $\mathcal{B}_{ab}^{44} \,(=\mathcal{A}_{ab}^{44})$ and $\mathcal{B}_{ab}^{55} \,(=\mathcal{A}_{ab}^{55})$ are already symmetric, since each direction must be universal for transversely isotropic solids. As a consequence, \eqref{eq:hyper-Cauchy elast-term-equiv-double index-incopm-ortho-1} becomes
\begin{equation}  \label{eq:hyper-Cauchy elast-term-equiv-double index-incopm-ortho-2}
\mathcal{A}_{ab}^{46} \equiv \mathcal{B}_{ab}^{74}\,, \qquad
\mathcal{A}_{ab}^{47} \equiv \mathcal{B}_{ab}^{84}\,, \qquad
\mathcal{A}_{ab}^{56} \equiv \mathcal{B}_{ab}^{75}\,, \qquad
\mathcal{A}_{ab}^{57} \equiv \mathcal{B}_{ab}^{85}\,.
\end{equation}

To complete the proof of the symmetry equivalence between the Cauchy elasticity terms and the hyperelasticity terms in this case, we need to show that the remaining terms in Cauchy elasticity, including $\mathcal{B}_{ab}^{94}$ and $\mathcal{B}_{ab}^{95}$, are trivially symmetric. Following \citep{YavariGoriely2021,Yavari2022Universal}, a direct computation shows that for all universal deformations of incompressible orthotropic hyperelastic solids, when $\mathbf{m} \perp \nabla I_i$, the expression ${\bar{\mathfrak{K}}^n_a}_{|n}$ vanishes identically. This, along with \eqref{eq:Kbar-incomp-ortho-case1-1}, leads to $\mathcal{B}_{ab}^{94}=\mathcal{B}_{ab}^{95}=0$. It follows that there are at most four independent terms in hyperelasticity ($\mathcal{A}^{46}$, $\mathcal{A}^{47}$, $\mathcal{A}^{56}$ and $\mathcal{A}^{57}$) and at most four independent terms in Cauchy elasticity ($\mathcal{B}^{74}$, $\mathcal{B}^{84}$, $\mathcal{B}^{75}$ and $\mathcal{B}^{85}$), which are related by \eqref{eq:hyper-Cauchy elast-term-equiv-double index-incopm-ortho-2}. Hence, the two groups of symmetry constraints are equivalent. This conclusion completes the proof of the symmetry equivalence in this case.

\subsubsection{Case 2: $\mathbf{n} \,(\mathbf{x})$ and $\mathbf{m} \,(\mathbf{x})$ are orthogonal to $\nabla I_i$} \label{Case2-incomp-ortho}
In this case, both $\mathbf{n}$ and $\mathbf{m}$ are orthogonal to $\nabla I_i \,(i=1,2,4,5,6,7)$, and accordingly
\begin{equation}  \label{eq:zero terms in case 2-incomp-ortho}
\begin{split}
&n^n \,I_{i,n}=m^n \,I_{i,n}=0 \,,\\
&\mathfrak{L}^n_a \,I_{i,n}=\bar{\mathfrak{L}}^n_a \,I_{i,n}=\mathfrak{K}^n_a \,I_{i,n}=\bar{\mathfrak{K}}^n_a \,I_{i,n}=0 \,,\\
&{\mathfrak{L}^n_a}_{|n}={\mathfrak{K}^n_a}_{|n}={\bar{\mathfrak{L}}^n_a}_{|n}={\bar{\mathfrak{K}}^n_a}_{|n}=0 \,.
\end{split}
\end{equation}
In view of the above relations, most of the hyperelasticity and Cauchy elasticity terms in \eqref{eq:universality terms-incom-Ortho-hyper-1}, \eqref{eq:universality terms-incom-Ortho-hyper-2}, \eqref{eq:universality terms-incom-Ortho-Cauchy elast-1} and \eqref{eq:universality terms-incom-Ortho-Cauchy elast-2} are identically zero. The remaining terms are related as follows (see \eqref{eq:hyper-Cauchy elast-term-relation-incopm-ortho-1} and \eqref{eq:hyper-Cauchy elast-term-relation-incopm-ortho-2})
\begin{equation}  \label{eq:hyper-Cauchy elast-term-relatin-incopm-ortho-Case2}
\begin{split}
&\mathcal{A}_{ab}^{46}=\mathcal{B}_{ab}^{46}+\mathcal{B}_{ab}^{74}\,, \\
&\mathcal{A}_{ab}^{47}=\mathcal{B}_{ab}^{47}\,, \\
&\mathcal{A}_{ab}^{56}=\mathcal{B}_{ab}^{75}\,, \\
&\mathcal{A}_{ab}^{146}=\mathcal{B}_{ab}^{146}\,, \\
&\mathcal{A}_{ab}^{147}=\mathcal{B}_{ab}^{147}\,, \\
&\mathcal{A}_{ab}^{156}=\mathcal{B}_{ab}^{156}\,, \\
&\mathcal{A}_{ab}^{157}=\mathcal{B}_{ab}^{157}\,, \\
&\mathcal{A}_{ab}^{246}=\mathcal{B}_{ab}^{246}\,, \\
&\mathcal{A}_{ab}^{247}=\mathcal{B}_{ab}^{247}\,, \\
&\mathcal{A}_{ab}^{256}=\mathcal{B}_{ab}^{256}\,, \\
&\mathcal{A}_{ab}^{257}=\mathcal{B}_{ab}^{257}\,.\end{split}
\end{equation}
In this case $\mathcal{B}^{46}_{ab}=\left(n_a \,n^n\right)_{|n} \,I_{6,b}$ and $\mathcal{B}^{44}_{ab}=\left(n_a \,n^n\right)_{|n} \,I_{4,b}$. Due to the functional dependence of $I_4$ and $I_{6}$, we have $\mathcal{B}^{46}_{ab} \equiv \mathcal{B}^{44}_{ab}$. Since $\mathcal{B}^{44}_{ab}$ is symmetric, it follows from \eqref{eq:hyper-Cauchy elast-term-relatin-incopm-ortho-Case2}$_1$ that
\begin{equation}  \label{eq:A46-B74-equiv-incomp-ortho-Case2}
\mathcal{A}_{ab}^{46} \equiv \mathcal{B}_{ab}^{74}  \,.
\end{equation}
Thus, the symmetry constraint terms in Cauchy elasticity correspond one to one with those in hyperelasticity in this case, showing that the two sets of constraints are equivalent.

We showed that for incompressible orthotropic solids, the symmetries of the terms corresponding to the coupling of $\mathbf{N}_1$ and $\mathbf{N}_2$ in hyperelasticity and those in Cauchy elasticity are equivalent for the two possible cases. Since each direction must be universal for transversely isotropic solids, the symmetry equivalence between the corresponding terms in hyperelasticity and Cauchy elasticity holds as demonstrated in \S\ref{Incomp-T iso-Elasticity}. In summary, we have proved the following result.
\begin{prop}
The universal deformations and material preferred directions of incompressible orthotropic Cauchy elasticity are identical to those of incompressible orthotropic hyperelasticity.
\end{prop}

\section{Universality Constraints in Monoclinic Elasticity} \label{Mono-Elasticity}
The material symmetry of a monoclinic solid is characterized by three unit vectors $\mathbf{N}_1$, $\mathbf{N}_2$ and $\mathbf{N}_3$ \citep{merodio2020finite}. The first two vectors, $\mathbf{N}_1$ and $\mathbf{N}_2$, are not mutually perpendicular, whereas $\mathbf{N}_3$ is oriented normal to the plane they span, i.e., $\mathbf{N}_1 \not\perp \mathbf{N}_2$, and $\mathbf{N}_3 \perp \operatorname{span}\{\mathbf{N}_1,\mathbf{N}_2\}$. The strain energy density function of monoclinic hyperelastic solids is defined in terms of nine invariants, i.e., $W=W(I_1,\cdots,I_9)$. The first seven invariants are the same as those introduced for orthotropic solids, while the two additional invariants are defined as  \citep{Spencer1986}
\begin{equation}  \label{eq:invariants-mono}
I_8=\mathcal{g} \,\mathbf{N}_1 \cdot \mathbf{C}\cdot \mathbf{N}_2 \,, \qquad
I_9=\mathcal{g}^2\,,
\end{equation}
where $\mathcal{g}= \mathbf{N}_1 \cdot \mathbf{N}_2$. The second Piola–Kirchhoff stress tensor for monoclinic solids is written as \citep{Yavari2022Universal,YavariGoriely2021}
\begin{equation} \label{eq:second piola-mono-hyper} 
\begin{split}
\mathbf{S} = &2\,W_1 \, \mathbf{G}^{\sharp}+2\,W_2 \,(I_2 \,\mathbf{C}^{-1}-I_3 \,\mathbf{C}^{-2})+2\,W_3 \,I_3 \,\mathbf{C}^{-1}\\
&+2 \, W_4 \,(\mathbf{N}_1 \otimes \mathbf{N}_1)+2\,W_5 \,[\mathbf{N}_1 \otimes (\mathbf{C} \cdot \mathbf{N}_1)+(\mathbf{C} \cdot \mathbf{N}_1) \otimes \mathbf{N}_1] \\
&+2 \, W_6 \,(\mathbf{N}_2 \otimes \mathbf{N}_2)+2\,W_7 \,[\mathbf{N}_2 \otimes (\mathbf{C} \cdot \mathbf{N}_2)+(\mathbf{C} \cdot \mathbf{N}_2) \otimes \mathbf{N}_2] 
+\mathcal{g} \,W_8 \, (\mathbf{N}_1 \otimes \mathbf{N}_2 +\mathbf{N}_2 \otimes \mathbf{N}_1) \,.  
\end{split}
\end{equation}
Note that $W_9$ does not contribute to the above equation because $\dfrac{\partial{I_9}}{\partial{\mathbf{C}^\flat}}=0$. The Cauchy stress is then written as
\begin{equation} \label{eq:Cauchy stress-mono-hyper} 
\begin{split}
\boldsymbol{\sigma}=&\frac{2}{\sqrt{I_3}} \,W_1 \,\mathbf{b}^\sharp+\frac{2}{\sqrt{I_3}} \,(I_2 \,W_2+I_3 \,W_3) \,\mathbf{g}^\sharp -2 \,\sqrt{I_3} \,W_2 \, \mathbf{c}^\sharp\\
&+\frac{2}{\sqrt{I_3}} \,W_4 \,(\mathbf{n}_1 \otimes \mathbf{n}_1)+\frac{2}{\sqrt{I_3}} \,W_5 \,[\mathbf{n}_1 \otimes (\mathbf{b} \cdot \mathbf{n}_1)+(\mathbf{b} \cdot \mathbf{n}_1) \otimes \mathbf{n}_1] \\
&+\frac{2}{\sqrt{I_3}} \,W_6 \,(\mathbf{n}_2 \otimes \mathbf{n}_2)+\frac{2}{\sqrt{I_3}} \,W_7 \,[\mathbf{n}_2 \otimes (\mathbf{b} \cdot \mathbf{n}_2)+(\mathbf{b} \cdot \mathbf{n}_2) \otimes \mathbf{n}_2] 
+\frac{1}{\sqrt{I_3}} \,\mathcal{g} \,W_8 \,(\mathbf{n}_1 \otimes \mathbf{n}_2+\mathbf{n}_2 \otimes \mathbf{n}_1)\,,
\end{split}
\end{equation}
where $W_i= \dfrac{\partial W}{\partial I_i} \,(i=1,\cdots,9)$, $\mathbf{n}_1=\mathbf{F} \cdot \mathbf{N}_1$ and $\mathbf{n}_2=\mathbf{F} \cdot \mathbf{N}_2$. In components, it takes the following form
\begin{equation} \label{eq:Cauchy stress-mono-hyper-components} 
\begin{split}
\sigma^{ab} =&\frac{2}{\sqrt{I_3}}\,[W_1 \,b^{ab}+(I_2 \,W_2+I_3 \,W_3)g^{ab}-I_3 \,W_2 \, c^{ab}\\
&+W_4 \,n_{1}^{a} \,n_{1}^{b}+W_5 \,(n_{1}^{a} \,b^{bc} \,n_{1}^{d} \,g_{cd} + n_{1}^{b} \,b^{ac} \,n_{1}^{d} \,g_{cd}) \\
&+W_6 \,n_{2}^{a} \,n_{2}^{b}+W_7 \,(n_{2}^{a} \,b^{bc} \,n_{2}^{d} \,g_{cd} + n_{2}^{b} \,b^{ac} \,n_{2}^{d} \,g_{cd}) 
+\frac{1}{2} \,\mathcal{g} \,W_8 \,(n_1^a \,n_2^b+n_2^a \,n_1^b)] \,,
\end{split}
\end{equation}
where $n_1^a=F^{a}{}_A \,N_{1}^{A}$ and $n_2^a=F^{a}{}_{A} \,N_{2}^{A}$.

Referring to \citep{YavariGoriely2025}, the second Piola–Kirchhoff stress tensor for monoclinic Cauchy elastic solids has the following representation
\begin{equation} \label{eq:second piola-mono-Cauchy elast} 
\begin{split}
\mathbf{S} = &a_0 \,\mathbf{G}^{\sharp}+ a_1 \,\mathbf{C}^{\sharp}+a_2 \, \mathbf{C}^{2\sharp}+a_3 \,(\mathbf{N}_1\otimes \mathbf{N}_1)+a_4 \,[ \mathbf{N}_1\otimes (\mathbf{C}\cdot \mathbf{N}_1)+( \mathbf{C}\cdot \mathbf{N}_1)\otimes \mathbf{N}_1] \\
&+a_5 \,[\mathbf{N}_1\otimes (\mathbf{C}^{2} \cdot \mathbf{N}_1)+(\mathbf{C}^2 \cdot \mathbf{N}_1)\otimes \mathbf{N}_1] \\
&+a_6 \,(\mathbf{N}_2\otimes \mathbf{N}_2)+a_7 \,[\mathbf{N}_2 \otimes (\mathbf{C}\cdot \mathbf{N}_2)+(\mathbf{C}\cdot \mathbf{N}_2)\otimes \mathbf{N}_2] \\
&+a_8 \,[\mathbf{N}_2 \otimes (\mathbf{C}^{2} \cdot \mathbf{N}_2)+(\mathbf{C}^2 \cdot \mathbf{N}_2)\otimes \mathbf{N}_2]+\mathcal{g} \,a_9 \,(\mathbf{N}_1 \otimes \mathbf{N}_2+\mathbf{N}_2 \otimes \mathbf{N}_1)\,, 
\end{split}
\end{equation}
where $a_i \,(I_1,\cdots,I_9) \,,i=0,\cdots,9$ are the response functions. The Cauchy stress tensor is similarly written as
\begin{equation} \label{eq:Cauchy stress-mono-Cauchy elast} 
\begin{split}
\boldsymbol{\sigma}=&\tilde{a}_0 \,\mathbf{g}^{\sharp}+\tilde{a}_1 \,\mathbf{b}^{\sharp}+\tilde{a}_2 \,\mathbf{c}^{\sharp} +\tilde{a}_3 \,(\mathbf{n}_1 \otimes \mathbf{n}_1)+\tilde{a}_4 \,[\mathbf{n}_1 \otimes (\mathbf{b} \cdot \mathbf{n}_1)+(\mathbf{b} \cdot \mathbf{n}_1) \otimes \mathbf{n}_1] \\
&+\tilde{a}_5 \,[\mathbf{n}_1 \otimes (\mathbf{c} \cdot \mathbf{n}_1)+(\mathbf{c} \cdot \mathbf{n}_1) \otimes \mathbf{n}_1] 
+\tilde{a}_6 \,(\mathbf{n}_2 \otimes \mathbf{n}_2)+\tilde{a}_7 \,[\mathbf{n}_2 \otimes (\mathbf{b} \cdot \mathbf{n}_2)+(\mathbf{b} \cdot \mathbf{n}_2) \otimes \mathbf{n}_2] \\
&+\tilde{a}_8 \,[\mathbf{n}_2 \otimes (\mathbf{c} \cdot \mathbf{n}_2)+(\mathbf{c} \cdot \mathbf{n}_2) \otimes \mathbf{n}_2] +\mathcal{g} \,\tilde{a}_9 \,(\mathbf{n}_1 \otimes \mathbf{n}_2+\mathbf{n}_2 \otimes \mathbf{n}_1 )\,,
\end{split}
\end{equation}
where $\tilde{a}_i \,(I_1,\cdots,I_9) \,,i=0,\cdots,9$ are the response functions.

Taking $I_3=1$, we have the following expression for the second Piola–Kirchhoff stress tensor of incompressible monoclinic hyperelastic solids \citep{Yavari2022Universal,YavariGoriely2021}
\begin{equation} \label{eq:second piola-incomp-mono-hyper} 
\begin{split}
\mathbf{S} =& -p \, \mathbf{C}^{-1}+2\,W_1 \, \mathbf{G}^{\sharp}+2\,W_2 \,(I_2 \,\mathbf{C}^{-1}-\mathbf{C}^{-2}) \\
&+2 \, W_4 \,(\mathbf{N}_1 \otimes \mathbf{N}_1)
+2\,W_5 \,[\mathbf{N}_1 \otimes (\mathbf{C} \cdot \mathbf{N}_1)+(\mathbf{C} \cdot \mathbf{N}_1) \otimes \mathbf{N}_1] \\
&+2 \, W_6 \,(\mathbf{N}_2 \otimes \mathbf{N}_2)
+2\,W_7 \,[\mathbf{N}_2 \otimes (\mathbf{C} \cdot \mathbf{N}_2)+(\mathbf{C} \cdot \mathbf{N}_2) \otimes \mathbf{N}_2] 
+\mathcal{g} \,W_8 \, (\mathbf{N}_1 \otimes \mathbf{N}_2 +\mathbf{N}_2 \otimes \mathbf{N}_1) \,,
\end{split}
\end{equation}
where $W=W \,(I_1,I_2,I_4,I_5,I_6,I_7,I_8,I_9)$. The Cauchy stress is similarly written as \citep{Yavari2022Universal,YavariGoriely2021}
\begin{equation} \label{eq:Cauchy stress-incomp-mono-hyper}
\begin{split}
\boldsymbol{\sigma} =& -p \,\mathbf{g}^\sharp+ 2 \,W_1 \, \mathbf{b}^\sharp- 2 \,W_2 \, \mathbf{c}^\sharp+ 2 \,W_4 \,(\mathbf{n}_1 \otimes \mathbf{n}_1) \\
&+ 2 \,W_5 \, [\mathbf{n}_1 \otimes (\mathbf{b} \cdot \mathbf{n}_1)+(\mathbf{b} \cdot \mathbf{n}_1) \otimes \mathbf{n}_1] + 2 \,W_6 \,(\mathbf{n}_2 \otimes \mathbf{n}_2) \\
&+ 2 \,W_7 \, [\mathbf{n}_2 \otimes (\mathbf{b} \cdot \mathbf{n}_2)+(\mathbf{b} \cdot \mathbf{n}_2) \otimes \mathbf{n}_2]+\mathcal{g} \,W_8 \,(\mathbf{n}_1 \otimes \mathbf{n}_2+\mathbf{n}_2 \otimes \mathbf{n}_1)\,.
\end{split}
\end{equation}
Hence, the Cauchy stress can be represented in components as
\begin{equation} \label{eq:Cauchy stress-incomp-mono-hyper-components}
\sigma^{ab} = -p \,g^{ab}+ 2 \,W_1 \, b^{ab}- 2 \,W_2 \, c^{ab}+ 2 \,W_4 \,n_1^{a} \,n_1^{b}+ 2 \,W_5 \, \ell_1^{ab} + 2 \,W_6 \,n_2^{a} \,n_2^{b} + 2 \,W_7 \, \ell_2^{ab}+ W_8 \, \ell_3^{ab} \,,
\end{equation}
where $\ell_1^{ab}=n_1^a \,b^{bc} \,n_1^d \,g_{cd} + n_1^b \,b^{ac} \,n_1^d \,g_{cd}$, $\ell_2^{ab}=n_2^a \,b^{bc} \,n_2^d \,g_{cd} + n_2^b \,b^{ac} \,n_2^d \,g_{cd}$, and $\ell_3^{ab}=\mathcal{g} \,(n_1^a \,n_2^b+n_1^b \,n_2^a)$.  

In the case of incompressible monoclinic Cauchy elastic solids, the second Piola–Kirchhoff stress tensor is written as
\begin{equation} \label{eq:second piola-incomp-mono-Cauchy elast} 
\begin{split}
\mathbf{S}= &-p \,\mathbf{C}^{-1}+\bar{a}_0 \,\mathbf{G}^{\sharp}+ \bar{a}_1 \,\mathbf{C}^{\sharp}+\bar{a}_2 \,(\mathbf{N}_1\otimes \mathbf{N}_1)+\bar{a}_4 \,[\mathbf{N}_1\otimes (\mathbf{C}\cdot \mathbf{N}_1)+( \mathbf{C}\cdot \mathbf{N}_1)\otimes \mathbf{N}_1] \\
&+\bar{a}_5 \,[\mathbf{N}_1\otimes (\mathbf{C}^{-1} \cdot \mathbf{N}_1)+(\mathbf{C}^{-1} \cdot \mathbf{N}_1)\otimes \mathbf{N}_1] \\
&+\bar{a}_6 \,(\mathbf{N}_2\otimes \mathbf{N}_2)+\bar{a}_7 \,[\mathbf{N}_2\otimes (\mathbf{C}\cdot \mathbf{N}_2)+(\mathbf{C}\cdot \mathbf{N}_2)\otimes \mathbf{N}_2] \\
&+\bar{a}_8 \,[\mathbf{N}_2\otimes (\mathbf{C}^{-1} \cdot \mathbf{N}_2)+(\mathbf{C}^{-1} \cdot \mathbf{N}_2)\otimes \mathbf{N}_2]+\mathcal{g} \,\bar{a}_9 \,(\mathbf{N}_1 \otimes \mathbf{N}_2 +\mathbf{N}_2 \otimes \mathbf{N}_1) \,,
\end{split}
\end{equation}
where $\bar{a}_i \,(I_1,I_2,I_4,I_5,I_6,I_7,I_8,I_9)$, $i=0,1,2,4,5,6,7,8,9$ are the response functions. Thus, the Cauchy stress tensor for incompressible monoclinic Cauchy elastic solids is represented as
\begin{equation} \label{eq:Cauchy stress-incomp-mono-Cauchy elast} 
\begin{split}
\boldsymbol{\sigma}=&-p \,\mathbf{g}^{\sharp}+\alpha_1 \,\mathbf{b}^{\sharp}+\alpha_2 \,\mathbf{c}^{\sharp} +\alpha_4 \,(\mathbf{n}_1 \otimes \mathbf{n}_1)+\alpha_5 \,[\mathbf{n}_1 \otimes (\mathbf{b} \cdot \mathbf{n}_1)+(\mathbf{b} \cdot \mathbf{n}_1) \otimes \mathbf{n}_1] \\
&+\alpha_6 \,[\mathbf{n}_1 \otimes (\mathbf{c} \cdot \mathbf{n}_1)+(\mathbf{c} \cdot \mathbf{n}_1) \otimes \mathbf{n}_1] 
+\alpha_7 \,(\mathbf{n}_2 \otimes \mathbf{n}_2)+\alpha_8 \,[\mathbf{n}_2 \otimes (\mathbf{b} \cdot \mathbf{n}_2)+(\mathbf{b} \cdot \mathbf{n}_2) \otimes \mathbf{n}_2] \\
&+\alpha_9 \,[\mathbf{n}_2 \otimes (\mathbf{c} \cdot \mathbf{n}_2)+(\mathbf{c} \cdot \mathbf{n}_2) \otimes \mathbf{n}_2] + \mathcal{g} \,\alpha_{10} \,(\mathbf{n}_1 \otimes \mathbf{n}_2 +\mathbf{n}_2 \otimes \mathbf{n}_1) \,,
\end{split}
\end{equation}
where $\alpha_{i}=\alpha_{i} \,(I_1, I_2, I_4,I_5,I_6,I_7,I_8,I_9)$, $i=1,2,4,5,6,7,8,9,10$ are arbitrary response functions.

\subsection{Compressible monoclinic solids}
The Cauchy stress for compressible monoclinic Cauchy elastic solids is given in components as (see \eqref{eq:Cauchy stress-mono-Cauchy elast})
\begin{equation} \label{eq:Cauchy stress-mono-Cauchy elast-components} 
\sigma^{ab}=\alpha_1 \,g^{ab}+\alpha_2 \,b^{ab}+\alpha_3 \,c^{ab} +\alpha_4 \,n_1^a \,n_1^b+\alpha_5 \,\ell_1^{ab}+\alpha_6 \,\bar{\ell}_1^{ab}+\alpha_7 \,n_2^a \,n_2^b+\alpha_8 \,\ell_2^{ab}+\alpha_9 \,\bar{\ell}_2^{ab}+\alpha_{10} \,\ell_3^{ab} \,,
\end{equation}
where $\bar{\ell}_1^{ab}=n_1^a \,c^{bc} \,n_1^d \,g_{cd} + n_1^b \,c^{ac} \,n_1^d \,g_{cd}$, $\bar{\ell}_2^{ab}=n_2^a \,c^{bc} \,n_2^d \,g_{cd} + n_2^b \,c^{ac} \,n_2^d \,g_{cd}$, and $\alpha_i \,(I_1,I_2,I_3,I_4,I_5,I_6,I_7,I_8,I_9)$, $i=1,\cdots,10$ are arbitrary response functions. Substituting \eqref{eq:Cauchy stress-mono-Cauchy elast-components} into the equilibrium equations~\eqref{eq:equilibrium} yields
\begin{equation} \label{eq:equilibrium-comp-mono-1}
\begin{split}
&\alpha_2 \,{b^{ab}}_{|b}+\alpha_3 \,{c^{ab}}_{|b}
+\alpha_4 \,\left(n_1^a \,n_1^b\right)_{|b}+\alpha_5 \,{\ell_1^{ab}}_{|b}+\alpha_6 \,{\bar{\ell}_1^{ab}}_{|b}+\alpha_7 \,\left(n_2^a \,n_2^b\right)_{|b}+\alpha_8 \,{\ell_2^{ab}}_{|b}+\alpha_9 \,{\bar{\ell}_2^{ab}}_{|b} \\
&+\alpha_{10} \,{\ell_3^{ab}}_{|b}+\alpha_{1,i} \,I_{i,b} \,g^{ab}+\alpha_{2,i} \,I_{i,b} \,b^{ab}+\alpha_{3,i} \,I_{i,b} \,c^{ab}+\alpha_{4,i} \,I_{i,b} \,(n_1^a \,n_1^b)+\alpha_{5,i} \,I_{i,b} \,\ell_1^{ab} \\
&+\alpha_{6,i} \,I_{i,b} \,\bar{\ell}_1^{ab}+\alpha_{7,i} \,I_{i,b} \,(n_2^a \,n_2^b)+\alpha_{8,i} \,I_{i,b} \,\ell_2^{ab}+\alpha_{9,i} \,I_{i,b} \,\bar{\ell}_2^{ab}+\alpha_{10,i} \,I_{i,b} \,\ell_3^{ab}=0 \,,
\end{split}
\end{equation}
where $i=1,\cdots,9$. Therefore, the following universality constraints are obtained
\begin{equation} \label{eq:constraints-comp-mono}
\begin{split}
&{b^{ab}}_{|b}={c^{ab}}_{|b}=0 \,, \\
&\left(n_1^a \,n_1^b\right)_{|b}=0 \,,\\
&\left(n_2^a \,n_2^b\right)_{|b}=0 \,, \\
&{\ell_1^{ab}}_{|b}=0 \,,\\
&{\ell_2^{ab}}_{|b}=0 \,, \\
&{\ell_3^{ab}}_{|b}=0 \,, \\
&I_{i,b} \,g^{ab}=0\,, \\
&I_{i,b} \,b^{ab}=0\,, \\
&I_{i,b} \,c^{ab}=0\,, \\
&I_{i,b} \,n_1^a \,n_1^b=0 \,,\\
&I_{i,b} \,n_2^a \,n_2^b=0\,, \\
&I_{i,b} \,\ell_1^{ab}=0 \,,\\
&I_{i,b} \,\ell_2^{ab}=0\,, \\
&I_{i,b} \,\ell_3^{ab}=0\,, \\
&{\bar{\ell}_1^{ab}}_{|b}=0 \,, \\
&{\bar{\ell}_2^{ab}}_{|b}=0 \,, \\
&I_{i,b} \,\bar{\ell}_1^{ab}=0\,, \\
&I_{i,b} \,\bar{\ell}_2^{ab}=0\,.
\end{split}
\end{equation}
The constraints~\eqref{eq:constraints-comp-mono}$_1$--\eqref{eq:constraints-comp-mono}$_{14}$ are identical to those for hyperelastic solids \citep{Yavari2022Universal,YavariGoriely2021}. Thus, similar to orthotropic solids, \eqref{eq:constraints-comp-mono}$_{15}$--\eqref{eq:constraints-comp-mono}$_{18}$ are the extra constraints in Cauchy elasticity.

Except for \eqref{eq:constraints-comp-mono}$_6$ and \eqref{eq:constraints-comp-mono}$_{14}$, the remaining constraints (for $i=1,\cdots,7$) are the same as those for orthotropic solids. Consequently, universal deformations are homogeneous, $I_i \,(i=1,\cdots,7)$ are constant, and $\mathbf{N}_1$ and $\mathbf{N}_2$ are constant unit vectors. The constraints ${\ell_3^{ab}}_{|b}=0$ and $I_{i,b} \,g^{ab}=0$ (for $i=8,9$) imply that $\mathbf{N}_3$ is also a constant vector and that $I_8$ and $I_9$ are constant (note that \eqref{eq:constraints-comp-mono}$_{14}$ is then trivially satisfied). Given these results, it can be shown that the additional constraints in Cauchy elasticity, i.e., \eqref{eq:constraints-comp-mono}$_{15}$–-\eqref{eq:constraints-comp-mono}$_{18}$, are trivially satisfied. In summary, we have proved the following result.
\begin{prop}
The universal deformations and material preferred directions of compressible monoclinic Cauchy elasticity are identical to those of compressible monoclinic hyperelasticity.
\end{prop}

\subsection{Incompressible monoclinic solids} \label{Incomp-Mono-Elasticity}
Let us substitute \eqref{eq:Cauchy stress-incomp-mono-hyper} into the equilibrium equations to get
\begin{equation} \label{eq:pa-mono-hyperelastic}
p_{|b} \,g^{ab}=2 \,{\left[W_1 \,b^{ab}-W_2 \,c^{ab} +W_4 \,n_1^a \,n_1^b+W_5 \,\ell_1^{ab}+W_6 \,n_2^a \,n_2^b+W_7 \,\ell_2^{ab} +\frac{1}{2} \,W_8 \,\ell_3^{ab}\right]}_{|b} \,,
\end{equation}
for hyperelastic solids. Similarly, substituting the Cauchy stress \eqref{eq:Cauchy stress-incomp-mono-Cauchy elast} into the equilibrium equations~\eqref{eq:equilibrium} gives the following equation for Cauchy elastic solids
\begin{equation} \label{eq:pa-mono-Cauchy elast}
p_{|b} \,g^{ab}={\left[\alpha_1 \,b^{ab}+\alpha_2 \,c^{ab}+\alpha_4 \,n_1^a \,n_1^b+\alpha_5 \,\ell_1^{ab}+\alpha_6 \,\bar{\ell}_1^{ab}+\alpha_7 \,n_2^a \,n_2^b+\alpha_8 \,\ell_2^{ab}+\alpha_9 \,\bar{\ell}_2^{ab}+\alpha_{10} \,\ell_3^{ab}\right]}_{|b} \,.
\end{equation}

In hyperelasticity, $p_{|ab}=\sum_{\kappa} \mathcal{A}_{ab}^{\kappa} \,W_{\kappa }$, and therefore the condition $p_{|ab}=p_{|ba}$ is identical to $\mathcal{A}_{ab}^{\kappa}=\mathcal{A}_{ba}^{\kappa}$. It was found that there are $78$ additional terms for monoclinic hyperelastic solids which are  \citep{Yavari2022Universal,YavariGoriely2021}
\begin{equation} \label{eq:universality terms-incom-mono-hyper-1} 
\begin{split}
&\mathcal{A}_{ab}^{8}={\mathfrak{q}^n_a}_{|nb} \,,\\
&\mathcal{A}_{ab}^{18}= {\mathfrak{q}^n_a}_{|n} \,I_{1,b}+\left(\mathfrak{q}^n_a \,I_{1,n}\right)_{|b}+ {b^n_a}_{|n} \,I_{8,b}+\left(b^n_a \,I_{8,n}\right)_{|b} \,,\\
&\mathcal{A}_{ab}^{19}={b^n_a}_{|n} \,I_{9,b}+\left(b^n_a \,I_{9,n}\right)_{|b} \,,\\
&\mathcal{A}_{ab}^{28}= {\mathfrak{q}^n_a}_{|n} \,I_{2,b}+\left(\mathfrak{q}^n_a \,I_{2,n}\right)_{|b}- {c^n_a}_{|n} \,I_{8,b}-\left(c^n_a \,I_{8,n}\right)_{|b} \,,\\
&\mathcal{A}_{ab}^{29}=-{c^n_a}_{|n} \,I_{9,b}-\left(c^n_a \,I_{9,n}\right)_{|b} \,,\\
&\mathcal{A}_{ab}^{48}= {\mathfrak{q}^n_a}_{|n} \,I_{4,b}+\left(\mathfrak{q}^n_a \,I_{4,n}\right)_{|b}+ \left(n_a \,n^n\right)_{|n} \,I_{8,b}+\left(n_a \,n^n \,I_{8,n}\right)_{|b} \,,\\
&\mathcal{A}_{ab}^{49}=\left(n_a \,n^n\right)_{|n} \,I_{9,b}+\left(n_a \,n^n \,I_{9,n}\right)_{|b} \,,\\
&\mathcal{A}_{ab}^{58}= {\mathfrak{q}^n_a}_{|n} \,I_{5,b}+\left(\mathfrak{q}^n_a \,I_{5,n}\right)_{|b}+ {\mathfrak{L}^n_a}_{|n} \,I_{8,b}+\left(\mathfrak{L}^n_a \,I_{8,n}\right)_{|b} \,,\\
&\mathcal{A}_{ab}^{59}={\mathfrak{L}^n_a}_{|n} \,I_{9,b}+\left(\mathfrak{L}^n_a \,I_{9,n}\right)_{|b} \,,\\
&\mathcal{A}_{ab}^{68}= {\mathfrak{q}^n_a}_{|n} \,I_{6,b}+\left(\mathfrak{q}^n_a \,I_{6,n}\right)_{|b}+ \left(m_a \,m^n\right)_{|n} \,I_{8,b}+\left(m_a \,m^n \,I_{8,n}\right)_{|b} \,,\\
&\mathcal{A}_{ab}^{69}=\left(m_a \,m^n\right)_{|n} \,I_{9,b}+\left(m_a \,m^n \,I_{9,n}\right)_{|b} \,,\\
&\mathcal{A}_{ab}^{78}= {\mathfrak{q}^n_a}_{|n} \,I_{7,b}+\left(\mathfrak{q}^n_a \,I_{7,n}\right)_{|b}+ {\mathfrak{K}^n_a}_{|n} \,I_{8,b}+\left(\mathfrak{K}^n_a \,I_{8,n}\right)_{|b} \,,\\
&\mathcal{A}_{ab}^{79}={\mathfrak{K}^n_a}_{|n} \,I_{9,b}+\left(\mathfrak{K}^n_a \,I_{9,n}\right)_{|b} \,,\\
&\mathcal{A}_{ab}^{88}= {\mathfrak{q}^n_a}_{|n} \,I_{8,b}+\left(\mathfrak{q}^n_a \,I_{8,n}\right)_{|b} \,,\\
&\mathcal{A}_{ab}^{89}= {\mathfrak{q}^n_a}_{|n} \,I_{9,b}+\left(\mathfrak{q}^n_a \,I_{9,n}\right)_{|b} \,,
\end{split}
\end{equation}
and
\begin{equation} \label{eq:universality terms-incom-mono-hyper-2} 
\begin{split}
&\mathcal{A}_{ab}^{118}=b_a^n \,(I_{1,n} \,I_{8,b}+I_{1,b} \,I_{8,n})\,,\\
&\mathcal{A}_{ab}^{119}=b_a^n \,(I_{1,n} \,I_{9,b}+I_{1,b} \,I_{9,n})\,,\\
&\mathcal{A}_{ab}^{128}=b_a^n \,(I_{2,n} \,I_{8,b}+I_{2,b} \,I_{8,n})-c_a^n \,(I_{1,n} \,I_{8,b}+I_{1,b} \,I_{8,n})\,,\\
&\mathcal{A}_{ab}^{129}=b_a^n \,(I_{2,n} \,I_{9,b}+I_{2,b} \,I_{9,n})-c_a^n \,(I_{1,n} \,I_{9,b}+I_{1,b} \,I_{9,n})\,,\\
&\mathcal{A}_{ab}^{148}=b_a^n \,(I_{4,n} \,I_{8,b}+I_{4,b} \,I_{8,n})+ n_a \,n^n \,(I_{1,n} \,I_{8,b}+I_{1,b} \,I_{8,n})\,,\\
&\mathcal{A}_{ab}^{149}=b_a^n \,(I_{4,n} \,I_{9,b}+I_{4,b} \,I_{9,n})+ n_a \,n^n \,(I_{1,n} \,I_{9,b}+I_{1,b} \,I_{9,n})\,,\\
&\mathcal{A}_{ab}^{158}=b_a^n \,(I_{5,n} \,I_{8,b}+I_{5,b} \,I_{8,n})+ \mathfrak{L}^n_a \,(I_{1,n} \,I_{8,b}+I_{1,b} \,I_{8,n})\,,\\
&\mathcal{A}_{ab}^{159}=b_a^n \,(I_{5,n} \,I_{9,b}+I_{5,b} \,I_{9,n})+ \mathfrak{L}^n_a \,(I_{1,n} \,I_{9,b}+I_{1,b} \,I_{9,n})\,,\\
&\mathcal{A}_{ab}^{168}=b_a^n \,(I_{6,n} \,I_{8,b}+I_{6,b} \,I_{8,n})+ m_a \,m^n \,(I_{1,n} \,I_{8,b}+I_{1,b} \,I_{8,n})\,,\\
&\mathcal{A}_{ab}^{169}=b_a^n \,(I_{6,n} \,I_{9,b}+I_{6,b} \,I_{9,n})+ m_a \,m^n \,(I_{1,n} \,I_{9,b}+I_{1,b} \,I_{9,n})\,,\\
&\mathcal{A}_{ab}^{178}=b_a^n \,(I_{7,n} \,I_{8,b}+I_{7,b} \,I_{8,n})+ \mathfrak{K}^n_a \,(I_{1,n} \,I_{8,b}+I_{1,b} \,I_{8,n})\,,\\
&\mathcal{A}_{ab}^{179}=b_a^n \,(I_{7,n} \,I_{9,b}+I_{7,b} \,I_{9,n})+ \mathfrak{K}^n_a \,(I_{1,n} \,I_{9,b}+I_{1,b} \,I_{9,n})\,,\\
&\mathcal{A}_{ab}^{188}=b_a^n \,I_{8,n} \,I_{8,b}+\mathfrak{q}^n_a \,(I_{1,n} \,I_{8,b}+I_{1,b} \,I_{8,n})\,,\\
&\mathcal{A}_{ab}^{189}=b_a^n \,(I_{8,n} \,I_{9,b}+I_{8,b} \,I_{9,n})+\mathfrak{q}^n_a \,(I_{1,n} \,I_{9,b}+I_{1,b} \,I_{9,n})\,,\\
&\mathcal{A}_{ab}^{199}=b_a^n \,I_{9,n} \,I_{9,b}\,,\\
\end{split}
\end{equation}
and
\begin{equation} \label{eq:universality terms-incom-mono-hyper-3} 
\begin{split}
&\mathcal{A}_{ab}^{228}=-c_a^n \,(I_{2,n} \,I_{8,b}+I_{2,b} \,I_{8,n})\,,\\
&\mathcal{A}_{ab}^{229}=-c_a^n \,(I_{2,n} \,I_{9,b}+I_{2,b} \,I_{9,n})\,,\\
&\mathcal{A}_{ab}^{248}=-c_a^n \,(I_{4,n} \,I_{8,b}+I_{4,b} \,I_{8,n})+n_a \,n^n\,(I_{2,n} \,I_{8,b}+I_{2,b} \,I_{8,n})\,,\\
&\mathcal{A}_{ab}^{249}=-c_a^n \,(I_{4,n} \,I_{9,b}+I_{4,b} \,I_{9,n})+n_a \,n^n\,(I_{2,n} \,I_{9,b}+I_{2,b} \,I_{9,n})\,,\\
&\mathcal{A}_{ab}^{258}=-c_a^n \,(I_{5,n} \,I_{8,b}+I_{5,b} \,I_{8,n})+\mathfrak{L}^n_a\,(I_{2,n} \,I_{8,b}+I_{2,b} \,I_{8,n})\,,\\
&\mathcal{A}_{ab}^{259}=-c_a^n \,(I_{5,n} \,I_{9,b}+I_{5,b} \,I_{9,n})+\mathfrak{L}^n_a \,(I_{2,n} \,I_{9,b}+I_{2,b} \,I_{9,n})\,,\\
&\mathcal{A}_{ab}^{268}=-c_a^n \,(I_{6,n} \,I_{8,b}+I_{6,b} \,I_{8,n})+m_a \,m^n \,(I_{2,n} \,I_{8,b}+I_{2,b} \,I_{8,n})\,,\\
&\mathcal{A}_{ab}^{269}=-c_a^n \,(I_{6,n} \,I_{9,b}+I_{6,b} \,I_{9,n})+m_a \,m^n \,(I_{2,n} \,I_{9,b}+I_{2,b} \,I_{9,n})\,,\\
&\mathcal{A}_{ab}^{278}=-c_a^n \,(I_{7,n} \,I_{8,b}+I_{7,b} \,I_{8,n})+\mathfrak{K}^n_a \,(I_{2,n} \,I_{8,b}+I_{2,b} \,I_{8,n})\,,\\
&\mathcal{A}_{ab}^{279}=-c_a^n \,(I_{7,n} \,I_{9,b}+I_{7,b} \,I_{9,n})+\mathfrak{K}^n_a \,(I_{2,n} \,I_{9,b}+I_{2,b} \,I_{9,n})\,,\\
&\mathcal{A}_{ab}^{288}=-c_a^n \,I_{8,n} \,I_{8,b}+\mathfrak{q}^n_a \,(I_{2,n} \,I_{8,b}+I_{2,b} \,I_{8,n})\,,\\
&\mathcal{A}_{ab}^{289}=-c_a^n \,(I_{8,n} \,I_{9,b}+I_{8,b} \,I_{9,n})+\mathfrak{q}^n_a \,(I_{2,n} \,I_{9,b}+I_{2,b} \,I_{9,n})\,,\\
&\mathcal{A}_{ab}^{299}=-c_a^n \,I_{9,n} \,I_{9,b}\,,\\
&\mathcal{A}_{ab}^{448}=n_a \,n^n \,(I_{4,n} \,I_{8,b} +I_{4,b} \,I_{8,n})\,,\\
&\mathcal{A}_{ab}^{449}=n_a \,n^n \,(I_{4,n} \,I_{9,b} +I_{4,b} \,I_{9,n})\,,\\
\end{split}
\end{equation}
and
\begin{equation} \label{eq:universality terms-incom-mono-hyper-4} 
\begin{split}
&\mathcal{A}_{ab}^{458}=n_a \,n^n \,(I_{5,n} \,I_{8,b} +I_{5,b} \,I_{8,n})+\mathfrak{L}^n_a \,(I_{4,n} \,I_{8,b} +I_{4,b} \,I_{8,n})\,,\\
&\mathcal{A}_{ab}^{459}=n_a \,n^n \,(I_{5,n} \,I_{9,b} +I_{5,b} \,I_{9,n})+\mathfrak{L}^n_a \,(I_{4,n} \,I_{9,b} +I_{4,b} \,I_{9,n})\,,\\
&\mathcal{A}_{ab}^{468}=n_a \,n^n \,(I_{6,n} \,I_{8,b} +I_{6,b} \,I_{8,n})+m_a \,m^n \,(I_{4,n} \,I_{8,b} +I_{4,b} \,I_{8,n})\,,\\
&\mathcal{A}_{ab}^{469}=n_a \,n^n \,(I_{6,n} \,I_{9,b} +I_{6,b} \,I_{9,n})+m_a \,m^n \,(I_{4,n} \,I_{9,b} +I_{4,b} \,I_{9,n})\,,\\
&\mathcal{A}_{ab}^{478}=n_a \,n^n \,(I_{7,n} \,I_{8,b} +I_{7,b} \,I_{8,n})+\mathfrak{K}^n_a \,(I_{4,n} \,I_{8,b} +I_{4,b} \,I_{8,n})\,,\\
&\mathcal{A}_{ab}^{479}=n_a \,n^n \,(I_{7,n} \,I_{9,b} +I_{7,b} \,I_{9,n})+\mathfrak{K}^n_a \,(I_{4,n} \,I_{9,b} +I_{4,b} \,I_{9,n})\,,\\
&\mathcal{A}_{ab}^{488}=n_a \,n^n \,I_{8,n} \,I_{8,b}+\mathfrak{q}^n_a \,(I_{4,n} \,I_{8,b}+I_{4,b} \,I_{8,n})\,,\\
&\mathcal{A}_{ab}^{489}=n_a \,n^n \,(I_{8,n} \,I_{9,b}+I_{8,b} \,I_{9,n})+\mathfrak{q}^n_a \,(I_{4,n} \,I_{9,b}+I_{4,b} \,I_{9,n})\,,\\
&\mathcal{A}_{ab}^{499}=n_a \,n^n \,I_{9,n} \,I_{9,b}\,,\\
&\mathcal{A}_{ab}^{558}=\mathfrak{L}^n_a \,(I_{5,n} \,I_{8,b} +I_{5,b} \,I_{8,n})\,,\\
&\mathcal{A}_{ab}^{559}=\mathfrak{L}^n_a \,(I_{5,n} \,I_{9,b} +I_{5,b} \,I_{9,n})\,,\\
&\mathcal{A}_{ab}^{568}=\mathfrak{L}^n_a \,(I_{6,n} \,I_{8,b} +I_{6,b} \,I_{8,n}) + m_a \,m^n \,(I_{5,n} \,I_{8,b} +I_{5,b} \,I_{8,n})\,,\\
&\mathcal{A}_{ab}^{569}=\mathfrak{L}^n_a \,(I_{6,n} \,I_{9,b} +I_{6,b} \,I_{9,n}) + m_a \,m^n \,(I_{5,n} \,I_{9,b} +I_{5,b} \,I_{9,n})\,,\\
&\mathcal{A}_{ab}^{578}=\mathfrak{L}^n_a \,(I_{7,n} \,I_{8,b} +I_{7,b} \,I_{8,n})+\mathfrak{K}^n_a \,(I_{5,n} \,I_{8,b} +I_{5,b} \,I_{8,n})\,,\\
&\mathcal{A}_{ab}^{579}=\mathfrak{L}^n_a \,(I_{7,n} \,I_{9,b} +I_{7,b} \,I_{9,n})+\mathfrak{K}^n_a \,(I_{5,n} \,I_{9,b} +I_{5,b} \,I_{9,n})\,,\\
\end{split}
\end{equation}
and
\begin{equation} \label{eq:universality terms-incom-mono-hyper-5} 
\begin{split}
&\mathcal{A}_{ab}^{588}=\mathfrak{L}^n_a \,I_{8,n} \,I_{8,b}+\mathfrak{q}^n_a \,(I_{5,n} \,I_{8,b}+I_{5,b} \,I_{8,n}) \,,\\
&\mathcal{A}_{ab}^{589}=\mathfrak{L}^n_a \,(I_{8,n} \,I_{9,b}+I_{8,b} \,I_{9,n})+\mathfrak{q}^n_a \,(I_{5,n} \,I_{9,b}+I_{5,b} \,I_{9,n}) \,,\\
&\mathcal{A}_{ab}^{599}=\mathfrak{L}^n_a \,I_{9,n} \,I_{9,b} \,,\\
&\mathcal{A}_{ab}^{668}=m_a \,m^n \,(I_{6,n} \,I_{8,b} +I_{6,b} \,I_{8,n})\,,\\
&\mathcal{A}_{ab}^{669}=m_a \,m^n \,(I_{6,n} \,I_{9,b} +I_{6,b} \,I_{9,n})\,,\\
&\mathcal{A}_{ab}^{678}=m_a \,m^n \,(I_{7,n} \,I_{8,b} +I_{7,b} \,I_{8,n}) +\mathfrak{K}^n_a \,(I_{6,n} \,I_{8,b}+I_{6,b} \,I_{8,n}) \,,\\
&\mathcal{A}_{ab}^{679}=m_a \,m^n \,(I_{7,n} \,I_{9,b} +I_{7,b} \,I_{9,n}) +\mathfrak{K}^n_a \,(I_{6,n} \,I_{9,b}+I_{6,b} \,I_{9,n}) \,,\\
&\mathcal{A}_{ab}^{688}=m_a \,m^n \,I_{8,n} \,I_{8,b}+\mathfrak{q}^n_a \,(I_{6,n} \,I_{8,b}+I_{6,b} \,I_{8,n}) \,,\\
&\mathcal{A}_{ab}^{689}=m_a \,m^n \,(I_{8,n} \,I_{9,b}+I_{8,b} \,I_{9,n})+\mathfrak{q}^n_a \,(I_{6,n} \,I_{9,b}+I_{6,b} \,I_{9,n}) \,,\\
&\mathcal{A}_{ab}^{699}=m_a \,m^n \,I_{9,n} \,I_{9,b} \,,\\
&\mathcal{A}_{ab}^{778}=\mathfrak{K}^n_a \,(I_{7,n} \,I_{8,b}+I_{7,b} \,I_{8,n}) \,,\\
&\mathcal{A}_{ab}^{779}=\mathfrak{K}^n_a \,(I_{7,n} \,I_{9,b}+I_{7,b} \,I_{9,n}) \,,\\
&\mathcal{A}_{ab}^{788}=\mathfrak{K}^n_a \,I_{8,n} \,I_{8,b}+\mathfrak{q}^n_a \,(I_{7,n} \,I_{8,b}+I_{7,b} \,I_{8,n}) \,,\\
&\mathcal{A}_{ab}^{789}=\mathfrak{K}^n_a \,(I_{8,n} \,I_{9,b}+I_{8,b} \,I_{9,n})+\mathfrak{q}^n_a \,(I_{7,n} \,I_{9,b}+I_{7,b} \,I_{9,n}) \,,\\
&\mathcal{A}_{ab}^{799}=\mathfrak{K}^n_a \,I_{9,n} \,I_{9,b} \,,\\
&\mathcal{A}_{ab}^{888}=\mathfrak{q}^n_a \,I_{8,n} \,I_{8,b} \,,\\
&\mathcal{A}_{ab}^{889}=\mathfrak{q}^n_a \,(I_{8,n} \,I_{9,b}+I_{8,b} \,I_{9,n}) \,,\\
&\mathcal{A}_{ab}^{999}=\mathfrak{q}^n_a \,I_{9,n} \,I_{9,b} \,,
\end{split}
\end{equation}
where $\mathbf{n}=\mathbf{n}_1$, $\mathbf{m}=\mathbf{n}_2$, $\mathfrak{L}^{ab}=\ell_{1}^{ab}$, $\mathfrak{K}^{ab}=\ell_{2}^{ab}$ and $\mathfrak{q}^{ab}=\ell_{3}^{ab}$.

Taking \eqref{eq:pa-mono-Cauchy elast} into account, one may determine $p_{|ab}$ in Cauchy elasticity as
\begin{equation} \label{eq:p_{|ab}-mono-Cauchy elastic}
\begin{aligned}
p_{|ab} &=\sum_{\kappa}\Big({\mathcal{B}_{ab}^{1\kappa} \,\alpha_{1\kappa}}+{\mathcal{B}_{ab}^{2\kappa} \,\alpha_{2\kappa}}+{\mathcal{B}_{ab}^{4\kappa} \,\alpha_{4\kappa}}+{\mathcal{B}_{ab}^{5\kappa} \,\alpha_{5\kappa}}+{\mathcal{B}_{ab}^{6\kappa} \,\alpha_{6\kappa}} \\
&\qquad\quad +{\mathcal{B}_{ab}^{7\kappa} \,\alpha_{7\kappa}}+{\mathcal{B}_{ab}^{8\kappa} \,\alpha_{8\kappa}}+{\mathcal{B}_{ab}^{9\kappa} \,\alpha_{9\kappa}}+{\mathcal{B}_{ab}^{10 \,\kappa} \,\alpha_{10 \,\kappa}}\Big)\,,
\end{aligned}
\end{equation}
where $\mathcal{B}_{ab}^{1\kappa}$, $\mathcal{B}_{ab}^{2\kappa}$, $\mathcal{B}_{ab}^{4\kappa}$, $\mathcal{B}_{ab}^{5\kappa}$, $\mathcal{B}_{ab}^{6\kappa}$, $\mathcal{B}_{ab}^{7\kappa}$, $\mathcal{B}_{ab}^{8\kappa}$, $\mathcal{B}_{ab}^{9\kappa}$ and $\mathcal{B}_{ab}^{10 \,\kappa}$ are the matrices of coefficients of $\alpha_{1\kappa}$, $\alpha_{2\kappa}$, $\alpha_{4\kappa}$, $\alpha_{5\kappa}$, $\alpha_{6\kappa}$, $\alpha_{7\kappa}$, $\alpha_{8\kappa}$, $\alpha_{9\kappa}$ and $\alpha_{10 \,\kappa}$, respectively ($\alpha_{i\kappa} = \partial \alpha_i/\partial I_\kappa$, where $\kappa$ is a multi-index). The extra terms in Cauchy elasticity are found to be
\begin{equation} \label{eq:universality terms-incom-mono-Cauchy elast-1} 
\begin{split}
&\mathcal{B}_{ab}^{10}={\mathfrak{q}^n_a}_{|nb} \,,\\
&\mathcal{B}_{ab}^{18}={b^n_a}_{|n} \,I_{8,b}+\left(b^n_a \,I_{8,n}\right)_{|b} \,,\\
&\mathcal{B}_{ab}^{19}={b^n_a}_{|n} \,I_{9,b}+\left(b^n_a \,I_{9,n}\right)_{|b} \,,\\
&\mathcal{B}_{ab}^{28}= -{c^n_a}_{|n} \,I_{8,b}-\left(c^n_a \,I_{8,n}\right)_{|b} \,,\\
&\mathcal{B}_{ab}^{29}=-{c^n_a}_{|n} \,I_{9,b}-\left(c^n_a \,I_{9,n}\right)_{|b} \,,\\
&\mathcal{B}_{ab}^{48}=\left(n_a \,n^n\right)_{|n} \,I_{8,b}+\left(n_a \,n^n \,I_{8,n}\right)_{|b} \,,\\
&\mathcal{B}_{ab}^{49}=\left(n_a \,n^n\right)_{|n} \,I_{9,b}+\left(n_a \,n^n \,I_{9,n}\right)_{|b} \,,\\
&\mathcal{B}_{ab}^{58}={\mathfrak{L}^n_a}_{|n} \,I_{8,b}+\left(\mathfrak{L}^n_a \,I_{8,n}\right)_{|b} \,,\\
&\mathcal{B}_{ab}^{59}={\mathfrak{L}^n_a}_{|n} \,I_{9,b}+\left(\mathfrak{L}^n_a \,I_{9,n}\right)_{|b} \,,\\
&\mathcal{B}_{ab}^{68}={\bar{\mathfrak{L}}^n_a}_{|n} \,I_{8,b}+\left(\bar{\mathfrak{L}}^n_a \,I_{8,n}\right)_{|b} \,,\\
&\mathcal{B}_{ab}^{69}={\bar{\mathfrak{L}}^n_a}_{|n} \,I_{9,b}+\left(\bar{\mathfrak{L}}^n_a \,I_{9,n}\right)_{|b} \,,\\
&\mathcal{B}_{ab}^{78}=\left(m_a \,m^n\right)_{|n} \,I_{8,b}+\left(m_a \,m^n \,I_{8,n}\right)_{|b} \,,\\
&\mathcal{B}_{ab}^{79}=\left(m_a \,m^n\right)_{|n} \,I_{9,b}+\left(m_a \,m^n \,I_{9,n}\right)_{|b} \,,\\
&\mathcal{B}_{ab}^{88}={\mathfrak{K}^n_a}_{|n} \,I_{8,b}+\left(\mathfrak{K}^n_a \,I_{8,n}\right)_{|b} \,,\\
&\mathcal{B}_{ab}^{89}={\mathfrak{K}^n_a}_{|n} \,I_{9,b}+\left(\mathfrak{K}^n_a \,I_{9,n}\right)_{|b} \,,\\
&\mathcal{B}_{ab}^{98}={\bar{\mathfrak{K}}^n_a}_{|n} \,I_{8,b}+\left(\bar{\mathfrak{K}}^n_a \,I_{8,n}\right)_{|b} \,,\\
&\mathcal{B}_{ab}^{99}={\bar{\mathfrak{K}}^n_a}_{|n} \,I_{9,b}+\left(\bar{\mathfrak{K}}^n_a \,I_{9,n}\right)_{|b} \,,\\
&\mathcal{B}_{ab}^{10 \,1}={\mathfrak{q}^n_a}_{|n} \,I_{1,b}+\left(\mathfrak{q}^n_a \,I_{1,n}\right)_{|b} \,,\\
&\mathcal{B}_{ab}^{10 \,2}={\mathfrak{q}^n_a}_{|n} \,I_{2,b}+\left(\mathfrak{q}^n_a \,I_{2,n}\right)_{|b} \,,\\
&\mathcal{B}_{ab}^{10 \,4}={\mathfrak{q}^n_a}_{|n} \,I_{4,b}+\left(\mathfrak{q}^n_a \,I_{4,n}\right)_{|b} \,,\\
&\mathcal{B}_{ab}^{10 \,5}={\mathfrak{q}^n_a}_{|n} \,I_{5,b}+\left(\mathfrak{q}^n_a \,I_{5,n}\right)_{|b}\,,\\
&\mathcal{B}_{ab}^{10 \,6}={\mathfrak{q}^n_a}_{|n} \,I_{6,b}+\left(\mathfrak{q}^n_a \,I_{6,n}\right)_{|b} \,,\\
&\mathcal{B}_{ab}^{10 \,7}={\mathfrak{q}^n_a}_{|n} \,I_{7,b}+\left(\mathfrak{q}^n_a \,I_{7,n}\right)_{|b} \,,\\
&\mathcal{B}_{ab}^{10 \,8}= {\mathfrak{q}^n_a}_{|n} \,I_{8,b}+\left(\mathfrak{q}^n_a \,I_{8,n}\right)_{|b} \,,\\
&\mathcal{B}_{ab}^{10 \,9}= {\mathfrak{q}^n_a}_{|n} \,I_{9,b}+\left(\mathfrak{q}^n_a \,I_{9,n}\right)_{|b} \,,
\end{split}
\end{equation}
and
\begin{equation} \label{eq:universality terms-incom-mono-Cauchy elast-2} 
\begin{split}
&\mathcal{B}_{ab}^{118}=b_a^n \,(I_{1,n} \,I_{8,b}+I_{1,b} \,I_{8,n})\,,\\
&\mathcal{B}_{ab}^{119}=b_a^n \,(I_{1,n} \,I_{9,b}+I_{1,b} \,I_{9,n})\,,\\
&\mathcal{B}_{ab}^{128}=b_a^n \,(I_{2,n} \,I_{8,b}+I_{2,b} \,I_{8,n})\,,\\
&\mathcal{B}_{ab}^{129}=b_a^n \,(I_{2,n} \,I_{9,b}+I_{2,b} \,I_{9,n})\,,\\
&\mathcal{B}_{ab}^{148}=b_a^n \,(I_{4,n} \,I_{8,b}+I_{4,b} \,I_{8,n})\,,\\
&\mathcal{B}_{ab}^{149}=b_a^n \,(I_{4,n} \,I_{9,b}+I_{4,b} \,I_{9,n})\,,\\
&\mathcal{B}_{ab}^{158}=b_a^n \,(I_{5,n} \,I_{8,b}+I_{5,b} \,I_{8,n})\,,\\
&\mathcal{B}_{ab}^{159}=b_a^n \,(I_{5,n} \,I_{9,b}+I_{5,b} \,I_{9,n})\,,\\
&\mathcal{B}_{ab}^{168}=b_a^n \,(I_{6,n} \,I_{8,b}+I_{6,b} \,I_{8,n})\,,\\
&\mathcal{B}_{ab}^{169}=b_a^n \,(I_{6,n} \,I_{9,b}+I_{6,b} \,I_{9,n})\,,\\
&\mathcal{B}_{ab}^{178}=b_a^n \,(I_{7,n} \,I_{8,b}+I_{7,b} \,I_{8,n})\,,\\
&\mathcal{B}_{ab}^{179}=b_a^n \,(I_{7,n} \,I_{9,b}+I_{7,b} \,I_{9,n})\,,\\
&\mathcal{B}_{ab}^{188}=b_a^n \,I_{8,n} \,I_{8,b}\,,\\
&\mathcal{B}_{ab}^{189}=b_a^n \,(I_{8,n} \,I_{9,b}+I_{8,b} \,I_{9,n})\,,\\
&\mathcal{B}_{ab}^{199}=b_a^n \,I_{9,n} \,I_{9,b}\,,\\
\end{split}
\end{equation}
and
\begin{equation} \label{eq:universality terms-incom-mono-Cauchy elast-3} 
\begin{split}
&\mathcal{B}_{ab}^{218}=-c_a^n \,(I_{1,n} \,I_{8,b}+I_{1,b} \,I_{8,n})\,,\\
&\mathcal{B}_{ab}^{219}=-c_a^n \,(I_{1,n} \,I_{9,b}+I_{1,b} \,I_{9,n})\,,\\
&\mathcal{B}_{ab}^{228}=-c_a^n \,(I_{2,n} \,I_{8,b}+I_{2,b} \,I_{8,n})\,,\\
&\mathcal{B}_{ab}^{229}=-c_a^n \,(I_{2,n} \,I_{9,b}+I_{2,b} \,I_{9,n})\,,\\
&\mathcal{B}_{ab}^{248}=-c_a^n \,(I_{4,n} \,I_{8,b}+I_{4,b} \,I_{8,n})\,,\\
&\mathcal{B}_{ab}^{249}=-c_a^n \,(I_{4,n} \,I_{9,b}+I_{4,b} \,I_{9,n})\,,\\
&\mathcal{B}_{ab}^{258}=-c_a^n \,(I_{5,n} \,I_{8,b}+I_{5,b} \,I_{8,n})\,,\\
&\mathcal{B}_{ab}^{259}=-c_a^n \,(I_{5,n} \,I_{9,b}+I_{5,b} \,I_{9,n})\,,\\
&\mathcal{B}_{ab}^{268}=-c_a^n \,(I_{6,n} \,I_{8,b}+I_{6,b} \,I_{8,n})\,,\\
&\mathcal{B}_{ab}^{269}=-c_a^n \,(I_{6,n} \,I_{9,b}+I_{6,b} \,I_{9,n})\,,\\
&\mathcal{B}_{ab}^{278}=-c_a^n \,(I_{7,n} \,I_{8,b}+I_{7,b} \,I_{8,n})\,,\\
&\mathcal{B}_{ab}^{279}=-c_a^n \,(I_{7,n} \,I_{9,b}+I_{7,b} \,I_{9,n})\,,\\
&\mathcal{B}_{ab}^{288}=-c_a^n \,I_{8,n} \,I_{8,b}\,,\\
&\mathcal{B}_{ab}^{289}=-c_a^n \,(I_{8,n} \,I_{9,b}+I_{8,b} \,I_{9,n})\,,\\
&\mathcal{B}_{ab}^{299}=-c_a^n \,I_{9,n} \,I_{9,b}\,,\\
\end{split}
\end{equation}
and
\begin{equation} \label{eq:universality terms-incom-mono-Cauchy elast-4} 
\begin{split}
&\mathcal{B}_{ab}^{418}=n_a \,n^n \,(I_{1,n} \,I_{8,b}+I_{1,b} \,I_{8,n})\,,\\
&\mathcal{B}_{ab}^{419}=n_a \,n^n \,(I_{1,n} \,I_{9,b}+I_{1,b} \,I_{9,n})\,,\\
&\mathcal{B}_{ab}^{428}=n_a \,n^n \,(I_{2,n} \,I_{8,b}+I_{2,b} \,I_{8,n})\,,\\
&\mathcal{B}_{ab}^{429}=n_a \,n^n \,(I_{2,n} \,I_{9,b}+I_{2,b} \,I_{9,n})\,,\\
&\mathcal{B}_{ab}^{448}=n_a \,n^n \,(I_{4,n} \,I_{8,b}+I_{4,b} \,I_{8,n})\,,\\
&\mathcal{B}_{ab}^{449}=n_a \,n^n \,(I_{4,n} \,I_{9,b}+I_{4,b} \,I_{9,n})\,,\\
&\mathcal{B}_{ab}^{458}=n_a \,n^n \,(I_{5,n} \,I_{8,b}+I_{5,b} \,I_{8,n})\,,\\
&\mathcal{B}_{ab}^{459}=n_a \,n^n \,(I_{5,n} \,I_{9,b}+I_{5,b} \,I_{9,n})\,,\\
&\mathcal{B}_{ab}^{468}=n_a \,n^n \,(I_{6,n} \,I_{8,b}+I_{6,b} \,I_{8,n})\,,\\
&\mathcal{B}_{ab}^{469}=n_a \,n^n \,(I_{6,n} \,I_{9,b}+I_{6,b} \,I_{9,n})\,,\\
&\mathcal{B}_{ab}^{478}=n_a \,n^n \,(I_{7,n} \,I_{8,b}+I_{7,b} \,I_{8,n})\,,\\
&\mathcal{B}_{ab}^{479}=n_a \,n^n \,(I_{7,n} \,I_{9,b}+I_{7,b} \,I_{9,n})\,,\\
&\mathcal{B}_{ab}^{488}=n_a \,n^n \,I_{8,n} \,I_{8,b}\,,\\
&\mathcal{B}_{ab}^{489}=n_a \,n^n \,(I_{8,n} \,I_{9,b}+I_{8,b} \,I_{9,n})\,,\\
&\mathcal{B}_{ab}^{499}=n_a \,n^n \,I_{9,n} \,I_{9,b}\,,\\
\end{split}
\end{equation}
and
\begin{equation} \label{eq:universality terms-incom-mono-Cauchy elast-5} 
\begin{split}
&\mathcal{B}_{ab}^{518}=\mathfrak{L}^n_a \,(I_{1,n} \,I_{8,b}+I_{1,b} \,I_{8,n})\,,\\
&\mathcal{B}_{ab}^{519}=\mathfrak{L}^n_a \,(I_{1,n} \,I_{9,b}+I_{1,b} \,I_{9,n})\,,\\
&\mathcal{B}_{ab}^{528}=\mathfrak{L}^n_a \,(I_{2,n} \,I_{8,b}+I_{2,b} \,I_{8,n})\,,\\
&\mathcal{B}_{ab}^{529}=\mathfrak{L}^n_a \,(I_{2,n} \,I_{9,b}+I_{2,b} \,I_{9,n})\,,\\
&\mathcal{B}_{ab}^{548}=\mathfrak{L}^n_a \,(I_{4,n} \,I_{8,b}+I_{4,b} \,I_{8,n})\,,\\
&\mathcal{B}_{ab}^{549}=\mathfrak{L}^n_a \,(I_{4,n} \,I_{9,b}+I_{4,b} \,I_{9,n})\,,\\
&\mathcal{B}_{ab}^{558}=\mathfrak{L}^n_a \,(I_{5,n} \,I_{8,b}+I_{5,b} \,I_{8,n})\,,\\
&\mathcal{B}_{ab}^{559}=\mathfrak{L}^n_a \,(I_{5,n} \,I_{9,b}+I_{5,b} \,I_{9,n})\,,\\
&\mathcal{B}_{ab}^{568}=\mathfrak{L}^n_a \,(I_{6,n} \,I_{8,b}+I_{6,b} \,I_{8,n})\,,\\
&\mathcal{B}_{ab}^{569}=\mathfrak{L}^n_a \,(I_{6,n} \,I_{9,b}+I_{6,b} \,I_{9,n})\,,\\
&\mathcal{B}_{ab}^{578}=\mathfrak{L}^n_a \,(I_{7,n} \,I_{8,b}+I_{7,b} \,I_{8,n})\,,\\
&\mathcal{B}_{ab}^{579}=\mathfrak{L}^n_a \,(I_{7,n} \,I_{9,b}+I_{7,b} \,I_{9,n})\,,\\
&\mathcal{B}_{ab}^{588}=\mathfrak{L}^n_a \,I_{8,n} \,I_{8,b}\,,\\
&\mathcal{B}_{ab}^{589}=\mathfrak{L}^n_a \,(I_{8,n} \,I_{9,b}+I_{8,b} \,I_{9,n})\,,\\
&\mathcal{B}_{ab}^{599}=\mathfrak{L}^n_a \,I_{9,n} \,I_{9,b}\,,\\
\end{split}
\end{equation}
and
\begin{equation} \label{eq:universality terms-incom-mono-Cauchy elast-6} 
\begin{split}
&\mathcal{B}_{ab}^{618}=\bar{\mathfrak{L}}^n_a \,(I_{1,n} \,I_{8,b}+I_{1,b} \,I_{8,n})\,,\\
&\mathcal{B}_{ab}^{619}=\bar{\mathfrak{L}}^n_a \,(I_{1,n} \,I_{9,b}+I_{1,b} \,I_{9,n})\,,\\
&\mathcal{B}_{ab}^{628}=\bar{\mathfrak{L}}^n_a \,(I_{2,n} \,I_{8,b}+I_{2,b} \,I_{8,n})\,,\\
&\mathcal{B}_{ab}^{629}=\bar{\mathfrak{L}}^n_a \,(I_{2,n} \,I_{9,b}+I_{2,b} \,I_{9,n})\,,\\
&\mathcal{B}_{ab}^{648}=\bar{\mathfrak{L}}^n_a \,(I_{4,n} \,I_{8,b}+I_{4,b} \,I_{8,n})\,,\\
&\mathcal{B}_{ab}^{649}=\bar{\mathfrak{L}}^n_a \,(I_{4,n} \,I_{9,b}+I_{4,b} \,I_{9,n})\,,\\
&\mathcal{B}_{ab}^{658}=\bar{\mathfrak{L}}^n_a \,(I_{5,n} \,I_{8,b}+I_{5,b} \,I_{8,n})\,,\\
&\mathcal{B}_{ab}^{659}=\bar{\mathfrak{L}}^n_a \,(I_{5,n} \,I_{9,b}+I_{5,b} \,I_{9,n})\,,\\
&\mathcal{B}_{ab}^{668}=\bar{\mathfrak{L}}^n_a \,(I_{6,n} \,I_{8,b}+I_{6,b} \,I_{8,n})\,,\\
&\mathcal{B}_{ab}^{669}=\bar{\mathfrak{L}}^n_a \,(I_{6,n} \,I_{9,b}+I_{6,b} \,I_{9,n})\,,\\
&\mathcal{B}_{ab}^{678}=\bar{\mathfrak{L}}^n_a \,(I_{7,n} \,I_{8,b}+I_{7,b} \,I_{8,n})\,,\\
&\mathcal{B}_{ab}^{679}=\bar{\mathfrak{L}}^n_a \,(I_{7,n} \,I_{9,b}+I_{7,b} \,I_{9,n})\,,\\
&\mathcal{B}_{ab}^{688}=\bar{\mathfrak{L}}^n_a \,I_{8,n} \,I_{8,b}\,,\\
&\mathcal{B}_{ab}^{689}=\bar{\mathfrak{L}}^n_a \,(I_{8,n} \,I_{9,b}+I_{8,b} \,I_{9,n})\,,\\
&\mathcal{B}_{ab}^{699}=\bar{\mathfrak{L}}^n_a \,I_{9,n} \,I_{9,b}\,,\\
\end{split}
\end{equation}
and
\begin{equation} \label{eq:universality terms-incom-mono-Cauchy elast-7} 
\begin{split}
&\mathcal{B}_{ab}^{718}=m_a \,m^n \,(I_{1,n} \,I_{8,b}+I_{1,b} \,I_{8,n})\,,\\
&\mathcal{B}_{ab}^{719}=m_a \,m^n \,(I_{1,n} \,I_{9,b}+I_{1,b} \,I_{9,n})\,,\\
&\mathcal{B}_{ab}^{728}=m_a \,m^n \,(I_{2,n} \,I_{8,b}+I_{2,b} \,I_{8,n})\,,\\
&\mathcal{B}_{ab}^{729}=m_a \,m^n \,(I_{2,n} \,I_{9,b}+I_{2,b} \,I_{9,n})\,,\\
&\mathcal{B}_{ab}^{748}=m_a \,m^n \,(I_{4,n} \,I_{8,b}+I_{4,b} \,I_{8,n})\,,\\
&\mathcal{B}_{ab}^{749}=m_a \,m^n \,(I_{4,n} \,I_{9,b}+I_{4,b} \,I_{9,n})\,,\\
&\mathcal{B}_{ab}^{758}=m_a \,m^n \,(I_{5,n} \,I_{8,b}+I_{5,b} \,I_{8,n})\,,\\
&\mathcal{B}_{ab}^{759}=m_a \,m^n \,(I_{5,n} \,I_{9,b}+I_{5,b} \,I_{9,n})\,,\\
&\mathcal{B}_{ab}^{768}=m_a \,m^n \,(I_{6,n} \,I_{8,b}+I_{6,b} \,I_{8,n})\,,\\
&\mathcal{B}_{ab}^{769}=m_a \,m^n \,(I_{6,n} \,I_{9,b}+I_{6,b} \,I_{9,n})\,,\\
&\mathcal{B}_{ab}^{778}=m_a \,m^n \,(I_{7,n} \,I_{8,b}+I_{7,b} \,I_{8,n})\,,\\
&\mathcal{B}_{ab}^{779}=m_a \,m^n \,(I_{7,n} \,I_{9,b}+I_{7,b} \,I_{9,n})\,,\\
&\mathcal{B}_{ab}^{788}=m_a \,m^n \,I_{8,n} \,I_{8,b}\,,\\
&\mathcal{B}_{ab}^{789}=m_a \,m^n \,(I_{8,n} \,I_{9,b}+I_{8,b} \,I_{9,n})\,,\\
&\mathcal{B}_{ab}^{799}=m_a \,m^n \,I_{9,n} \,I_{9,b}\,,\\
\end{split}
\end{equation}
and
\begin{equation} \label{eq:universality terms-incom-mono-Cauchy elast-8} 
\begin{split}
&\mathcal{B}_{ab}^{818}=\mathfrak{K}^n_a \,(I_{1,n} \,I_{8,b}+I_{1,b} \,I_{8,n})\,,\\
&\mathcal{B}_{ab}^{819}=\mathfrak{K}^n_a \,(I_{1,n} \,I_{9,b}+I_{1,b} \,I_{9,n})\,,\\
&\mathcal{B}_{ab}^{828}=\mathfrak{K}^n_a \,(I_{2,n} \,I_{8,b}+I_{2,b} \,I_{8,n})\,,\\
&\mathcal{B}_{ab}^{829}=\mathfrak{K}^n_a \,(I_{2,n} \,I_{9,b}+I_{2,b} \,I_{9,n})\,,\\
&\mathcal{B}_{ab}^{848}=\mathfrak{K}^n_a \,(I_{4,n} \,I_{8,b}+I_{4,b} \,I_{8,n})\,,\\
&\mathcal{B}_{ab}^{849}=\mathfrak{K}^n_a \,(I_{4,n} \,I_{9,b}+I_{4,b} \,I_{9,n})\,,\\
&\mathcal{B}_{ab}^{858}=\mathfrak{K}^n_a \,(I_{5,n} \,I_{8,b}+I_{5,b} \,I_{8,n})\,,\\
&\mathcal{B}_{ab}^{859}=\mathfrak{K}^n_a \,(I_{5,n} \,I_{9,b}+I_{5,b} \,I_{9,n})\,,\\
&\mathcal{B}_{ab}^{868}=\mathfrak{K}^n_a \,(I_{6,n} \,I_{8,b}+I_{6,b} \,I_{8,n})\,,\\
&\mathcal{B}_{ab}^{869}=\mathfrak{K}^n_a \,(I_{6,n} \,I_{9,b}+I_{6,b} \,I_{9,n})\,,\\
&\mathcal{B}_{ab}^{878}=\mathfrak{K}^n_a \,(I_{7,n} \,I_{8,b}+I_{7,b} \,I_{8,n})\,,\\
&\mathcal{B}_{ab}^{879}=\mathfrak{K}^n_a \,(I_{7,n} \,I_{9,b}+I_{7,b} \,I_{9,n})\,,\\
&\mathcal{B}_{ab}^{888}=\mathfrak{K}^n_a \,I_{8,n} \,I_{8,b}\,,\\
&\mathcal{B}_{ab}^{889}=\mathfrak{K}^n_a \,(I_{8,n} \,I_{9,b}+I_{8,b} \,I_{9,n})\,,\\
&\mathcal{B}_{ab}^{899}=\mathfrak{K}^n_a \,I_{9,n} \,I_{9,b}\,,\\
\end{split}
\end{equation}
and
\begin{equation} \label{eq:universality terms-incom-mono-Cauchy elast-9} 
\begin{split}
&\mathcal{B}_{ab}^{918}=\bar{\mathfrak{K}}^n_a \,(I_{1,n} \,I_{8,b}+I_{1,b} \,I_{8,n})\,,\\
&\mathcal{B}_{ab}^{919}=\bar{\mathfrak{K}}^n_a \,(I_{1,n} \,I_{9,b}+I_{1,b} \,I_{9,n})\,,\\
&\mathcal{B}_{ab}^{928}=\bar{\mathfrak{K}}^n_a \,(I_{2,n} \,I_{8,b}+I_{2,b} \,I_{8,n})\,,\\
&\mathcal{B}_{ab}^{929}=\bar{\mathfrak{K}}^n_a \,(I_{2,n} \,I_{9,b}+I_{2,b} \,I_{9,n})\,,\\
&\mathcal{B}_{ab}^{948}=\bar{\mathfrak{K}}^n_a \,(I_{4,n} \,I_{8,b}+I_{4,b} \,I_{8,n})\,,\\
&\mathcal{B}_{ab}^{949}=\bar{\mathfrak{K}}^n_a \,(I_{4,n} \,I_{9,b}+I_{4,b} \,I_{9,n})\,,\\
&\mathcal{B}_{ab}^{958}=\bar{\mathfrak{K}}^n_a \,(I_{5,n} \,I_{8,b}+I_{5,b} \,I_{8,n})\,,\\
&\mathcal{B}_{ab}^{959}=\bar{\mathfrak{K}}^n_a \,(I_{5,n} \,I_{9,b}+I_{5,b} \,I_{9,n})\,,\\
&\mathcal{B}_{ab}^{968}=\bar{\mathfrak{K}}^n_a \,(I_{6,n} \,I_{8,b}+I_{6,b} \,I_{8,n})\,,\\
&\mathcal{B}_{ab}^{969}=\bar{\mathfrak{K}}^n_a \,(I_{6,n} \,I_{9,b}+I_{6,b} \,I_{9,n})\,,\\
&\mathcal{B}_{ab}^{978}=\bar{\mathfrak{K}}^n_a \,(I_{7,n} \,I_{8,b}+I_{7,b} \,I_{8,n})\,,\\
&\mathcal{B}_{ab}^{979}=\bar{\mathfrak{K}}^n_a \,(I_{7,n} \,I_{9,b}+I_{7,b} \,I_{9,n})\,,\\
&\mathcal{B}_{ab}^{988}=\bar{\mathfrak{K}}^n_a \,I_{8,n} \,I_{8,b}\,,\\
&\mathcal{B}_{ab}^{989}=\bar{\mathfrak{K}}^n_a \,(I_{8,n} \,I_{9,b}+I_{8,b} \,I_{9,n})\,,\\
&\mathcal{B}_{ab}^{999}=\bar{\mathfrak{K}}^n_a \,I_{9,n} \,I_{9,b}\,,\\
\end{split}
\end{equation}
and
\begin{equation} \label{eq:universality terms-incom-mono-Cauchy elast-10} 
\begin{split}
&\mathcal{B}_{ab}^{10 \,18}=\mathfrak{q}^n_a \,(I_{1,n} \,I_{8,b}+I_{1,b} \,I_{8,n})\,,\\
&\mathcal{B}_{ab}^{10 \,19}=\mathfrak{q}^n_a \,(I_{1,n} \,I_{9,b}+I_{1,b} \,I_{9,n})\,,\\
&\mathcal{B}_{ab}^{10 \,28}=\mathfrak{q}^n_a \,(I_{2,n} \,I_{8,b}+I_{2,b} \,I_{8,n})\,,\\
&\mathcal{B}_{ab}^{10 \,29}=\mathfrak{q}^n_a \,(I_{2,n} \,I_{9,b}+I_{2,b} \,I_{9,n})\,,\\
&\mathcal{B}_{ab}^{10 \,48}=\mathfrak{q}^n_a \,(I_{4,n} \,I_{8,b}+I_{4,b} \,I_{8,n})\,,\\
&\mathcal{B}_{ab}^{10 \,49}=\mathfrak{q}^n_a \,(I_{4,n} \,I_{9,b}+I_{4,b} \,I_{9,n})\,,\\
&\mathcal{B}_{ab}^{10 \,58}=\mathfrak{q}^n_a \,(I_{5,n} \,I_{8,b}+I_{5,b} \,I_{8,n})\,,\\
&\mathcal{B}_{ab}^{10 \,59}=\mathfrak{q}^n_a \,(I_{5,n} \,I_{9,b}+I_{5,b} \,I_{9,n})\,,\\
&\mathcal{B}_{ab}^{10 \,68}=\mathfrak{q}^n_a \,(I_{6,n} \,I_{8,b}+I_{6,b} \,I_{8,n})\,,\\
&\mathcal{B}_{ab}^{10 \,69}=\mathfrak{q}^n_a \,(I_{6,n} \,I_{9,b}+I_{6,b} \,I_{9,n})\,,\\
&\mathcal{B}_{ab}^{10 \,78}=\mathfrak{q}^n_a \,(I_{7,n} \,I_{8,b}+I_{7,b} \,I_{8,n})\,,\\
&\mathcal{B}_{ab}^{10 \,79}=\mathfrak{q}^n_a \,(I_{7,n} \,I_{9,b}+I_{7,b} \,I_{9,n})\,,\\
&\mathcal{B}_{ab}^{10 \,88}=\mathfrak{q}^n_a \,I_{8,n} \,I_{8,b}\,,\\
&\mathcal{B}_{ab}^{10 \,89}=\mathfrak{q}^n_a \,(I_{8,n} \,I_{9,b}+I_{8,b} \,I_{9,n})\,,\\
&\mathcal{B}_{ab}^{10 \,99}=\mathfrak{q}^n_a \,I_{9,n} \,I_{9,b}\,,
\end{split}
\end{equation}
where $\bar{\mathfrak{L}}^{ab}=\bar{\ell}_{1}^{ab}$ and $\bar{\mathfrak{K}}^{ab}=\bar{\ell}_{2}^{ab}$. Thus, there are $160$ additional universality terms for incompressible monoclinic Cauchy elastic solids.

First consider the terms $\mathcal{A}_{ab}^ \kappa$ and $\mathcal{B}_{ab}^ \kappa$, where $\kappa$ is a three-component index. From \eqref{eq:universality terms-incom-mono-hyper-2}--\eqref{eq:universality terms-incom-mono-hyper-5} and \eqref{eq:universality terms-incom-mono-Cauchy elast-2}--\eqref{eq:universality terms-incom-mono-Cauchy elast-10}, we have
\begin{equation} \label{eq:A199,B199-relation-incomp-mono} 
\begin{split}
&\mathcal{A}_{ab}^{199}=\mathcal{B}_{ab}^{199} \,,\\
&\mathcal{A}_{ab}^{119}=\mathcal{B}_{ab}^{119} \,, \\
&\mathcal{A}_{ab}^{499}=\mathcal{B}_{ab}^{499} \,,\\
&\mathcal{A}_{ab}^{699}=\mathcal{B}_{ab}^{799} \,, \\
&\mathcal{A}_{ab}^{489}=\mathcal{B}_{ab}^{489}+\mathcal{B}_{ab}^{10 \,49} \,,\\
&\mathcal{A}_{ab}^{689}=\mathcal{B}_{ab}^{789}+\mathcal{B}_{ab}^{10 \,69} \,, \\
&\mathcal{A}_{ab}^{188}=\mathcal{B}_{ab}^{188}+\mathcal{B}_{ab}^{10 \,18} \,, \\
&\mathcal{A}_{ab}^{118}=\mathcal{B}_{ab}^{118} \,.
\end{split}
\end{equation}
The symmetry of $\mathcal{A}_{ab}^{199}$ or $\mathcal{B}_{ab}^{199}$ asserts that $\nabla I_9$ is an eigenvector of $b_a^n$ (we assume that $I_i \,(i=1,2,4,5,6,7,8,9)$ are not constant), i.e., $b^n_a \,I_{9,n}=\lambda_9 \,I_{9,a}$, where $\lambda_9$ is the corresponding eigenvalue. Substituting this expression into $\mathcal{A}_{ab}^{119}$ or $\mathcal{B}_{ab}^{119}$ gives
\begin{equation} \label{eq:A199-incomp-mono} 
\mathcal{A}_{ab}^{119}=\mathcal{B}_{ab}^{119}= \lambda_1 \,I_{1,a} \,I_{9,b}+\lambda_9 \,I_{9,a} \,I_{1,b} \,,
\end{equation}
which is symmetric only when $\lambda_1=\lambda_9$, or equivalently, $\nabla I_1$ and $\nabla I_9$ are parallel. Thus, $I_9$ and $I_i \,(i=1,2,4,5,6,7)$ are functionally dependent. Moreover, the symmetry of the term $\mathcal{A}_{ab}^{499}$ or $\mathcal{B}_{ab}^{499}$ is preserved if $\mathbf{n}$ and $\nabla I_9$ are either parallel or orthogonal. Referring to \citep{YavariGoriely2021}, we find that the former case is inadmissible, and thus $\mathbf{n} \perp \nabla I_9$. A similar line of reasoning shows that $\mathbf{m}$ is orthogonal to $\nabla I_9$, ensuring that the term $\mathcal{A}_{ab}^{699}$ or $\mathcal{B}_{ab}^{799}$ remains symmetric. Let us now consider the term $\mathcal{B}_{ab}^{10 \,49}$. Since $n^n \,I_{9,n}=m^n \,I_{9,n}=0$ and $\nabla I_9$ and $\nabla I_4$ are parallel, it follows that $n^n \,I_{4,n}=m^n \,I_{4,n}=0$. Hence 
\begin{equation} \label{eq:B1049-incomp-mono} 
\mathcal{B}_{ab}^{10 \,49}= \mathcal{g} \,(n_a \,m^n \,I_{4,n} \,I_{9,b} +n_a \,m^n \,I_{9,n} \,I_{4,b}+m_a \,n^n \,I_{4,n} \,I_{9,b} +m_a \,n^n \,I_{9,n} \,I_{4,b})=0 \,, 
\end{equation}
in which we used $\mathfrak{q}^n_a=\mathcal{g} \,(n_a \,m^n + m_a \,n^n)$. Therefore, \eqref{eq:A199,B199-relation-incomp-mono}$_5$ becomes
\begin{equation} \label{eq:A489,B489-relation-incomp-mono} 
\mathcal{A}_{ab}^{489}=\mathcal{B}_{ab}^{489}=n_a \,n^n \,I_{8,n} \,I_{9,b}\,.
\end{equation}
Because $\mathbf{n}$ and $\nabla I_9$ are not parallel, the term $\mathcal{A}_{ab}^{489}$ or $\mathcal{B}_{ab}^{489}$ is symmetric only if $n^n \,I_{8,n}=0$, i.e., $\mathbf{n} \perp \nabla I_8$. Similarly, $\mathcal{B}_{ab}^{10 \,69}=0$ and \eqref{eq:A199,B199-relation-incomp-mono}$_6$ is written as
\begin{equation} \label{eq:A689,B789-relation-incomp-mono} 
\mathcal{A}_{ab}^{689}=\mathcal{B}_{ab}^{789}=m_a \,m^n \,I_{8,n} \,I_{9,b} \,.
\end{equation}
The symmetry of $\mathcal{A}_{ab}^{689}$ or $\mathcal{B}_{ab}^{789}$ then gives $\mathbf{m} \perp \nabla I_8$. Therefore, $\mathfrak{q}^n_a \,I_{8,n}=0$, and consequently $\mathcal{B}_{ab}^{10 \,18}=0$. Hence \eqref{eq:A199,B199-relation-incomp-mono}$_7$ is simplified to read
\begin{equation} \label{eq:A188,B188-relation-incomp-mono} 
\mathcal{A}_{ab}^{188}=\mathcal{B}_{ab}^{188}=b_a^n \,I_{8,n} \,I_{8,b} \,,
\end{equation}
which implies that $b^n_a \,I_{8,n}=\lambda_8 \,I_{8,a}$, where $\lambda_8$ is the corresponding eigenvalue. Given this result, we have
\begin{equation} \label{eq:A118,B118-relation-incomp-mono} 
\mathcal{A}_{ab}^{118}=\mathcal{B}_{ab}^{118}=\lambda_1 \,I_{1,a} \,I_{8,b}+\lambda_8 \,I_{8,a} \,I_{1,b} \,,
\end{equation}
which is symmetric only if $\nabla I_1$ and $\nabla I_8$ are parallel. 

To summarize, we showed that the symmetries of the terms $\{\mathcal{A}_{ab}^{199},\mathcal{A}_{ab}^{119},\mathcal{A}_{ab}^{499},\mathcal{A}_{ab}^{699},\mathcal{A}_{ab}^{489},\mathcal{A}_{ab}^{689},\mathcal{A}_{ab}^{188},\mathcal{A}_{ab}^{118}\}$ in hyperelasticity are equivalent to those of $\{\mathcal{B}_{ab}^{199},\mathcal{B}_{ab}^{119},\mathcal{B}_{ab}^{499},\mathcal{B}_{ab}^{799},\mathcal{B}_{ab}^{489},\mathcal{B}_{ab}^{689},\mathcal{B}_{ab}^{188},\mathcal{B}_{ab}^{118}\}$ in Cauchy elasticity. 
Both sets are symmetric if $\nabla I_8$ and $\nabla I_9$ are parallel to $\nabla I_i \,(i=1,2,4,5,6,7)$, and are orthogonal to $\mathbf{n}$ and $\mathbf{m}$. Therefore, $\nabla I_i \,(i=1,2,4,5,6,7,8,9)$ are mutually parallel and all orthogonal to $\mathbf{n}$ and $\mathbf{m}$. This conclusion, in turn, leads to
\begin{equation} \label{eq:LIi,a=0-incomp-mono} 
\mathfrak{L}^n_a \,I_{i,n}=\bar{\mathfrak{L}}^n_a \,I_{i,n}=\mathfrak{K}^n_a \,I_{i,n}=\bar{\mathfrak{K}}^n_a \,I_{i,n}=\mathfrak{q}^n_a \,I_{i,n}=0 \,,
\end{equation}
where $i=1,2,4,5,6,7,8,9$. With this conclusion, most of the remaining terms presented in \eqref{eq:universality terms-incom-mono-hyper-2}--\eqref{eq:universality terms-incom-mono-hyper-5} and \eqref{eq:universality terms-incom-mono-Cauchy elast-2}--\eqref{eq:universality terms-incom-mono-Cauchy elast-10} are zero. The non-vanishing terms are described by $b^n_a \,I_{i,n} \,I_{8,b}=\lambda_1 \,I_{i,a} \,I_{8,b}$ (or $b^n_a \,I_{i,n} \,I_{9,b}=\lambda_1 \,I_{i,a} \,I_{9,b}$), and $c^n_a \,I_{i,n} \,I_{8,b}=\dfrac{1}{\lambda_1} \,I_{i,a} \,I_{8,b}$ (or $c^n_a \,I_{i,n} \,I_{9,b}=\dfrac{1}{\lambda_1} \,I_{i,a} \,I_{9,b}$) which are symmetric owing to the functional dependence of $I_i \,(i=1,2,4,5,6,7,8,9)$. Therefore, the symmetries of the remaining terms trivially hold, and the Cauchy elasticity and hyperelasticity terms are equivalent in this case.

We now turn our attention to the terms $\mathcal{A}_{ab}^\kappa$ and $\mathcal{B}_{ab}^\kappa$, where $\kappa$ is a two-component index, as given by \eqref{eq:universality terms-incom-mono-hyper-1}$_2$--\eqref{eq:universality terms-incom-mono-hyper-1}$_{15}$ and \eqref{eq:universality terms-incom-mono-Cauchy elast-1}$_2$--\eqref{eq:universality terms-incom-mono-Cauchy elast-1}$_{25}$ for hyperelastic and Cauchy elastic solids, respectively. Referring to \citep{YavariGoriely2021}, one may find that 
\begin{equation}  \label{eq:elln=0-incomp-mono}
{\mathfrak{L}^{n}_{a}}_{|n}={\bar{\mathfrak{L}}^{n}_{a}}_{|n}={\mathfrak{K}^{n}_{a}}_{|n}={\bar{\mathfrak{K}}^{n}_{a}}_{|n}={\mathfrak{q}^n_a}_{|n}=0 \,. 
\end{equation}
Based on \eqref{eq:LIi,a=0-incomp-mono} and \eqref{eq:elln=0-incomp-mono}, the terms $\mathcal{B}_{ab}^{58}$, $\mathcal{B}_{ab}^{59}$, $\mathcal{B}_{ab}^{68}$, $\mathcal{B}_{ab}^{69}$, $\mathcal{B}_{ab}^{88}$, $\mathcal{B}_{ab}^{89}$, $\mathcal{B}_{ab}^{98}$, $\mathcal{B}_{ab}^{99}$ and $\mathcal{B}_{ab}^{10 \,i} \,(i=1,2,4,5,6,7,8,9)$ in Cauchy elasticity, as well as $\mathcal{A}_{ab}^{58}$, $\mathcal{A}_{ab}^{59}$, $\mathcal{A}_{ab}^{78}$, $\mathcal{A}_{ab}^{79}$, $\mathcal{A}_{ab}^{88}$ and $\mathcal{A}_{ab}^{89}$ in hyperelasticity vanish. Subsequently, eight terms remain in Cauchy elasticity and eight in hyperelasticity, which correspond to each other through the following relations (see \eqref{eq:universality terms-incom-mono-hyper-1} and \eqref{eq:universality terms-incom-mono-Cauchy elast-1})
\begin{equation} \label{eq:hyper-Cauchy elast-term-relation-incomp-mono} 
\begin{split}
&\mathcal{A}_{ab}^{18}=\mathcal{B}_{ab}^{18} \,,\\
&\mathcal{A}_{ab}^{19}=\mathcal{B}_{ab}^{19} \,,\\
&\mathcal{A}_{ab}^{28}=\mathcal{B}_{ab}^{28} \,,\\
&\mathcal{A}_{ab}^{29}=\mathcal{B}_{ab}^{29} \,,\\
&\mathcal{A}_{ab}^{48}=\mathcal{B}_{ab}^{48} \,,\\
&\mathcal{A}_{ab}^{49}=\mathcal{B}_{ab}^{49} \,,\\
&\mathcal{A}_{ab}^{68}=\mathcal{B}_{ab}^{78} \,,\\
&\mathcal{A}_{ab}^{69}=\mathcal{B}_{ab}^{79} \,.
\end{split}
\end{equation}
Regarding $\mathcal{A}_{ab}^{8} = \mathcal{B}_{ab}^{10} = 0$, one now concludes that the Cauchy elasticity terms in this case are equivalent to those in hyperelasticity. Therefore, for incompressible monoclinic solids, the additional terms appearing in hyperelasticity are equivalent to those in Cauchy elasticity.

In summary, we have proved the following result.
\begin{prop}
The universal deformations and material preferred directions of incompressible monoclinic Cauchy elasticity are identical to those of incompressible monoclinic hyperelasticity.
\end{prop}

\section{Conclusions}  \label{Sec:Conclusions}

In this paper, we analyzed universal deformations in compressible and incompressible anisotropic Cauchy elastic solids. We showed that for transversely isotropic, orthotropic, and monoclinic materials, the sets of universal deformations and universal material preferred directions coincide with those previously obtained in the hyperelastic case. Thus, the existence of an energy function does not affect the form or characterization of universal deformations and material preferred directions in Cauchy elasticity. This result establishes that universal deformations and material preferred directions are independent of whether the constitutive law is derived from a potential. These findings extend and generalize earlier results for isotropic solids to the anisotropic setting. The present analysis provides a foundation for further exploration of universal deformations in more general material frameworks, including non-Cauchy elastic solids, materials with residual stress or microstructure, and generalized continua where additional internal variables appear.

\bibliographystyle{abbrvnat}
\bibliography{ref,ref1}

\begin{thebibliography}{47}
\providecommand{\natexlab}[1]{#1}
\providecommand{\url}[1]{\texttt{#1}}
\expandafter\ifx\csname urlstyle\endcsname\relax
  \providecommand{\doi}[1]{doi: #1}\else
  \providecommand{\doi}{doi: \begingroup \urlstyle{rm}\Url}\fi

\bibitem[Beatty(1978)]{Beatty1978}
M.~F. Beatty.
\newblock General solutions in the equilibrium theory of inextensible elastic
  materials.
\newblock \emph{Acta Mechanica}, 29\penalty0 (1):\penalty0 119--126, 1978.

\bibitem[Beatty(1989)]{Beatty1989}
M.~F. Beatty.
\newblock A class of universal relations for constrained, isotropic elastic
  materials.
\newblock \emph{Acta Mechanica}, 80\penalty0 (3):\penalty0 299--312, 1989.

\bibitem[Beskos(1972)]{Beskos1972}
D.~E. Beskos.
\newblock Universal solutions for fiber-reinforced compressible isotropic
  elastic materials.
\newblock \emph{Journal of Elasticity}, 2\penalty0 (3):\penalty0 153--168,
  1972.

\bibitem[Beskos(1973)]{Beskos1973}
D.~E. Beskos.
\newblock Universal solutions for fiber-reinforced incompressible isotropic
  elastic materials.
\newblock \emph{International Journal of Solids and Structures}, 9\penalty0
  (4):\penalty0 553--567, 1973.

\bibitem[Boehler(1979)]{Boehler1979}
J.-P. Boehler.
\newblock A simple derivation of representations for non-polynomial
  constitutive equations in some cases of anisotropy.
\newblock \emph{Zeitschrift f{\"u}r Angewandte Mathematik und Mechanik (ZAMM)},
  59\penalty0 (4):\penalty0 157--167, 1979.

\bibitem[Boehler(1987)]{Boehler1987}
J.-P. Boehler.
\newblock Representations for isotropic and anisotropic non-polynomial tensor
  functions.
\newblock In \emph{Applications of Tensor Functions in Solid Mechanics}, pages
  31--53. Springer, 1987.

\bibitem[Cauchy(1828)]{Cauchy1828}
A.-L. Cauchy.
\newblock Sur les {\'e}quations qui expriment les conditions d'{\'e}quilibre ou
  les lois du mouvement int{\'e}rieur d'un corps solide, {\'e}lastique ou non
  {\'e}lastique.
\newblock \emph{Exercises de Math{\'e}matiques}, 3:\penalty0 160--187, 1828.

\bibitem[Erdemir and Carroll(2007)]{Erdemir2007}
E.~Erdemir and M.~M. Carroll.
\newblock Finite deformations and motions of radially inextensible hollow
  spheres.
\newblock \emph{Journal of Elasticity}, 88\penalty0 (3):\penalty0 193--205,
  2007.

\bibitem[Ericksen(1954)]{Ericksen1954}
J.~L. Ericksen.
\newblock Deformations possible in every isotropic, incompressible, perfectly
  elastic body.
\newblock \emph{Zeitschrift f{\"u}r Angewandte Mathematik und Physik},
  5\penalty0 (6):\penalty0 466--489, 1954.

\bibitem[Ericksen(1955)]{Ericksen1955}
J.~L. Ericksen.
\newblock Deformations possible in every compressible, isotropic, perfectly
  elastic material.
\newblock \emph{Studies in Applied Mathematics}, 34\penalty0 (1-4):\penalty0
  126--128, 1955.

\bibitem[Fosdick(1966)]{Fosdick1966}
R.~L. Fosdick.
\newblock Remarks on {C}ompatibility.
\newblock \emph{Modern Developments in the Mechanics of Continua}, pages
  109--127, 1966.

\bibitem[Goodbrake et~al.(2020)Goodbrake, Yavari, and Goriely]{Goodbrake2020}
C.~Goodbrake, A.~Yavari, and A.~Goriely.
\newblock The anelastic {E}ricksen problem: {U}niversal deformations and
  universal eigenstrains in incompressible nonlinear anelasticity.
\newblock \emph{Journal of Elasticity}, 142\penalty0 (2):\penalty0 291--381,
  2020.

\bibitem[Goriely(2017)]{Goriely2017}
A.~Goriely.
\newblock \emph{The Mathematics and Mechanics of Biological Growth}, volume~45.
\newblock Springer, 2017.

\bibitem[Gurtin(1972)]{Gurtin1972}
M.~E. Gurtin.
\newblock The linear theory of elasticity.
\newblock In \emph{Handbuch der Physik, Band VIa/2.} Springer-Verlag, Berlin,
  1972.

\bibitem[Klingbeil and Shield(1966)]{KlingbeilShield1966}
W.~W. Klingbeil and R.~T. Shield.
\newblock On a class of solutions in plane finite elasticity.
\newblock \emph{Zeitschrift f{\"u}r angewandte Mathematik und Physik},
  17\penalty0 (4):\penalty0 489--511, 1966.

\bibitem[Merodio and Ogden(2020)]{merodio2020finite}
J.~Merodio and R.~W. Ogden.
\newblock Finite deformation elasticity theory.
\newblock In \emph{Constitutive Modelling of Solid Continua}, pages 17--52.
  Springer, 2020.

\bibitem[Morgan(1966)]{Morgan1966}
A.~J.~A. Morgan.
\newblock Some properties of media defined by constitutive equations in
  implicit form.
\newblock \emph{International Journal of Engineering Science}, 4\penalty0
  (2):\penalty0 155--178, 1966.

\bibitem[Pipkin(1974)]{Pipkin1974}
A.~C. Pipkin.
\newblock Generalized plane deformations of ideal fiber-reinforced materials.
\newblock \emph{Quarterly of Applied Mathematics}, 32\penalty0 (3):\penalty0
  253--263, 1974.

\bibitem[Pipkin(1979)]{Pipkin1979}
A.~C. Pipkin.
\newblock Stress analysis for fiber-reinforced materials.
\newblock In \emph{Advances in Applied Mechanics}, volume~19, pages 1--51.
  Elsevier, 1979.

\bibitem[Pipkin and Rogers(1971)]{Pipkin1971}
A.~C. Pipkin and T.~G. Rogers.
\newblock Plane deformations of incompressible fiber-reinforced materials.
\newblock \emph{Journal of Applied Mechanics}, 38\penalty0 (3):\penalty0
  634--640, 1971.

\bibitem[Rajagopal(2003)]{Rajagopal2003}
K.~R. Rajagopal.
\newblock On implicit constitutive theories.
\newblock \emph{Applications of Mathematics}, 48:\penalty0 279--319, 2003.

\bibitem[Rajagopal(2007)]{Rajagopal2007}
K.~R. Rajagopal.
\newblock The elasticity of elasticity.
\newblock \emph{Zeitschrift f{\"u}r angewandte Mathematik und Physik},
  58:\penalty0 309--317, 2007.

\bibitem[Rivlin(1948)]{Rivlin1948}
R.~S. Rivlin.
\newblock Large elastic deformations of isotropic materials {IV}. {F}urther
  developments of the general theory.
\newblock \emph{Philosophical Transactions of the Royal Society of London A},
  241\penalty0 (835):\penalty0 379--397, 1948.

\bibitem[Rivlin(1949{\natexlab{a}})]{Rivlin1949a}
R.~S. Rivlin.
\newblock Large elastic deformations of isotropic materials. {V}. {T}he problem
  of flexure.
\newblock \emph{Proceedings of the Royal Society of London A}, 195\penalty0
  (1043):\penalty0 463--473, 1949{\natexlab{a}}.

\bibitem[Rivlin(1949{\natexlab{b}})]{Rivlin1949b}
R.~S. Rivlin.
\newblock A note on the torsion of an incompressible highly elastic cylinder.
\newblock In \emph{Mathematical Proceedings of the Cambridge Philosophical
  Society}, volume~45, pages 485--487. Cambridge University Press,
  1949{\natexlab{b}}.

\bibitem[Rivlin and Saunders(1951)]{Rivlin1951}
R.~S. Rivlin and D.~W. Saunders.
\newblock Large elastic deformations of isotropic materials {VII}.
  {E}xperiments on the deformation of rubber.
\newblock \emph{Philosophical Transactions of the Royal Society of London A},
  243\penalty0 (865):\penalty0 251--288, 1951.

\bibitem[Saccomandi and Beatty(2002)]{Saccomandi2002}
G.~Saccomandi and M.~F. Beatty.
\newblock Universal relations for fiber-reinforced elastic materials.
\newblock \emph{Mathematics and Mechanics of Solids}, 7\penalty0 (1):\penalty0
  95--110, 2002.

\bibitem[Singh and Pipkin(1965)]{SinghPipkin1965}
M.~Singh and A.~C. Pipkin.
\newblock Note on {E}ricksen's problem.
\newblock \emph{Zeitschrift f{\"u}r angewandte Mathematik und Physik},
  16\penalty0 (5):\penalty0 706--709, 1965.

\bibitem[Spencer(1970)]{Spencer1970}
A.~J.~M. Spencer.
\newblock A note on the decomposition of tensors into traceless symmetric
  tensors.
\newblock \emph{International Journal of Engineering Science}, 8\penalty0
  (6):\penalty0 475--481, 1970.
\newblock \doi{10.1016/0020-7225(70)90031-5}.

\bibitem[Spencer(1986)]{Spencer1986}
A.~J.~M. Spencer.
\newblock Modelling of finite deformations of anisotropic materials.
\newblock In \emph{Large Deformations of Solids: Physical Basis and
  Mathematical Modelling}, pages 41--52. Springer, 1986.

\bibitem[Tadmor et~al.(2012)Tadmor, Miller, and Elliott]{Tadmor2012}
E.~B. Tadmor, R.~E. Miller, and R.~S. Elliott.
\newblock \emph{Continuum Mechanics and Thermodynamics: From Fundamental
  Concepts to Governing Equations}.
\newblock Cambridge University Press, 2012.

\bibitem[Truesdell(1952)]{Truesdell1952}
C.~Truesdell.
\newblock The mechanical foundations of elasticity and fluid dynamics.
\newblock \emph{Journal of Rational Mechanics and Analysis}, 1\penalty0
  (1):\penalty0 125--300, 1952.

\bibitem[Truesdell(1966)]{Truesdell1966}
C.~Truesdell.
\newblock \emph{The Elements of Continuum Mechanics}.
\newblock Springer-Verlag, 1966.

\bibitem[Truesdell and Noll(2004)]{TruesdellNoll2004}
C.~Truesdell and W.~Noll.
\newblock \emph{The Non-Linear Field Theories of Mechanics}.
\newblock Springer, 2004.

\bibitem[Yavari(2021)]{Yavari2021}
A.~Yavari.
\newblock Universal deformations in inhomogeneous isotropic nonlinear elastic
  solids.
\newblock \emph{Proceedings of the Royal Society A}, 477\penalty0
  (2253):\penalty0 20210547, 2021.

\bibitem[Yavari(2024)]{Yavari2024}
A.~Yavari.
\newblock Universal deformations and inhomogeneities in isotropic {C}auchy
  elasticity.
\newblock \emph{Proceedings of the Royal Society A}, 2024.
\newblock URL \url{https://arxiv.org/abs/2404.06235}.

\bibitem[Yavari(2025)]{Yavari2025FibersCauchy}
A.~Yavari.
\newblock On universal deformations of compressible {C}auchy elastic solids
  reinforced by inextensible fibers.
\newblock \emph{Journal of the Mechanics and Physics of Solids}, 188:\penalty0
  106340, 2025.
\newblock \doi{10.1016/j.jmps.2025.106340}.

\bibitem[Yavari and Goriely(2016)]{YavariGoriely2016}
A.~Yavari and A.~Goriely.
\newblock The anelastic {E}ricksen problem: {U}niversal eigenstrains and
  deformations in compressible isotropic elastic solids.
\newblock \emph{Proc. R. Soc. A}, 472\penalty0 (2196):\penalty0 20160690, 2016.

\bibitem[Yavari and Goriely(2021)]{YavariGoriely2021}
A.~Yavari and A.~Goriely.
\newblock Universal deformations in anisotropic nonlinear elastic solids.
\newblock \emph{Journal of the Mechanics and Physics of Solids}, 156:\penalty0
  104598, 2021.

\bibitem[Yavari and
  Goriely(2022{\natexlab{a}})]{Yavari2022Anelastic-Universality}
A.~Yavari and A.~Goriely.
\newblock Universality in anisotropic linear anelasticity.
\newblock \emph{Journal of Elasticity}, 150\penalty0 (2):\penalty0 241--259,
  2022{\natexlab{a}}.

\bibitem[Yavari and Goriely(2022{\natexlab{b}})]{YavariGoriely2022}
A.~Yavari and A.~Goriely.
\newblock The universal program of linear elasticity.
\newblock \emph{Mathematics and Mechanics of Solids}, 2022{\natexlab{b}}.

\bibitem[Yavari and Goriely(2023{\natexlab{a}})]{Yavari2022Universal}
A.~Yavari and A.~Goriely.
\newblock The universal program of nonlinear hyperelasticity.
\newblock \emph{Journal of Elasticity}, 154\penalty0 (1):\penalty0 91--146,
  2023{\natexlab{a}}.

\bibitem[Yavari and Goriely(2023{\natexlab{b}})]{YavariGoriely2023Universal}
A.~Yavari and A.~Goriely.
\newblock The universal program of nonlinear hyperelasticity.
\newblock \emph{Journal of Elasticity}, 154\penalty0 (1):\penalty0 91--146,
  2023{\natexlab{b}}.

\bibitem[Yavari and Goriely(2024)]{Yavari2024ImplicitElasticity}
A.~Yavari and A.~Goriely.
\newblock Controllable deformations in compressible isotropic implicit
  elasticity.
\newblock \emph{Zeitschrift f{\"u}r angewandte Mathematik und Physik},
  75\penalty0 (5):\penalty0 169, 2024.

\bibitem[Yavari and Goriely(2025)]{YavariGoriely2025}
A.~Yavari and A.~Goriely.
\newblock Nonlinear {C}auchy elasticity.
\newblock \emph{Archive for Rational Mechanics and Analysis}, 249\penalty0
  (57), 2025.
\newblock \doi{10.1007/s00205-025-02120-0}.

\bibitem[Yavari and Sfyris(2025)]{YavariSfyris2025}
A.~Yavari and D.~Sfyris.
\newblock Universal displacements in anisotropic linear {C}auchy elasticity.
\newblock \emph{Journal of Elasticity}, 157\penalty0 (1):\penalty0 1--15, 2025.

\bibitem[Yavari et~al.(2020)Yavari, Goodbrake, and Goriely]{Yavari2020}
A.~Yavari, C.~Goodbrake, and A.~Goriely.
\newblock Universal displacements in linear elasticity.
\newblock \emph{Journal of the Mechanics and Physics of Solids}, 135:\penalty0
  103782, 2020.

\end{thebibliography}

\end{document}